\documentclass[a4paper,10pt]{article}

\usepackage{geometry}
\usepackage[utf8x]{inputenc}
\usepackage[T1]{fontenc}
\usepackage{amsmath}
\usepackage{amstext}
\usepackage{amsfonts}
\usepackage{amssymb}
\usepackage[fixamsmath]{mathtools}
\usepackage{bm}
\usepackage{dsfont}
\usepackage{units}
\usepackage{graphicx}
\usepackage{booktabs}
\usepackage{multirow}
\usepackage[capitalize]{cleveref}
\usepackage[subrefformat=parens]{subcaption}

\usepackage[numbers,sort&compress]{natbib}
\usepackage{authblk}

\newcommand{\perfect}[1]{\tilde{#1}}
\newcommand{\cond}[1][\:\!]{{#1}|{#1}}
\newcommand{\without}{\sim}
\newcommand{\coeffM}{a}
\newcommand{\coeffL}{b}
\newcommand{\coeffH}{c}
\newcommand{\coeffMu}{d}
\newcommand{\coeffVar}{e}
\newcommand{\basis}{\Psi}
\newcommand{\basisU}{\psi}
\newcommand{\scale}{Z}
\newcommand{\dimParam}{M}
\newcommand{\dimData}{N}
\newcommand{\auxQuantity}{\mathcal{G}}
\newcommand{\newBase}{g}
\newcommand{\coeffBaseL}{b^{g}}
\newcommand{\coeffBaseH}{c^{g}}
\newcommand{\basisBase}{\Psi^{g}}

\newlength{\figHeight}	
\newlength{\figWidth}	
\newlength{\femWidth}	
\newlength{\subWidth}	

\title{Spectral likelihood expansions for Bayesian inference}
\author[1]{Joseph B.\ Nagel\thanks{nagel@ibk.baug.ethz.ch}}
\author[1]{Bruno Sudret\thanks{sudret@ibk.baug.ethz.ch}}
\affil[1]{
ETH Z\"{u}rich, Institute of Structural Engineering \par
Chair of Risk, Safety \& Uncertainty Quantification \par
Stefano-Franscini-Platz 5, CH-8093 Z\"{u}rich, Switzerland
}
\date{\today}

\begin{document}

\maketitle

\begin{abstract}
A spectral approach to Bayesian inference is presented.
It pursues the emulation of the posterior probability density.
The starting point is a series expansion of the likelihood function in terms of orthogonal polynomials.
From this spectral likelihood expansion all statistical quantities of interest can be calculated semi-analytically.
The posterior is formally represented as the product of a reference density and a linear combination of polynomial basis functions.
Both the model evidence and the posterior moments are related to the expansion coefficients.
This formulation avoids Markov chain Monte Carlo simulation and allows one to make use of linear least squares instead.
The pros and cons of spectral Bayesian inference are discussed and demonstrated on the basis of simple applications from classical statistics and inverse modeling.
\\[1em] \textit{Keywords:} Uncertainty quantification, Bayesian inference, Inverse problem, Surrogate modeling, Polynomial chaos, Posterior expansion
\end{abstract}

\section{Introduction}
In view of inverse modeling \cite{Inversion:Tarantola2005,Inversion:Kaipio2005} and uncertainty quantification \cite{Uncertainty:Smith2014,Uncertainty:Sullivan2015},
Bayesian inference establishes a convenient framework for the data analysis of engineering systems \cite{Bayesian:Hadidi2008,Bayesian:Beck2010,Bayesian:Yuen2011}.
It adopts probability theory in order to represent, propagate and update epistemic parameter uncertainties.
The prior distribution captures the uncertainty of unknown model parameters before the data are analyzed.
A posterior is then constructed as an updated distribution that captures the remaining uncertainty after the data have been processed.
The computation of this posterior is the primary task in Bayesian inference.
\par 
Some simplified statistical models admit closed-form expressions of the posterior density.
Beyond these so-called conjugate cases,
computational approaches either aim at evaluating expectation values under the posterior or drawing samples from it \cite{Bayesian:Evans1995}.
This is usually accomplished with stochastic methods such as Markov chain Monte Carlo \cite{MCMC:Gilks1996,MCMC:Brooks2011}.
Nowadays this class of techniques constitutes the mainstay of Bayesian computations.
The posterior is explored by realizing an appropriate Markov chain over the prior support that exhibits the posterior as its long-run distribution.
In turn, the obtained sample is used to empirically approximate the statistical quantities of interest.
These include the characteristics of the posterior and the predictive distribution.
This way of proceeding suffers from some inherent deficiencies.
The presence of sample autocorrelation and the absence of a convergence criterion cause severe practical problems.
Moreover, Markov chain Monte Carlo typically requires a large number of serial forward model runs.
Since in engineering applications even a single model run can be computationally taxing, this may be prohibitive.
In the recent past, numerous enhancements have been proposed in order to accelerate Markov chain Monte Carlo for Bayesian inverse problems.
This includes the implementation of more efficient sampling algorithms,
e.g.\ transitional Markov chain Monte Carlo \cite{MCMC:Beck2002,MCMC:Ching2007} or Hamiltonian Monte Carlo \cite{MCMC:Cheung2009,MCMC:Boulkaibet2015,Nagel:JRUES2016},
and the substitution of the forward model with an inexpensive metamodel,
e.g.\ based on Gaussian process models \cite{Kriging:Higdon2004,Kriging:Higdon2015} or polynomial chaos expansions \cite{PCE:Marzouk2007,PCE:Marzouk2009:a,PCE:Marzouk2009:b}.
Although these approaches promise significant speedups, they still inherit all principle shortcomings of sample-based posterior representations.
\par 
Unfortunately there are only few fundamental alternatives to stochastic sampling.
Variational Bayesian inference establishes such an alternative where the posterior is sought through deterministic optimization \cite{Bayesian:Ormerod2010,Bayesian:Fox2012,Bayesian:Sun2013}.
In particular, a member from a simple parametric family of probability densities is selected such that some distance to the posterior is minimized.
In this regard, the Kullback-Leibler divergence is often chosen as a relative measure of the dissimilarity of two probability densities.
The procedure commonly rests upon some simplifying independence assumptions.
Variational methods are regarded as less computing intensive than Markov chain Monte Carlo, yet they are only approximate.
They are prominently used in machine learning and computer science \cite{Bayesian:Jordan1999,Bayesian:Jaakkola2000},
and since recently such methods are applied to inverse problems \cite{Bayesian:Chappell2009,Bayesian:Jin2010}, too.
A particularly interesting implementation of variational Bayesian inference has been proposed in \cite{Mapping:ElMoselhy2012}.
The posterior is parametrized as a transformation of the prior density and can be computed based on the corresponding back-transformation.
More specifically, a random variable transformation is sought in polynomial form
such that the Kullback-Leibler divergence between the prior density and the back-transformed posterior density is minimized.
This formulation is supported by arguments from optimal transport theory which also allows for a practical regularization of the problem.
Finally, samples from the posterior distribution are obtained by independently sampling from the prior and applying the polynomial map.
Another approach to certain Bayesian inverse problems has been recently devised in \cite{Bayesian:Schwab2012,Bayesian:Schillings2013}.
Based on monomial Taylor expansions of the forward model and of the posterior density, the computation of expectation values under the posterior is tackled by sparse numerical quadrature.
\par 
In this paper we propose a novel approach to surrogate the posterior probability density in itself.
The main idea is to decompose the likelihood function into a series of polynomials that are orthogonal with respect to the prior density.
It is shown that all statistical quantities of interest can then be easily extracted from this spectral likelihood expansion.
Emulators of the joint posterior density and its marginals are derived as the product of the prior,
that functions as the reference density, and a linear combination of polynomials, that acts as an adjustment.
In doing so, the model evidence simply emerges as the coefficient of the constant term in the expansion.
Moreover, closed-form expressions for the first posterior moments in terms of the low-order expansion coefficients are given.
The propagation of the posterior uncertainty through physical models can be easily accomplished based on a further postprocessing of the expansion coefficients.
In this sense, spectral Bayesian inference is semi-analytic.
While the corrections required for an expansion of the posterior with respect to the prior as the reference density may be large,
they can be small for an expansion around a properly chosen auxiliary density.
A change of the reference density is therefore suggested in order to increase the efficiency of computing a posterior surrogate.
The devised formulation entirely avoids Markov chain Monte Carlo.
Instead it draws on the machinery of spectral methods \cite{Math:Boyd2001,Math:Kopriva2009,Math:Shen2011} and approximation theory \cite{Math:Christensen2004,Math:Trigub2004,Math:Trefethen2013}.
It is proposed to compute the expansion coefficients via linear least squares \cite{Statistics:Lawson1995,Statistics:Bjorck1996}.
This allows one to make use of the wealth of statistical learning methods \cite{Statistics:Vapnik2000,Statistics:Hastie2009} that are designed for this type of problems.
The approach features a natural convergence criterion and it is amenable to parallel computing.
\par 
The scope of applicability of the proposed approach covers problems from Bayesian inference for which the likelihood function can be evaluated
and for which polynomials can be constructed that are orthogonal with respect to the prior, possibly after a carefully chosen variable transformation.
This excludes statistical models that involve intractable likelihoods \cite{Bayesian:Marin2012,Bayesian:Sunnaker2013}, i.e.\ the likelihood cannot be exactly evaluated.
It also excludes improper prior distributions \cite{Bayesian:Taraldsen2010,Bayesian:Kitanidis2012}, i.e.\ the prior does not integrate to one or any finite value,
and models with pathologic priors such as the Cauchy distribution for which the moments are not defined \cite{Bayesian:Gelman2008,Bayesian:Fuquene2009}.
Many hierarchical Bayesian models \cite{Nagel:JAIS2015,Nagel:PEM2016} are not covered by the devised problem formulation.
They are either based on conditional priors, which does not allow for orthogonal polynomials, or on integrated likelihoods, which can only be evaluated subject to noise.
\par 
Spectral likelihood expansions complement the existing array of Bayesian methods with a way of surrogating the posterior density directly.
They have the potential to remedy at least some of the shortcomings of Markov chain Monte Carlo.
Yet, their practical implementation poses challenges.
Hence, the goal of this paper is to discuss and investigate the possibilities and limitations of the approach.
The method of spectral likelihood expansions is therefore applied to well-known calibration problems from classical statistics and inverse heat conduction.
We restrict the analysis to low-dimensional problems.
The final results are compared with corresponding results from Markov chain Monte Carlo simulations.
\par 
The manuscript is structured as follows.
The principles of Bayesian inference are summarized in \cref{sec:Bayesian}.
Surrogate forward modeling with polynomial chaos expansions is reviewed in \cref{sec:PCE}.
After that, spectral likelihood expansions are introduced as an alternative approach to Bayesian inference in \cref{sec:SLE}.
Two well-known Gaussian fitting examples and the identification of thermal properties of a composite material serve as numerical demonstrations in \cref{sec:Examples}.
Finally, it is summarized and concluded in \cref{sec:Conclusion}.

\section{Bayesian inference} \label{sec:Bayesian}
Let \(\bm{x} = (x_1,\ldots,x_\dimParam)^\top \in \mathcal{D}_{\bm{x}}\)
with \(\mathcal{D}_{\bm{x}} = \mathcal{D}_{x_1} \times \ldots \times \mathcal{D}_{x_\dimParam} \subset \mathds{R}^\dimParam\) be a vector of unknown parameters.
The goal of statistical inference is to deduce these unknowns from the observed data \(\bm{y} = (y_1,\ldots,y_\dimData)^\top \in \mathds{R}^\dimData\).
In Bayesian statistics one adapts probabilistic models for representing uncertainty.
Hence, let \((\Omega,\mathcal{F},\mathcal{P})\) be a suitable probability space with a sample space \(\Omega\), a \(\sigma\)-field \(\mathcal{F}\) and a probability measure \(\mathcal{P}\).
On this space one can define a prior model of the unknowns and an observational model of the data
that represent the encountered parameter uncertainties and the experimental situation, respectively.
\par 
The epistemic uncertainty of the parameter values is cast as a \(\mathcal{D}_{\bm{x}}\)-valued random vector \(\bm{X} \colon \Omega \rightarrow \mathcal{D}_{\bm{x}} \subset \mathds{R}^\dimParam\).
Here, the components of \(\bm{X} = (X_1,\ldots,X_\dimParam)^\top\) are \(\mathcal{D}_{x_i}\)-valued random variables
\(X_i \colon \Omega \rightarrow \mathcal{D}_{x_i} \subset \mathds{R}\) for \(i = 1,\ldots,\dimParam\).
Since the data have not been processed at this stage, the joint density of \(\bm{X} \sim \pi(\bm{x})\) is called the \emph{prior density}.
Similarly, a \(\mathds{R}^\dimData\)-valued random vector \(\bm{Y} \colon \Omega \rightarrow \mathds{R}^\dimData\) represents the observables.
The components of \(\bm{Y} = (Y_1,\ldots,Y_\dimData)^\top\) are real-valued random variables \(Y_i \colon \Omega \rightarrow \mathds{R}\) for \(i = 1,\ldots,\dimData\).
In order to draw inferences from the data about the unknowns, one has to formulate an observational model that establishes a relationship between those quantities.
Commonly this is a probabilistic representation of the observables \(\bm{Y} \cond \bm{x} \sim f(\bm{y} \cond \bm{x})\) that is conditional on the unknown parameters.
For the actually acquired data \(\bm{Y} = \bm{y}\), the \emph{likelihood function} \(\mathcal{L}(\bm{x}) = f(\bm{y} \cond \bm{x})\)
is defined by interpreting the conditional density \(f(\bm{y} \cond \bm{x})\) as a function of the unknowns \(\bm{x}\).
\par 
Given this setup, one can formulate an updated probability density \(\pi(\bm{x} \cond \bm{y})\) of the unknowns that is conditioned on the realized data.
This so-called \emph{posterior density} results from \emph{Bayes' law}
\begin{equation} \label{eq:Bayesian:Posterior}
  \pi(\bm{x} \cond \bm{y}) = \frac{\mathcal{L}(\bm{x}) \pi(\bm{x})}{\scale}.
\end{equation}
It completely summarizes the available information about the unknowns after the data have been analyzed.
The \emph{model evidence} \(\scale\) properly normalizes the posterior density.
It can be written as
\begin{equation} \label{eq:Bayesian:ScaleFactor}
  \scale = \int\limits_{\mathcal{D}_{\bm{x}}} \mathcal{L}(\bm{x}) \, \pi(\bm{x}) \, \mathrm{d} \bm{x}.
\end{equation}
\par 
One is often interested in the marginals and moments of the posterior.
The posterior marginal \(\pi(x_j \cond \bm{y})\) of a single unknown \(x_j\) with \(j \in \{1,\ldots,\dimParam\}\) is defined as
\begin{equation} \label{eq:Bayesian:Marginal1D}
  \pi(x_j \cond \bm{y}) = \int\limits_{\mathcal{D}_{\bm{x}_{\without j}}} \pi(\bm{x} \cond \bm{y}) \, \mathrm{d} \bm{x}_{\without j}.
\end{equation}
Here, the simplifying notation \(\bm{x}_{\without j} = (x_1,\ldots,x_{j-1},x_{j+1},\ldots,x_\dimParam)^\top\) is introduced.
For the mean \(\mathds{E}[\bm{X} \cond \bm{y}]\) and the covariance matrix \(\mathrm{Cov}[\bm{X} \cond \bm{y}]\) of the posterior one has
\begin{gather} 
  \mathds{E}[\bm{X} \cond \bm{y}] = \int\limits_{\mathcal{D}_{\bm{x}}} \bm{x} \, \pi(\bm{x} \cond \bm{y}) \, \mathrm{d} \bm{x}, \label{eq:Bayesian:PosteriorMean} \\
  \mathrm{Cov}[\bm{X} \cond \bm{y}] = \int\limits_{\mathcal{D}_{\bm{x}}} \left( \bm{x} - \mathds{E}[\bm{X} \cond \bm{y}]) (\bm{x} - \mathds{E}[\bm{X} \cond \bm{y}] \right)^\top
  \pi(\bm{x} \cond \bm{y}) \, \mathrm{d} \bm{x}. \label{eq:Bayesian:PosteriorCovariance}
\end{gather}
\par 
More generally, Bayesian inference focuses the computation of posterior expectation values of the \emph{quantities of interest} (QoI) \(h \colon \mathcal{D}_{\bm{x}} \rightarrow \mathds{R}\).
These expectations may be formally expressed as
\begin{equation} \label{eq:Bayesian:QoI}
  \mathds{E}[h(\bm{X}) \cond \bm{y}] = \int\limits_{\mathcal{D}_{\bm{x}}} h(\bm{x}) \, \pi(\bm{x} \cond \bm{y}) \, \mathrm{d} \bm{x}.
\end{equation}
For later considerations, it is remarked that this integration over the posterior density can be interpreted as a reweighted integration over the prior density
\begin{equation} \label{eq:Baysian:KeyIdentity}
  \mathds{E}[h(\bm{X}) \cond \bm{y}]
  = \int\limits_{\mathcal{D}_{\bm{x}}} h(\bm{x}) \frac{\mathcal{L}(\bm{x})}{\scale} \pi(\bm{x}) \, \mathrm{d} \bm{x}
  = \frac{1}{\scale} \mathds{E}[h(\bm{X}) \mathcal{L}(\bm{X})].
\end{equation}

\subsection{Bayesian inverse problems}
The Bayesian framework described above can be applied to a vast range of scenarios from classical statistics
\cite{Bayesian:Jackman2009,Bayesian:Gelman2014:3rd} and inverse modeling \cite{Bayesian:Stuart2010,Bayesian:Ernst2014}.
In inverse problems, a so-called \emph{forward model} \(\mathcal{M}\) establishes a mathematical representation of the physical system under consideration.
It is the function
\begin{equation} \label{eq:Bayesian:Inverse:ForwardModel}
  \begin{aligned}
    \mathcal{M} \colon \mathcal{D}_{\bm{x}} &\rightarrow \mathds{R}^\dimData \\
    \bm{x} &\mapsto \mathcal{M}(\bm{x})
  \end{aligned}
\end{equation}
that maps model inputs \(\bm{x} \in \mathcal{D}_{\bm{x}} \subset \mathds{R}^\dimParam\) to outputs \(\perfect{\bm{y}} = \mathcal{M}(\bm{x}) \in \mathds{R}^\dimData\).
\emph{Inversion} is the process of inferring the unknown forward model parameters \(\bm{x}\) with the measured data \(\bm{y}\) of its response.
\par 
A probabilistic model of the observables is commonly constructed supposing that they can be represented as the sum \(\bm{Y} = \perfect{\bm{Y}} + \bm{E}\)
of the model response vector \(\perfect{\bm{Y}} = \mathcal{M}(\bm{X}) \colon \Omega \rightarrow \mathds{R}^\dimData\) and another random vector \(\bm{E} \colon \Omega \rightarrow \mathds{R}^\dimData\).
The latter accounts for measurement noise and forward model inadequacy.
It is assumed that the \emph{residual vector} \(\bm{E}\) is statistically independent from \(\bm{X}\).
An unknown realization \(\bm{E} = \bm{\varepsilon}\) measures the discrepancy between the actually measured data
\(\bm{y} = \perfect{\bm{y}} + \bm{\varepsilon}\) and the model response \(\perfect{\bm{y}} = \mathcal{M}(\bm{x})\) at the true value \(\bm{x}\).
Typically one starts from the premise that the residual \(\bm{E} \sim \mathcal{N}(\bm{\varepsilon} \cond \bm{0},\bm{\Sigma})\) follows a Gaussian distribution.
Here, \(\bm{\Sigma}\) is a symmetric and positive-definite covariance matrix.
The observational model is then simply given as \(\bm{Y} \cond \bm{x} \sim \mathcal{N}(\bm{y} \cond \mathcal{M}(\bm{x}),\bm{\Sigma})\).
For the likelihood this implies
\begin{equation} \label{eq:Bayesian:Inverse:Likelihood}
  \mathcal{L}(\bm{x}) = \frac{1}{\sqrt{(2\pi)^{\dimData} \det(\bm{\Sigma})}}
  \exp \left( - \frac{1}{2} \left( \bm{y}-\mathcal{M}(\bm{x}) \right)^\top \bm{\Sigma}^{-1} \left( \bm{y}-\mathcal{M}(\bm{x}) \right) \right).
\end{equation}
For the actually acquired data \(\bm{y}\), this is understood as a function of the unknowns \(\bm{x}\).
The Bayesian solution to the inverse problem posed is then the posterior in \cref{eq:Bayesian:Posterior} where the likelihood is given as in \cref{eq:Bayesian:Inverse:Likelihood}.
It summarizes the collected information about the unknown forward model inputs.

\subsection{Markov chain Monte Carlo}
Apart from some exceptional cases, the posterior density in \cref{eq:Bayesian:Posterior} does not exhibit a closed-form expression.
Thus one settles either for computing expectation values under the posterior or for sampling from the posterior.
The former can be accomplished through stochastic integration techniques such as \emph{Monte Carlo} (MC) \cite{MCMC:Caflisch1998} or \emph{importance sampling} \cite{MCMC:Tokdar2010}.
For the latter one usually has to resort to \emph{Markov chain Monte Carlo} (MCMC) sampling \cite{MCMC:Gilks1996,MCMC:Brooks2011}.
An ergodic Markov chain \(\bm{X}^{(1)},\bm{X}^{(2)},\ldots\) over the support \(\mathcal{D}_{\bm{x}}\) is constructed in such a way that the posterior arises as the invariant distribution
\begin{equation} \label{eq:Bayesian:InvariantDistribution}
  \pi(\bm{x}^{(t+1)} \cond \bm{y}) = \int\limits_{\mathcal{D}_{\bm{x}}} \pi(\bm{x}^{(t)} \cond \bm{y}) \, \mathcal{K}(\bm{x}^{(t)},\bm{x}^{(t+1)}) \, \mathrm{d} \bm{x}^{(t)}.
\end{equation}
Here, \(\mathcal{K}(\bm{x}^{(t)},\bm{x}^{(t+1)})\) denotes the density of the transition probability from the state \(\bm{x}^{(t)}\)
of the Markov chain at a time \(t\) to its state \(\bm{x}^{(t+1)}\) at time \(t+1\).
The \emph{Metropolis-Hastings} (MH) algorithm \cite{MCMC:Metropolis1953,MCMC:Hastings1970} suggests an easy principle
for the construction of a Markov kernel \(\mathcal{K}\) that satisfies \cref{eq:Bayesian:InvariantDistribution}.
It is based on sampling candidates from a proposal distribution and a subsequent accept/reject decision.
The transition kernel defined this ways satisfies detailed balance,
i.e.\ time reversibility \(\pi(\bm{x}^{(t)} \cond \bm{y}) \, \mathcal{K}(\bm{x}^{(t)},\bm{x}^{(t+1)}) = \pi(\bm{x}^{(t+1)} \cond \bm{y}) \, \mathcal{K}(\bm{x}^{(t+1)},\bm{x}^{(t)})\).
This is a sufficient condition for \cref{eq:Bayesian:InvariantDistribution} to apply.
In practice, one initializes the Markov chain at some \(\bm{x}^{(1)} \in \mathcal{D}_{\bm{x}}\) and then
iteratively applies the MH updates from \(\bm{x}^{(t)}\) to \(\bm{x}^{(t+1)}\) for a finite number of times \(T\).
The \emph{ergodic theorem} then ensures that one can approximate the population average in \cref{eq:Bayesian:QoI} in an asymptotically consistent way as the time average
\begin{equation} \label{eq:Bayesian:ErgodicTheorem}
  \mathds{E}[h(\bm{X}) \cond \bm{y}] \approx \frac{1}{T} \sum\limits_{t=1}^T h(\bm{x}^{(t)}).
\end{equation}
\par 
A whole string a unpleasant consequences is entailed by the fact that MCMC updates are typically local and serially dependent.
The quality of the posterior approximation is governed by the MCMC sample autocorrelation.
In order to ensure an efficient posterior exploration one has to carefully design and tune the proposal distribution.
This is an extremely tedious and problem-dependent task.
Yet, even for comparably efficient MCMC updates, a large number of MCMC iterations may be required in order to achieve an acceptable degree of fidelity of the final results.
In inverse modeling this requires an even larger number of serial forward solves which can be prohibitively expensive for demanding models.
Another intrinsic MCMC weakness is that it lacks a clear convergence and stopping criterion,
i.e.\ for diagnosing when the chain has forgotten its initialization and has converged to the target distribution in \cref{eq:Bayesian:InvariantDistribution},
and for the assessment of when the MC error in \cref{eq:Bayesian:ErgodicTheorem} has become sufficiently small.
Even though there are more or less sophisticated convergence diagnostics \cite{MCMC:Cowles1996,MCMC:Brooks1998:Roberts},
those heuristic checks may very well fail, e.g.\ when separated posterior modes have not yet been detected.
\par 
The model evidence in \cref{eq:Bayesian:ScaleFactor} is important in the context of model comparison and selection \cite{Bayesian:Vehtari2012}.
In engineering applications it often happens that one wants to judge the performance of various competing models against measured data \cite{Bayesian:Beck2004,Bayesian:Yuen2010}.
While in variational Bayesian inference at least a lower bound of the model evidence is implicitly computed as a side product, in the MH algorithm it is not computed at all.
Avoiding the explicit computation of the model evidence is beneficial for parameter estimation, but it does not allow for model selection.
To this effect one has to rely on dedicated methods \cite{Bayesian:Han2001,Bayesian:Dellaportas2002}.

\section{Surrogate forward modeling} \label{sec:PCE}
In the analysis of engineering systems it has become a widespread practice to substitute expensive computer models with inexpensive \emph{metamodels} or \emph{surrogate models}.
Those approximations mimic the functional relationship between the inputs and the outputs of the original model in \cref{eq:Bayesian:Inverse:ForwardModel}.
Metamodeling promises significant gains in situations that require a large number of forward model runs, e.g.\ for optimization problems, uncertainty analysis and inverse modeling.
Important classes of metamodels are based on Gaussian process models or Kriging \cite{Kriging:OHagan1978,Kriging:Sacks1989} and polynomial chaos expansions \cite{PCE:Ghanem1991}.
More recent introductions to these subjects can be found in \cite{Kriging:Santner2003,Kriging:Rasmussen2006} and \cite{PCE:LeMaitre2010,PCE:Xiu2010}, respectively.
Nowadays the application of Kriging \cite{Kriging:Higdon2004,Kriging:Higdon2015} and polynomial chaos surrogates \cite{PCE:Marzouk2007,PCE:Marzouk2009:a,PCE:Marzouk2009:b}
is commonplace in Bayesian inverse problems.
\par 
We focus on polynomial chaos metamodels next.
The idea is to decompose the forward model response into polynomial terms that are orthogonal with respect to a weight function.
In stochastic analysis this weight is often identified with a probability density in order to facilitate uncertainty propagation.
In inverse analysis it is commonly equated with the prior in order to enhance MCMC posterior sampling.
The formalism of polynomial chaos expansions is rooted in spectral methods and functional approximations with orthogonal polynomials.
Hence, the function space point of view is emphasized in this section.
We also concentrate on linear least squares for the practical computation of the expansions coefficients.

\subsection{\(L_{\pi}^2\) function space}
From here on it is assumed that the components of the uncertain parameter vector \(\bm{X} = (X_1,\ldots,X_\dimParam)^\top\) are independent random variables \(X_i\).
Thus their joint density can be written as
\begin{equation} \label{eq:PCE:Prior}
  \pi(\bm{x}) = \prod\limits_{i=1}^\dimParam \pi_i(x_i).
\end{equation}
Let \(L_{\pi}^2(\mathcal{D}_{\bm{x}}) = \{u \colon \mathcal{D}_{\bm{x}} \rightarrow \mathds{R} \cond \int_{\mathcal{D}_{\bm{x}}} u^2(\bm{x}) \, \pi(\bm{x}) \, \mathrm{d} \bm{x} < \infty\}\)
be the Hilbert space of functions that are square integrable with respect to the prior density in \cref{eq:PCE:Prior}.
For \(u,v \in L_{\pi}^2(\mathcal{D}_{\bm{x}})\) a weighted inner product \(\langle \cdot,\cdot \rangle_{L_{\pi}^2}\) and its associated norm \(\lVert \cdot \rVert_{L_{\pi}^2}\) are defined as
\begin{align}
  \left\langle u,v \right\rangle_{L_{\pi}^2} &= \int\limits_{\mathcal{D}_{\bm{x}}} u(\bm{x}) v(\bm{x}) \, \pi(\bm{x}) \, \mathrm{d} \bm{x}, \label{eq:PCE:Product} \\
  \left\lVert u \right\rVert_{L_{\pi}^2} &= \left\langle u,u \right\rangle_{L_{\pi}^2}^{1/2}. \label{eq:PCE:Norm}
\end{align}
Given that \(u,v \in L_{\pi}^2(\mathcal{D}_{\bm{x}})\),
the real-valued random variables \(u(\bm{X}),v(\bm{X}) \colon \Omega \rightarrow \mathds{R}\) on the probability space \((\Omega,\mathcal{F},\mathcal{P})\) have a finite variance.
One can then write the inner product in \cref{eq:PCE:Product} as the expectation
\begin{equation} \label{eq:PCE:KeyIdentity}
  \left\langle u, v \right\rangle_{L_{\pi}^2} = \mathds{E}[u(\bm{X}) v(\bm{X})].
\end{equation}
In the further course of the presentation, the identity in \cref{eq:PCE:KeyIdentity} is frequently used
in order to switch back and forth between expectation values under the prior distribution and weighted inner products.

\subsection{Orthonormal polynomials}
Now a basis of the space \(L_{\pi}^2(\mathcal{D}_{\bm{x}})\) is constructed with orthogonal polynomials \cite{Math:Stahl1992,Math:Gautschi2004,Math:Jackson2004}.
Let \(\{\basis^{(i)}_{\alpha_i}\}_{\alpha_i \in \mathds{N}}\) be a family of univariate polynomials in the input variable \(x_i \in \mathcal{D}_{x_i}\).
Each member is characterized by its polynomial degree \(\alpha_i \in \mathds{N}\).
The polynomials are required to be orthonormal in the sense that
\begin{equation} \label{eq:PCE:Orthonormality:Univariate}
  \left\langle \basis^{(i)}_{\alpha_i}, \basis^{(i)}_{\beta_i} \right\rangle_{L_{\pi_i}^2} = \delta_{\alpha_i \beta_i}
  = \begin{cases} 1 &\text{if} \;\, \alpha_i = \beta_i, \\
                  0 &\text{if} \;\, \alpha_i \neq \beta_i.
    \end{cases}
\end{equation}
These polynomials form a complete orthogonal system in \(L_{\pi_i}^2(\mathcal{D}_{x_i})\).
Next, a set of multivariate polynomials \(\{\basis_{\bm{\alpha}}\}_{\bm{\alpha} \in \mathds{N}^\dimParam}\)
in the input variables \(\bm{x} \in \mathcal{D}_{\bm{x}}\) is constructed as the tensor product
\begin{equation} \label{eq:PCE:MultivariatePolynomial}
  \basis_{\bm{\alpha}}(\bm{x}) = \prod\limits_{i=1}^\dimParam \basis^{(i)}_{\alpha_i}(x_i).
\end{equation}
Here, \(\bm{\alpha} = (\alpha_1,\ldots,\alpha_\dimParam) \in \mathds{N}^\dimParam\) is a multi-index that characterizes the polynomials.
By construction, namely due to \cref{eq:PCE:Prior,eq:PCE:Orthonormality:Univariate,eq:PCE:MultivariatePolynomial}, they are orthonormal in the sense that
\begin{equation} \label{eq:PCE:Orthonormality:Multivariate}
  \langle \basis_{\bm{\alpha}}, \basis_{\bm{\beta}} \rangle_{L_{\pi}^2} = \delta_{\bm{\alpha} \bm{\beta}}
  = \begin{cases} 1 &\text{if} \;\, \bm{\alpha} = \bm{\beta}, \\
                  0 &\text{if} \;\, \bm{\alpha} \neq \bm{\beta}.
    \end{cases}
\end{equation}
These polynomials establish a complete orthogonal basis in \(L_{\pi}^2(\mathcal{D}_{\bm{x}})\).
Note that the constant term is always given as \(\basis^{(i)}_{0} = 1\) in the univariate case.
This ensures the proper normalization \(\lVert \basis^{(i)}_{0} \rVert_{L_{\pi_i}^2} = 1\).
In the multivariate case one similarly has \(\basis_{\bm{0}} = 1\) with \(\lVert \basis_{\bm{0}} \rVert_{L_{\pi}^2} = 1\).

\subsection{Hermite and Legendre polynomials}
Two classical univariate families are the \emph{Hermite} and the \emph{Legendre polynomials} \(\{H_{\alpha_i}(x_i)\}_{\alpha_i \in \mathds{N}}\) for \(x_i \in \mathds{R}\)
and \(\{P_{\alpha_i}(x_i)\}_{\alpha_i \in \mathds{N}}\) for \(x_i \in [-1,1]\), respectively.
The former are orthogonal with respect to the weight function \(\mathcal{N}(x_i \cond 0,1) = (2 \pi)^{-1/2} \exp(-x_i^2/2) \),
the latter with respect to \(\mathcal{U}(x_i \cond -1,1) = \mathrm{I}_{[-1,1]}(x_i) / 2\).
Here, \(\mathrm{I}_{[-1,1]}\) denotes the indicator function of the interval \([-1,1]\).
A short summary of these two univariate families is given in \cref{tab:PCE:UnivariateFamilies}.
Over the respective domains, their first members are defined as given in the \cref{sec:App:Polynomials}.
Classical orthogonal polynomials \(\{\basisU_{\alpha_i}^{(i)}(x_i)\}_{\alpha_i \in \mathds{N}}\) are typically not normalized, e.g.\ the aforementioned Hermite or Legendre families.
An orthonormal family \(\{\basis^{(i)}_{\alpha_i}\}_{\alpha_i \in \mathds{N}}\) is then obtained
through an appropriate normalization with \(\basis^{(i)}_{\alpha_i} = \basisU^{(i)}_{\alpha_i} / \lVert \basisU^{(i)}_{\alpha_i} \rVert_{L_{\pi_i}^2}\).
\begin{table}[htb]
  \caption{Two families of orthogonal polynomials.}
  \label{tab:PCE:UnivariateFamilies}
  \centering
    \begin{tabular}{llllll}
      \toprule
      Input type & Polynomials & \multicolumn{1}{l}{\(\mathcal{D}_{x_i}\)} & \multicolumn{1}{l}{\(\pi_i(x_i)\)}
      & \multicolumn{1}{l}{\(\basisU^{(i)}_{\alpha_i}(x_i)\)} & \multicolumn{1}{l}{\(\lVert \basisU^{(i)}_{\alpha_i} \rVert_{L_{\pi_i}^2}\)} \\
      \midrule
      Gaussian & Hermite  & \(\mathds{R}\) & \(\mathcal{N}(x_i \cond 0,1)\)  & \(H_{\alpha_i}(x_i)\) & \(\sqrt{\alpha_i !}\)         \\
      Uniform  & Legendre & \([-1,1]\)     & \(\mathcal{U}(x_i \cond -1,1)\) & \(P_{\alpha_i}(x_i)\) & \(\sqrt{1/(2 \alpha_i + 1)}\) \\
      \bottomrule
    \end{tabular}
\end{table}
\par 
In practice, the parameter space \(\mathcal{D}_{\bm{x}}\) and the input distribution \(\pi(\bm{x})\)
are often not directly suitable for an expansion based on the two standardized families in \cref{tab:PCE:UnivariateFamilies}.
One possibility is then to employ suitably chosen or constructed polynomials \cite{PCE:Xiu2002:b,PCE:Witteveen2007}.
Another possibility is to use an invertible function \(\mathcal{T} \colon \mathds{R}^\dimParam \rightarrow \mathds{R}^\dimParam\), sufficiently well-behaved and as linear as possible,
in order to transform the physical variables \(\bm{x}\) into \emph{standardized variables} \(\bm{\xi} = \mathcal{T}(\bm{x})\),
i.e.\ the image \(\mathcal{D}_{\bm{\xi}} = \mathcal{T}(\mathcal{D}_{\bm{x}})\) and the transformed weight function
\(\pi_{\mathcal{T}}(\bm{\xi}) = \pi(\mathcal{T}^{-1}(\bm{\xi})) \left\lvert \det J_{\mathcal{T}^{-1}}(\bm{\xi}) \right\rvert\) are of a standard form.
Here, \(J_{\mathcal{T}^{-1}} = \mathrm{d} \mathcal{T}^{-1} / \mathrm{d} \bm{\xi}\) is the Jacobian matrix.
If such a change of variables is needed, the considerations that are given below for \(\mathcal{M}\) and \(h\) in the variables \(\bm{x} \in \mathcal{D}_{\bm{x}}\)
can be straightforwardly repeated for \(\mathcal{M} \circ \mathcal{T}^{-1}\) and \(h \circ \mathcal{T}^{-1}\) in the variables \(\bm{\xi} \in \mathcal{D}_{\bm{\xi}}\).
In this case, the expectation in \cref{eq:Bayesian:QoI} follows the integration by substitution
\begin{equation} \label{eq:PCE:IntegrationBySubstitution}
  \mathds{E}[h(\bm{X}) \cond \bm{y}]
  = \frac{1}{\scale} \int\limits_{\mathcal{D}_{\bm{\xi}}} h \left( \mathcal{T}^{-1}(\bm{\xi}) \right)
  \mathcal{L} \left( \mathcal{T}^{-1}(\bm{\xi}) \right) \, \pi_{\mathcal{T}}(\bm{\xi}) \, \mathrm{d} \bm{\xi}.
\end{equation}

\subsection{Polynomial chaos expansions}
For simplicity, herein the presentation is restricted to scalar-valued models \(\mathcal{M} \colon \mathcal{D}_{\bm{x}} \rightarrow \mathds{R}\).
The extension to the multivariate case is straightforward.
It is supposed that the forward model \(\mathcal{M} \in L_{\pi}^2(\mathcal{D}_{\bm{x}})\) is mean-square integrable.
This is a reasonable assumption as seen from a physical perspective.
In \(L_{\pi}^2(\mathcal{D}_{\bm{x}})\) the \emph{generalized Fourier expansion} of \(\mathcal{M}\)
in terms of the orthogonal polynomials \(\{\basis_{\bm{\alpha}}\}_{\bm{\alpha} \in \mathds{N}^\dimParam}\) is then given as
\begin{gather}
  \mathcal{M} = \sum\limits_{\bm{\alpha} \in \mathds{N}^\dimParam} \coeffM_{\bm{\alpha}} \basis_{\bm{\alpha}}, \quad \text{with} \label{eq:PCE:PCE} \\
  \coeffM_{\bm{\alpha}}
  = \langle \mathcal{M}, \basis_{\bm{\alpha}} \rangle_{L_{\pi}^2}
  = \int\limits_{\mathcal{D}_{\bm{x}}} \mathcal{M}(\bm{x}) \basis_{\bm{\alpha}}(\bm{x}) \, \pi(\bm{x}) \, \mathrm{d} \bm{x}. \label{eq:PCE:Projection}
\end{gather}
The \emph{generalized Fourier coefficients} \(\{\coeffM_{\bm{\alpha}}\}_{\bm{\alpha} \in \mathds{N}^\dimParam}\) of the series expansion in \cref{eq:PCE:PCE}
are defined as the orthogonal projection of \(\mathcal{M}\) onto the basis elements in \cref{eq:PCE:Projection}.
The corresponding Fourier series of the second-order random variable \(\perfect{Y} = \mathcal{M}(\bm{X})\) on \((\Omega,\mathcal{F},\mathcal{P})\) is a so-called
\emph{polynomial chaos expansion} (PCE) \(\mathcal{M}(\bm{X}) = \sum_{\bm{\alpha} \in \mathds{N}^\dimParam} \coeffM_{\bm{\alpha}} \basis_{\bm{\alpha}}(\bm{X})\).
\par 
PCEs have been popularized in the context of uncertainty propagation where the goal is the quantification of the distribution of \(\perfect{Y} = \mathcal{M}(\bm{X})\).
For this purpose it comes in handy that the mean and the variance of this random variable can be easily determined from its PCE coefficients.
Indeed, with \cref{eq:PCE:Orthonormality:Multivariate} it is easy to verify that they are simply given as
\begin{align}
  \mathds{E}[\mathcal{M}(\bm{X})] = \left\langle \basis_{\bm{0}}, \sum\limits_{\bm{\alpha} \in \mathds{N}^\dimParam} \coeffM_{\bm{\alpha}} \basis_{\bm{\alpha}} \right\rangle_{L_{\pi}^2}
  &= \coeffM_{\bm{0}}, \label{eq:PCE:Mean} \\
  \mathrm{Var}[\mathcal{M}(\bm{X})]
  = \left\lVert \sum\limits_{\bm{\alpha} \in \mathds{N}^\dimParam} \coeffM_{\bm{\alpha}} \basis_{\bm{\alpha}} - \coeffM_{\bm{0}} \basis_{\bm{0}} \right\rVert_{L_{\pi}^2}^2
  &= \sum\limits_{\bm{\alpha} \in \mathds{N}^\dimParam  \setminus \{\bm{0}\}} \coeffM_{\bm{\alpha}}^2. \label{eq:PCE:Variance}
\end{align}
The simple identities in \cref{eq:PCE:Mean,eq:PCE:Variance} follow from the definitions of the inner product and the associated norm in \cref{eq:PCE:Product,eq:PCE:Norm}, respectively.

\subsection{Truncated series}
For a practical computation one has to truncate the infinite series in \cref{eq:PCE:PCE}.
Let the total degree of a multivariate polynomial \(\basis_{\bm{\alpha}}\) be defined as \(\lVert \bm{\alpha} \rVert_1 = \sum_{i=1}^\dimParam \lvert \alpha_i \rvert\).
A \emph{standard truncation scheme} is then adopted by limiting the terms in \cref{eq:PCE:PCE} to the finite set of multi-indices
\begin{equation} \label{eq:PCE:StandardTruncation}
  \mathcal{A}_p = \left\{ \bm{\alpha} \in \mathds{N}^\dimParam \colon \lVert \bm{\alpha} \rVert_1 \leq p \right\}.
\end{equation}
This specifies a set of polynomials \(\{\basis_{\bm{\alpha}}\}_{\bm{\alpha} \in \mathcal{A}_p}\)
such that their total degree \(\lVert \bm{\alpha} \rVert_1\) is smaller than or equal to a chosen \(p\).
The total number of terms retained in the set \(\mathcal{A}_p\) is given as
\begin{equation} \label{eq:PCE:Cardinality}
  P = \begin{pmatrix}
        \dimParam + p \\
        p
      \end{pmatrix} = \frac{(\dimParam + p)!}{\dimParam! \, p!}.
\end{equation}
The dramatic increase of the total number of terms \(P\) with the input dimensionality \(\dimParam\) and the maximal polynomial degree \(p\),
that is described by \cref{eq:PCE:Cardinality}, is commonly referred to as the \emph{curse of dimensionality}.
A simple idea to limit the number of regressors relies on hyperbolic truncation sets.
For \(0 < q < 1\) a quasinorm is defined as \(\lVert \bm{\alpha} \rVert_q = (\sum_{i=1}^\dimParam \lvert \alpha_i^q \rvert)^{1/q}\).
The corresponding \emph{hyperbolic truncation scheme} is then given as \(\mathcal{A}_p^q = \{\bm{\alpha} \in \mathds{N}^\dimParam \cond \lVert \bm{\alpha} \rVert_q \leq p\}\).
Adopting the standard scheme in \cref{eq:PCE:StandardTruncation}, a finite version of \cref{eq:PCE:PCE} can be written as
\begin{equation} \label{eq:PCE:TruncatedPCE}
  \hat{\mathcal{M}}_p(\bm{x}) = \sum\limits_{\bm{\alpha} \in \mathcal{A}_p} \coeffM_{\bm{\alpha}} \basis_{\bm{\alpha}}(\bm{x}).
\end{equation}
\par 
In engineering problems, one uses \(\hat{\mathcal{M}}_p(\bm{x})\) as a functional approximation of \(\mathcal{M}(\bm{x})\),
i.e.\ as a \emph{polynomial response surface} \cite{Statistics:Box2007}.
This is justified because the approximation converges in the mean-square sense
\begin{equation} \label{eq:PCE:Convergence}
  \left\lVert \mathcal{M}-\hat{\mathcal{M}}_p \right\rVert_{L_{\pi}^2}^2
  = \mathds{E} \left[ \left( \mathcal{M}(\bm{X}) - \hat{\mathcal{M}}_p(\bm{X}) \right)^2 \right]
  = \sum\limits_{\bm{\alpha} \in \mathds{N}^\dimParam  \setminus \mathcal{A}_p} \coeffM_{\bm{\alpha}}^2
  \rightarrow 0, \quad \text{for} \;\, p \rightarrow \infty.
\end{equation}
The rate of the convergence in \cref{eq:PCE:Convergence} depends on the regularity of \(\mathcal{M}\).
On top of that, the response surface in \cref{eq:PCE:TruncatedPCE} is also optimal in the mean-square sense
\begin{equation} \label{eq:PCE:Optimality}
  \left\lVert \mathcal{M}-\hat{\mathcal{M}}_p \right\rVert_{L_{\pi}^2}^2
  = \operatorname*{inf}_{\mathcal{M}^\star \in \mathds{P}_p} \left\lVert \mathcal{M}-\mathcal{M}^\star \right\rVert_{L_{\pi}^2}^2,
  \quad \text{where} \;\, \mathds{P}_p = \mathrm{span} \left( \{\basis_{\bm{\alpha}}\}_{\bm{\alpha} \in \mathcal{A}_p} \right).
\end{equation}
According to \cref{eq:PCE:Optimality}, the response surface in \cref{eq:PCE:TruncatedPCE} minimizes the mean-square error over the space of polynomials
\(\mathds{P}_p = \mathrm{span} (\{\basis_{\bm{\alpha}}\}_{\bm{\alpha} \in \mathcal{A}_p})\) having a total degree of at most \(p\).

\subsection{Least squares}
In order to find a metamodel of the form \cref{eq:PCE:TruncatedPCE}, one computes approximations of the exact expansion coefficients in \cref{eq:PCE:Projection}.
Broadly speaking, one distinguishes between \emph{intrusive} and \emph{non-intrusive} computations.
While the former class of techniques is based on manipulations of the governing equations, the latter is exclusively build upon calls to the forward model at chosen input values.
Stochastic Galerkin methods belong to the class of intrusive techniques \cite{PCE:Babuska2004,PCE:Xiu2009:b},
whereas stochastic collocation \cite{PCE:Xiu2005,PCE:Xiu2007} and projection through numerical quadrature \cite{PCE:LeMaitre2001,PCE:LeMaitre2002} are non-intrusive approaches.
Herein we focus on another non-intrusive formulation that is based on least squares regression analysis \cite{PCE:Berveiller2006,PCE:Blatman2008}.
This formulation is based on linear least squares \cite{Statistics:Lawson1995,Statistics:Bjorck1996}
and related ideas from statistical learning theory \cite{Statistics:Vapnik2000,Statistics:Hastie2009}.
Since this includes sparsity-promoting fitting techniques from high-dimensional statistics \cite{Statistics:Giraud2015,Statistics:Hastie2015},
recently least squares projection methods receive considerable attention.
This includes frequentist \cite{PCE:Blatman2011,PCE:Doostan2011,PCE:Yan2012,PCE:Mathelin2012}
and Bayesian implementations \cite{PCE:Sargsyan2014,PCE:Ray2015,PCE:Karagiannis2014,PCE:Karagiannis2015} of shrinkage estimators.
Current results regarding the convergence behavior of such regression methods can be found in \cite{PCE:Cohen2013,PCE:Migliorati2014,PCE:Chkifa2015}.
\par 
We introduce a simplifying vector notation such that \(\bm{\coeffM} = (\coeffM_1,\ldots,\coeffM_P)^\top\) and
\(\bm{\basis} = (\basis_1,\ldots,\basis_P)^\top\) gather and order the coefficients and the polynomials for all \(\bm{\alpha} \in \mathcal{A}_p\).
For the truncated expression in \cref{eq:PCE:TruncatedPCE} one thus has \(\hat{\mathcal{M}}_p = \bm{\coeffM}^\top \bm{\basis}\).
The problem of finding \(\hat{\mathcal{M}}_p \in \mathds{P}_p\) that minimizes the mean-square error in \cref{eq:PCE:Optimality} may then be equivalently rephrased as
\begin{equation} \label{eq:PCE:Minimization}
  \bm{\coeffM} = \operatorname*{arg\,min}_{\bm{\coeffM}^\star \in \mathds{R}^P} \mathds{E} \left[ \left( \mathcal{M}(\bm{X}) - \bm{\coeffM}^{\star\top} \bm{\basis}(\bm{X}) \right)^2 \right].
\end{equation}
The stochastic optimization objective in \cref{eq:PCE:Minimization} establishes an alternative to the orthogonal projection in \cref{eq:PCE:Projection}.
This formulation may be more amenable to a numerical computation.
At the very least it allows one the draw on the machinery of linear least squares in order to compute an approximation \(\hat{\bm{\coeffM}}\) of the exact coefficients \(\bm{\coeffM}\).
To that end one discretizes \cref{eq:PCE:Minimization} as
\begin{equation} \label{eq:PCE:LeastSquares}
  \hat{\bm{\coeffM}} = \operatorname*{arg\,min}_{\bm{\coeffM}^\star \in \mathds{R}^P}
  \frac{1}{K} \sum\limits_{k=1}^K \left( \mathcal{M}(\bm{x}^{(k)}) - \bm{\coeffM}^{\star\top} \bm{\basis}(\bm{x}^{(k)}) \right)^2.
\end{equation}
Here, the \emph{experimental design} \(\mathcal{X} = (\bm{x}^{(1)},\ldots,\bm{x}^{(K)})\) is a representative sample of \(K\) forward model inputs,
e.g.\ randomly drawn from the input distribution in \cref{eq:PCE:Prior}.
It is then required to compute the corresponding model responses \(\mathcal{Y} = (\mathcal{M}(\bm{x}^{(1)}),\ldots,\mathcal{M}(\bm{x}^{(K)}))^\top\) in \(K\) training runs.
\par 
Now let \(\bm{A} \in \mathds{R}^{K \times P}\) be the matrix with entries \(A_{k,l} = \basis_l(\bm{x}^{(k)})\) for \(k=1,\ldots,K\) and \(l=1,\ldots,P\).
Moreover, let the system \(\mathcal{Y} = \bm{A} \hat{\bm{a}}\) be overdetermined with \(K \geq P\).
The linear least squares problem in \cref{eq:PCE:LeastSquares} may then be written as
\begin{equation} \label{eq:PCE:LeastSquaresMatrix}
  \hat{\bm{\coeffM}} = \operatorname*{arg\,min}_{\bm{\coeffM}^\star \in \mathds{R}^P} \lVert \mathcal{Y} - \bm{A} \bm{\coeffM}^{\star} \rVert^2
\end{equation}
The \emph{normal equations} \((\bm{A}^\top \bm{A}) \hat{\bm{\coeffM}} = \bm{A}^\top \mathcal{Y}\) establish the first-order condition for \cref{eq:PCE:LeastSquaresMatrix} to apply.
Given that the matrix \(\bm{A}^\top \bm{A}\) is non-singular, this linear system is solved by
\begin{equation} \label{eq:PCE:LeastSquaresSolution}
  \hat{\bm{\coeffM}} = (\bm{A}^\top \bm{A})^{-1} \bm{A}^\top \mathcal{Y}.
\end{equation}
The positive definiteness of \(\bm{A}^\top \bm{A}\), i.e.\ the columns of \(\bm{A}\) are linearly independent,
is the second-order condition for \cref{eq:PCE:LeastSquaresSolution} to be the minimum of \cref{eq:PCE:LeastSquaresMatrix}.
The \emph{ordinary least squares} (OLS) solution \cref{eq:PCE:LeastSquaresSolution} is commonly computed by means of linear algebra methods.
An alternative to OLS is \emph{least angle regression} (LAR) \cite{Statistics:Efron2004,Statistics:Hesterberg2008}.
LAR is well-suited for high-dimensional PCE regression problems \cite{PCE:Blatman2011} and can be even applied in the underdetermined case.
It is based on selecting only the most dominant regressors from a possibly large candidate set.
The resulting predictor is thus sparse as compared to the OLS solution.

\subsection{Prediction errors}
After the computation of a metamodel, one typically wants to assess its prediction accuracy.
Moreover, when a number of candidate surrogates is computed, one wants to compare their performances in order to eventually select the best.
Hence, one needs to define an appropriate criterion that allows for an accurate and efficient quantification of the approximation errors.
The natural measure of the mismatch between the forward model \(\mathcal{M}\) and an approximation \(\hat{\mathcal{M}}_p\)
is the \emph{generalization error} \(E_{\mathrm{Gen}} = \mathds{E}[(\mathcal{M}(\bm{X})-\hat{\mathcal{M}}_p(\bm{X}))^2]\).
This is exactly the error the minimization of which is posed by \cref{eq:PCE:Minimization}.
Since it remains unknown, it cannot be used as a performance measure.
One could estimate \(E_{\mathrm{Gen}}\) based on MC simulation, though.
However, this is not very efficient since it requires the execution of additional forward model runs.
In contrast, the \emph{empirical error} \(E_{\mathrm{Emp}} = K^{-1} \sum_{k=1}^K (\mathcal{M}(\bm{x}^{(k)})-\hat{\mathcal{M}}_p(\bm{x}^{(k)}))^2\)
is the quantity that is practically minimized according to \cref{eq:PCE:LeastSquares}.
This error indicator is obtained for free, however, it does not account for overfitting and thus tends to severely underestimate the real generalization error \(E_{\mathrm{Gen}}\).
\par 
In order to construct an estimate of \(E_{\mathrm{Gen}}\) that is more efficient than the MC estimate and more accurate than \(E_{\mathrm{Emp}}\),
one sometimes resorts to leave-one-out (LOO) cross validation \cite{Statistics:Arlot2010}.
Let \(\hat{\mathcal{M}}_{\without k}\) be the surrogate model that is obtained from the reduced experimental design
\(\mathcal{X}_{\without k} = (\bm{x}^{(1)},\ldots,\bm{x}^{(k-1)},\bm{x}^{(k+1)},\ldots,\bm{x}^{(K)})\), i.e.\ a single input \(\bm{x}^{(k)}\) has been dropped.
The \emph{LOO error} is then defined as
\begin{equation} \label{eq:PCE:ErrorLOO}
  E_{\mathrm{LOO}} = \frac{1}{K} \sum_{k=1}^K \left( \mathcal{M}(\bm{x}^{(k)})-\hat{\mathcal{M}}_{\without k}(\bm{x}^{(k)}) \right)^2.
\end{equation}
Without the need for re-running the forward model, this error allows for a fair assessment of how well
the performance of a metamodel \(\hat{\mathcal{M}}_p\) generalizes beyond the used experimental design.
Yet, \cref{eq:PCE:ErrorLOO} calls for conducting \(K\) separate regressions for finding \(\hat{\mathcal{M}}_{\without k}\) with an experimental design of the size \(K-1\).
A remarkably simple result from linear regression analysis states that \(E_{\mathrm{LOO}}\) can be also computed as
\begin{equation} \label{eq:PCE:SimpleLOO}
  E_{\mathrm{LOO}} = \frac{1}{K} \sum\limits_{k = 1}^K \left( \frac{\mathcal{M}(\bm{x}^{(k)}) - \hat{\mathcal{M}}_p(\bm{x}^{(k)})}{1 - h_k} \right)^2.
\end{equation}
Here, \(\hat{\mathcal{M}}_p\) is computed from the full experimental design \(\mathcal{X}\) and
\(h_k\) denotes the \(k\)-th diagonal entry of the matrix \(\bm{A} (\bm{A}^\top \bm{A})^{-1} \bm{A}^\top\).
This is more efficient since it does not require repeated fits.
A derivation of the formula in \cref{eq:PCE:SimpleLOO} can be found in \cite{Statistics:Seber2003}.
\par 
One may define \(\epsilon_{\mathrm{Emp}} = E_{\mathrm{Emp}} / \mathrm{Var}[\mathcal{Y}]\) and \(\epsilon_{\mathrm{LOO}} = E_{\mathrm{LOO}} / \mathrm{Var}[\mathcal{Y}]\)
as normalized versions of the empirical and the LOO error, respectively.
Here, \(\mathrm{Var}[\mathcal{Y}]\) is the empirical variance of the response sample \(\mathcal{Y}\).
These normalized errors can be used in order to judge and compare the performance of metamodels.
In turn, this enables a practical convergence analysis,
e.g.\ by monitoring the errors over repeated metamodel computations for an increasing experimental design size \(K\) and expansion order \(p\).

\section{Spectral Bayesian inference} \label{sec:SLE}
Different types of probability density approximations are encountered in statistical inference.
This includes series expansions for the population density of a random data sample in nonparametric distribution fitting.
Here, the unknown density of the data is either directly represented as a linear combination of polynomial basis functions \cite{Statistics:Efromovich2010}
or as the product of a base density times a superposition of polynomials \cite{Statistics:Jiang2011}.
The latter type of expansion is also encountered in parametric density estimation of random data
where one-dimensional posterior densities of the unknown parameter are expanded about a Gaussian baseline.
This can be based on a Taylor sum at a maximum likelihood estimate \cite{Bayesian:Johnson1967,Bayesian:Johnson1970}
or on Stein's lemma and numerical integration \cite{Bayesian:Weng2010,Bayesian:Weng2013}.
Moreover, different types of likelihood approximations are encountered in inverse modeling.
This includes direct approaches where the likelihood is approximated itself \cite{Inversion:Orlande2008,Kriging:Dietzel2014}
and indirect methods where the likelihood is approximated based on a surrogate of the forward model \cite{PCE:Marzouk2007,PCE:Marzouk2009:a,PCE:Marzouk2009:b}.
These techniques facilitate Bayesian inference within the limits of MCMC sampling.
\par 
By linking likelihood approximations to density expansions,
now we present a spectral formulation of Bayesian inference which targets the emulation of the posterior density.
Based on the theoretical and computational machinery of PCEs, the likelihood function itself is decomposed into polynomials that are orthogonal with respect to the prior distribution.
This spectral likelihood expansion enables semi-analytic Bayesian inference.
Simple formulas are derived for the joint posterior density and its marginals.
They are regarded as expansions of the posterior about the prior as the reference density.
The model evidence is shown to be the coefficient of the constant expansion term.
General QoI-expectations under the posterior and the first posterior moments are obtained through a mere postprocessing of the spectral coefficients.
After a discussion of the advantages and shortcomings of the spectral method, a change of the reference density is proposed in order to improve the efficacy.

\subsection{Spectral likelihood expansions} \label{sec:SLE:SLE}
The authors C.\ Soize and R.\ Ghanem start their paper \cite{PCE:Soize2004} with the following sentence:
``Characterizing the membership of a mathematical function in the most suitable functional space is a critical step toward analyzing it and identifying sequences of efficient approximants to it.''
As discussed in \cref{sec:Bayesian}, given the data \(\bm{y}\) and the statistical model \(f(\bm{y} \cond \bm{x})\), the likelihood is the function
\begin{equation} \label{eq:SLE:LikelihoodFunction}
  \begin{aligned}
    \mathcal{L} \colon \mathcal{D}_{\bm{x}} &\rightarrow \mathds{R}^{+} \\
    \bm{x} &\mapsto f(\bm{y} \cond \bm{x}).
  \end{aligned}
\end{equation}
It maps the parameter space \(\mathcal{D}_{\bm{x}}\) into the set of non-negative real numbers \(\mathds{R}^+\).
At this place we assume that the likelihood \(\mathcal{L} \in L_{\pi}^2(\mathcal{D}_{\bm{x}})\) is square integrable with respect to the prior.
From a statistical point of view, this is a reasonable supposition which is necessary in order to invoke the theory of \cref{sec:PCE}.
On condition that the likelihood is bounded from above, the mean-square integrability follows immediately from the axioms of probability.
Note that maximum likelihood estimates (MLE) implicitly rest upon this presumption.
If \(\bm{x}_{\mathrm{MLE}} \in \operatorname*{arg\,max}_{\bm{x} \in \mathcal{D}_{\bm{x}}} \mathcal{L}(\bm{x})\) is a MLE,
i.e.\ for all \(\bm{x} \in \mathcal{D}_{\bm{x}}\) it applies that \(\mathcal{L}(\bm{x}) \leq \mathcal{L}(\bm{x}_{\mathrm{MLE}}) < \infty\), then one trivially has
\(\int_{\mathcal{D}_{\bm{x}}} \mathcal{L}^2(\bm{x}) \, \pi(\bm{x}) \, \mathrm{d} \bm{x}
\leq \mathcal{L}^2(\bm{x}_{\mathrm{MLE}}) \int_{\mathcal{D}_{\bm{x}}} \pi(\bm{x}) \, \mathrm{d} \bm{x} = \mathcal{L}^2(\bm{x}_{\mathrm{MLE}})< \infty\).
\par 
Having identified \(L_{\pi}^2(\mathcal{D}_{\bm{x}})\) as a suitable function space for characterizing the likelihood,
one can represent the likelihood with respect to the orthonormal basis \(\{\basis_{\bm{\alpha}}\}_{\bm{\alpha} \in \mathds{N}^\dimParam}\).
This representation is
\begin{gather}
  \mathcal{L} = \sum\limits_{\bm{\alpha} \in \mathds{N}^\dimParam} \coeffL_{\bm{\alpha}} \basis_{\bm{\alpha}}, \quad \text{with} \label{eq:SLE:SLE} \\
  \coeffL_{\bm{\alpha}}
  = \langle \mathcal{L}, \basis_{\bm{\alpha}} \rangle_{L_{\pi}^2}
  = \int\limits_{\mathcal{D}_{\bm{x}}} \mathcal{L}(\bm{x}) \basis_{\bm{\alpha}}(\bm{x}) \, \pi(\bm{x}) \, \mathrm{d} \bm{x}. \label{eq:SLE:Projection}
\end{gather}
We refer to \cref{eq:SLE:SLE,eq:SLE:Projection} as a \emph{spectral likelihood expansion} (SLE).
Notice that the SLE coefficients \(\{\coeffL_{\bm{\alpha}}\}_{\bm{\alpha} \in \mathds{N}^\dimParam}\) are data-dependent.
This reflects the fact that the likelihood in \cref{eq:SLE:LikelihoodFunction} depends on the data.
With the truncation scheme in \cref{eq:PCE:StandardTruncation}, one can limit the infinite series in \cref{eq:SLE:SLE} to the finite number of terms for which \(\bm{\alpha} \in \mathcal{A}_p\).
A mean-square convergent response surface of the likelihood is then given as
\begin{equation} \label{eq:SLE:TruncatedSLE}
  \hat{\mathcal{L}}_p(\bm{x}) = \sum\limits_{\bm{\alpha} \in \mathcal{A}_p} \coeffL_{\bm{\alpha}} \basis_{\bm{\alpha}}(\bm{x}).
\end{equation}
\par 
For the time being we assume that the coefficients of the SLE in \cref{eq:SLE:SLE} or its response surface in \cref{eq:SLE:TruncatedSLE} are already known.
One can then accomplish Bayesian inference by extracting the joint posterior density or a \emph{posterior density surrogate},
its marginals and the corresponding QoI-expectations directly from the SLE.

\subsection{Joint posterior density}
We begin with the joint posterior density function and the model evidence.
By plugging \cref{eq:SLE:SLE} in \cref{eq:Bayesian:Posterior} one simply obtains the ``nonparametric'' expression
\begin{equation} \label{eq:SLE:Posterior}
  \pi(\bm{x} \cond \bm{y}) = \frac{1}{\scale} \left( \sum\limits_{\bm{\alpha} \in \mathds{N}^\dimParam} \coeffL_{\bm{\alpha}} \basis_{\bm{\alpha}}(\bm{x}) \right) \pi(\bm{x}).
\end{equation}
Due to the orthonormality of the basis, the model evidence is simply found as the coefficient of the constant SLE term.
This is easily verified by writing \cref{eq:Bayesian:ScaleFactor} as
\begin{equation} \label{eq:SLE:ScaleFactor}
  \scale = \left\langle 1, \mathcal{L} \right\rangle_{L_{\pi}^2}
  = \left\langle \basis_{\bm{0}}, \sum\limits_{\bm{\alpha} \in \mathds{N}^\dimParam} \coeffL_{\bm{\alpha}} \basis_{\bm{\alpha}} \right\rangle_{L_{\pi}^2} = \coeffL_{\bm{0}}.
\end{equation}
The remarkably simple result in \cref{eq:SLE:ScaleFactor} completes the expression of the posterior density in \cref{eq:SLE:Posterior}.
It is interesting to note that the posterior density is of the form
\begin{equation} \label{eq:SLE:EdgeworthExpansion}
  \pi(\bm{x} \cond \bm{y}) = \pi(\bm{x})
  \left( 1 + \sum\limits_{\bm{\alpha} \in \mathds{N}^\dimParam \setminus \{\bm{0}\}} \coeffL_{\bm{0}}^{-1} \coeffL_{\bm{\alpha}} \basis_{\bm{\alpha}}(\bm{x}) \right).
\end{equation}
In essence, the posterior density is here represented as a ``perturbation'' around the prior.
The latter establishes the leading term of the expansion and acts as the \emph{reference density}.
The expression in \cref{eq:SLE:EdgeworthExpansion} is reminiscent of an \emph{Edgeworth expansion} for a density function
in asymptotic statistics \cite{Statistics:BarndorffNielsen1989,Statistics:Kolassa2006,Statistics:Small2010}.

\subsection{Quantities of interest}
Based on the joint posterior density one can calculate the corresponding QoI-expectations in \cref{eq:Bayesian:QoI}.
At this point, the identities in \cref{eq:Baysian:KeyIdentity,eq:PCE:KeyIdentity} finally play a key role.
They allow one to express and treat the posterior expectation of a QoI with \(h \in  L_{\pi}^2(\mathcal{D}_{\bm{x}})\) as the weighted projection onto the likelihood
\begin{equation} \label{eq:SLE:KeyIdentity}
  \mathds{E}[h(\bm{X}) \cond \bm{y}]
  = \frac{1}{\scale} \mathds{E}[h(\bm{X}) \mathcal{L}(\bm{X})]
  = \frac{1}{\scale} \left\langle h, \mathcal{L} \right\rangle_{L_{\pi}^2}.
\end{equation}
Let \(h = \sum_{\bm{\alpha} \in \mathds{N}^\dimParam} \coeffH_{\bm{\alpha}} \basis_{\bm{\alpha}}\)
with \(\coeffH_{\bm{\alpha}} = \langle h, \basis_{\bm{\alpha}} \rangle_{L_{\pi}^2}\) be the QoI representation in the polynomial basis used.
The general posterior expectation in \cref{eq:Bayesian:QoI} follows then from \cref{eq:SLE:KeyIdentity} by \emph{Parseval's theorem}
\begin{equation} \label{eq:SLE:QoI}
  \mathds{E}[h(\bm{X}) \cond \bm{y}]
  = \frac{1}{\coeffL_{\bm{0}}} \left\langle \sum\limits_{\bm{\alpha} \in \mathds{N}^\dimParam} \coeffH_{\bm{\alpha}} \basis_{\bm{\alpha}}
  ,\sum\limits_{\bm{\alpha} \in \mathds{N}^\dimParam} \coeffL_{\bm{\alpha}} \basis_{\bm{\alpha}} \right\rangle_{L_{\pi}^2}
  = \frac{1}{\coeffL_{\bm{0}}} \sum\limits_{\bm{\alpha} \in \mathds{N}^\dimParam} \coeffH_{\bm{\alpha}} \coeffL_{\bm{\alpha}}.
\end{equation}
If the QoI \(h \in \mathds{P}_p\) is known by a multivariate monomial representation of finite order,
then its representation in the orthogonal basis can always be recovered by a change of basis.
Examples that relate to the first posterior moments are given shortly hereafter.
In a more complex case the QoI is a computational model itself and one would have to numerically compute a PCE surrogate.
While \cref{eq:PCE:Mean} facilitates the propagation of the prior uncertainty, \cref{eq:SLE:QoI} promotes the propagation of the posterior uncertainty.

\subsection{Posterior marginals}
Now the posterior marginals in \cref{eq:Bayesian:Marginal1D} are derived.
For some \(j \in \{1,\ldots,\dimParam\}\) let us introduce the new set of multi-indices \(\mathcal{A}^{(j)} = \{(\alpha_1,\ldots,\alpha_\dimParam) \cond \alpha_i = 0 \text{ for } i \neq j\}\).
With a slight abuse of the notation, the \emph{sub-expansion} of the SLE that only contains terms with \(\bm{\alpha} \in \mathcal{A}^{(j)}\) is denoted as
\begin{equation} \label{eq:SLE:SubExpansion1D}
  \mathcal{L}_j(x_j)
  = \sum\limits_{\bm{\alpha} \in \mathcal{A}^{(j)}} \coeffL_{\bm{\alpha}} \basis_{\bm{\alpha}}(\bm{x})
  = \sum\limits_{\mu \in \mathds{N}} \coeffL^{(j)}_{\mu} \basis_{\mu}^{(j)}(x_j),
  \quad \text{where} \;\, \coeffL^{(j)}_{\mu} = \coeffL_{(0,\ldots,0,\mu,0,\ldots,0)}.
\end{equation}
It collects all the polynomials that are constant in all variables \(\bm{x}_{\without j}\), i.e.\ they are non-constant in the single variable \(x_j\) only.
In this sense the sub-expansion in \cref{eq:SLE:SubExpansion1D} is virtually a function of \(x_j\) only.
For the posterior marginal of the single unknown \(x_j\) in \cref{eq:Bayesian:Marginal1D} one can derive
\begin{equation} \label{eq:SLE:Marginal1D}
  \pi(x_j \cond \bm{y})
  = \frac{\pi_j(x_j)}{\scale} \int\limits_{\mathcal{D}_{\bm{x}_{\without j}}} \mathcal{L}(\bm{x}) \, \pi(\bm{x}_{\without j}) \, \mathrm{d} \bm{x}_{\without j}
  = \frac{1}{\coeffL_{\bm{0}}} \mathcal{L}_j(x_j) \pi_j(x_j).
\end{equation}
These equalities apply due to the independent prior \(\pi(\bm{x}) = \pi(\bm{x}_{\without j}) \pi_j(x_j)\) in \cref{eq:PCE:Prior},
the orthonormality of the univariate polynomials in \cref{eq:PCE:Orthonormality:Univariate} and the tensor structure of the multivariate ones in \cref{eq:PCE:MultivariatePolynomial}.
For a pair \(j,k \in \{1,\ldots,\dimParam\}\) with \(j \neq k\) let us introduce yet another set of multi-indices
\(\mathcal{A}^{(j,k)} = \{(\alpha_1,\ldots,\alpha_\dimParam) \cond \alpha_i = 0 \text{ for } i \neq j,k\}\).
The sub-expansion of the full SLE that only contains terms with \(\bm{\alpha} \in \mathcal{A}^{(j,k)}\) is denoted as
\begin{equation} \label{eq:SLE:SubExpansion2D}
  \begin{gathered}
  \mathcal{L}_{j,k}(x_j,x_k)
  = \sum\limits_{\bm{\alpha} \in \mathcal{A}^{(j,k)}} \coeffL_{\bm{\alpha}} \basis_{\bm{\alpha}}(\bm{x})
  = \sum\limits_{\mu,\nu \in \mathds{N}} \coeffL^{(j,k)}_{\mu,\nu} \basis_{\mu}^{(j)}(x_j) \basis_{\nu}^{(k)}(x_k), \\
  \text{where} \;\, \coeffL^{(j,k)}_{\mu,\nu} = \coeffL_{(0,\ldots,0,\mu,0,\ldots,0,\nu,0,\ldots,0)}.
  \end{gathered}
\end{equation}
Since it only contains terms that are constant in \(\bm{x}_{\without j,k}\), the sub-expansion in \cref{eq:SLE:SubExpansion2D} can be seen as a function of \(x_j\) and \(x_k\).
The posterior density can then be marginalized as follows
\begin{equation} \label{eq:SLE:Marginal2D}
  \pi(x_j,x_k \cond \bm{y})
  = \int\limits_{\mathcal{D}_{\bm{x}_{\without j,k}}} \pi(\bm{x} \cond \bm{y}) \, \mathrm{d} \bm{x}_{\without j,k}
  = \frac{1}{\coeffL_{\bm{0}}} \mathcal{L}_{j,k}(x_j,x_k) \pi_j(x_j) \pi_k(x_k).
\end{equation}
Note that the dependency structure of \(\pi(x_j,x_k \cond \bm{y})\) in \cref{eq:SLE:Marginal2D} is induced by those terms of \(\mathcal{L}_{j,k}(x_j,x_k)\)
that are not present in \(\mathcal{L}_j(x_j)\) and \(\mathcal{L}_k(x_k)\), i.e.\ the terms \(\coeffL^{(j,k)}_{\mu,\nu} \basis_{\mu}^{(j)}(x_j) \basis_{\nu}^{(k)}(x_k)\) with \(\mu,\nu \neq 0\).

\subsection{First posterior moments}
With the two marginalizations of the posterior density in \cref{eq:SLE:Marginal1D,eq:SLE:Marginal2D}
one can calculate the entries of the posterior mean in \cref{eq:Bayesian:PosteriorMean} and the covariance matrix in \cref{eq:Bayesian:PosteriorCovariance}.
Let \(\{\coeffMu^{(j)}_0,\coeffMu^{(j)}_1\}\) be defined such that \(x_j = \coeffMu^{(j)}_{0} \basis^{(j)}_{0}(x_j) + \coeffMu^{(j)}_{1} \basis^{(j)}_{1}(x_j)\).
With this univariate representation and \cref{eq:SLE:Marginal1D} one easily obtains
\begin{equation} \label{eq:SLE:PosteriorMargMean}
  \mathds{E}[X_j \cond \bm{y}]
  = \frac{1}{\coeffL_{\bm{0}}} \left\langle x_j, \mathcal{L}_j \right\rangle_{L_{\pi_j}^2}
  = \frac{1}{\coeffL_{\bm{0}}} \left( \coeffMu^{(j)}_{0} \coeffL^{(j)}_{0} + \coeffMu^{(j)}_{1} \coeffL^{(j)}_{1} \right).
\end{equation}
Note that one actually has \(\coeffL^{(j)}_{0} = \coeffL_{\bm{0}}\) in this notation.
Diagonal entries of the covariance matrix in \cref{eq:Bayesian:PosteriorCovariance} can be similarly deduced.
Let \(\{\coeffVar_0^{(j)},\coeffVar_1^{(j)},\coeffVar_2^{(j)}\}\) the coefficients of the univariate representation
\((x_j - \mathds{E}[X_j \cond \bm{y}])^2 = \sum_{\mu=0}^2 \coeffVar^{(j)}_{\mu} \basis^{(j)}_{\mu}(x_j)\).
Then one simply has
\begin{equation} \label{eq:SLE:PosteriorMargVariance}
  \mathrm{Var}[X_j \cond \bm{y}]
  = \frac{1}{\coeffL_{\bm{0}}} \left\langle \left( x_j - \mathds{E}[X_j \cond \bm{y}] \right)^2, \mathcal{L}_j \right\rangle_{L_{\pi_j}^2}
  = \frac{1}{\coeffL_{\bm{0}}} \sum\limits_{\mu=0}^2 \coeffVar^{(j)}_{\mu} \coeffL^{(j)}_{\mu}.
\end{equation}
Finally, let \(\{\coeffVar_{0,0}^{(j,k)},\coeffVar_{0,1}^{(j,k)},\coeffVar_{1,0}^{(j,k)},\coeffVar_{1,1}^{(j,k)}\}\) be the coefficients of the bivariate PCE with
\((x_j - \mathds{E}[X_j \cond \bm{y}]) (x_k - \mathds{E}[X_k \cond \bm{y}]) = \sum_{\mu,\nu=0}^1 \coeffVar^{(j,k)}_{\mu,\nu} \basis^{(j)}_{\mu}(x_j) \basis^{(k)}_{\nu}(x_k)\).
For an off-diagonal entry of \cref{eq:Bayesian:PosteriorCovariance} one then finds
\begin{equation} \label{eq:SLE:PosteriorCovariance}
  \mathrm{Cov}[X_j,X_k \cond \bm{y}]
  = \frac{1}{\coeffL_{\bm{0}}} \sum\limits_{\mu,\nu=0}^1 \coeffVar^{(j,k)}_{\mu,\nu} \coeffL^{(j,k)}_{\mu,\nu}.
\end{equation}
Notation-wise, \cref{eq:SLE:PosteriorMargMean,eq:SLE:PosteriorMargVariance,eq:SLE:PosteriorCovariance} may seem to be somewhat cumbersome.
Nevertheless, they establish simple recipes of how to obtain the first posterior moments by a postprocessing of the low-degree SLE terms in closed-form.
Higher-order moments could be obtained similarly.
Some examples of how the corresponding QoIs can be represented in terms of orthogonal polynomials can be found in the \cref{sec:App:QoIs}.

\subsection{Discussion of the advantages}
In spectral Bayesian inference the posterior is genuinely characterized through the SLE and its coefficients.
The essential advantage of this approach is that all quantities of inferential relevance can be computed semi-analytically.
Simple formulas for the joint posterior density and the model evidence emerge in \cref{eq:SLE:Posterior,eq:SLE:ScaleFactor}.
They allow to establish \cref{eq:SLE:EdgeworthExpansion} as the posterior density surrogate.
General QoI-expectations under the posterior are then calculated via Parseval's formula in \cref{eq:SLE:QoI}.
The posterior marginals are obtained based on sub-expansions of the full SLE in \cref{eq:SLE:Marginal1D} and the first posterior moments
have closed-form expressions in \cref{eq:SLE:PosteriorMargMean,eq:SLE:PosteriorMargVariance,eq:SLE:PosteriorCovariance}.
These striking characteristics clearly distinguish spectral inference from integration and sampling approaches
where the posterior is summarized by expected values or random draws only.
As for the latter, one has to rely on kernel estimates of the posterior density and on empirical sample approximations of the QoI-expectations.
Also, the model evidence is not computed explicitly.
\par 
The practical computation of the SLE in \cref{eq:SLE:SLE} can be accomplished analogously to finding the PCE approximation of the forward model in \cref{eq:PCE:PCE},
e.g.\ by solving a linear least squares problem as in \cref{eq:PCE:LeastSquares}.
This allows one to draw on the vast number of tools that were developed for carrying out this well-known type of regression analysis.
An attractive feature of this procedure is that the prediction error of the obtained SLE acts as a natural convergence indicator.
We recall that the LOO error in \cref{eq:PCE:ErrorLOO} can be efficiently evaluated as per \cref{eq:PCE:SimpleLOO},
i.e.\ without the need for additional forward model runs or regression analyses.
The existence of an intrinsic convergence criterion is an advantage over traditional MCMC techniques.
Another advantage of the formulation its amenability to parallel computations.
While the workload posed by MCMC is inherently serial, running the forward model for each input in the experimental design is embarrassingly parallel.
Parallelization is also possible on the level of the linear algebra operations that are necessary in order to solve the normal equations.

\subsection{Discussion of the shortcomings}
The approximate nature of SLE computations is twofold, i.e.\ only a finite number of terms are kept in the expansion and the coefficients are inexact.
Unfortunately, a number of inconveniences may arise from these inevitable approximations.
The SLE and the correspondingly computed posterior density could spuriously take on negative values.
Also the estimated model evidence in \cref{eq:SLE:ScaleFactor} could take on negative values \(\scale < 0\).
Still note that the approximate posterior in \cref{eq:SLE:Posterior} always integrates to one.
For reasonably adequate SLEs, we expect that negative values only occur in the distributional tails.
Even so, the presence of negative values hampers the interpretation of the obtained posterior surrogate as a proper probability density,
e.g.\ it leads to finite negative probabilities that are somehow irritating.
From a more practical rather than a technical or philosophical perspective,
densities are ultimately instrumental to the evaluation of more concrete quantities such as the expectations in \cref{eq:SLE:QoI}.
The severity of negative densities has thus to be judged with respect to the distortions of these relevant values.
As long as their accurate approximation is guaranteed,
the possibility of negative density values is an unavoidable artifact that can be regarded as a minor blemish.
And the obtained surrogate density still proves to be expedient to effectively characterize the posterior distribution.
In this light, it is more unpleasant that the a posteriori estimates of the model parameters may violate the restrictions that were imposed a priori.
In \cref{eq:SLE:PosteriorMargMean} it could indeed happen that \(\mathds{E}[X_j \cond \bm{y}] \notin \mathcal{D}_{x_j}\).
Estimations of the second order moments in \cref{eq:SLE:PosteriorMargVariance} could result in unnatural values \(\mathrm{Var}[X_j \cond \bm{y}] < 0\), too.
Although these problems cannot be remedied unless one solves an appropriately constrained version of \cref{eq:PCE:LeastSquares}, they unlikely occur if the SLE is sufficiently accurate.
Anticipating outcomes from later numerical demonstrations, we remark that the occurrence of negative density values is observed,
while unphysical or unnatural estimates of the first posterior moments are not found.
\par 
The SLE decomposition into a globally smooth basis of tensorized polynomials suffers from some other intrinsic problems.
Generally, there is the curse of dimensionality, i.e.\ the increase of the number of regressors in \cref{eq:PCE:Cardinality}.
Furthermore, the SLE convergence rate in \cref{eq:PCE:Convergence} depends on the regularity of the underlying likelihood function.
For discontinuous forward or error models the SLE approximation with smooth polynomials converges only slowly.
Likelihood functions often show a peaked structure around the posterior modes and a vanishing behavior elsewhere.
Hence, any adequate superposition of polynomials has to capture those two different behavioral patterns
through some kind of ``constructive'' and ``destructive'' interaction between its terms, respectively.
Due to their global nature, the employed polynomial basis may not admit sparse likelihood representations.
In turn, a high number of terms might be necessary in order to accurately represent even simple likelihood functions.
Especially in the case of high-dimensional and unbounded parameter spaces this may cause severe practical problems.
Of course, in \cref{eq:SLE:SLE,eq:SLE:Projection} one could expand the likelihood in a different basis.
Yet, note that the QoIs in \cref{eq:SLE:QoI} would also have to be expanded in that basis.
\par 
The role of the prior for spectral inference is manifold.
Initially the posterior expectations in \cref{eq:Bayesian:QoI} have been rewritten as the weighted prior expectations in \cref{eq:Baysian:KeyIdentity}.
This formulation is accompanied by difficulties in computing posteriors that strongly deviate from the prior.
The same situation arises for crude MC integration and eventually motivates importance or MCMC sampling that allow to focus on localized regions of high posterior mass.
Those difficulties become manifest if the prior acts as the sampling distribution for the experimental design.
In this case, the SLE is only accurate over the regions of the parameter space that accumulate considerable shares of the total prior probability mass.
Approximation errors of the SLE in regions that are less supported by the prior then induce errors in the computed posterior surrogate.
Note that this difficulty is also encountered in MCMC posterior exploration with prior-based PCE metamodels.
Also, the error estimate in \cref{eq:PCE:SimpleLOO} then only measures the SLE accuracy with respect to the prior which may be misleading for the assessment of the posterior accuracy.
It is not clear how the errors of the likelihood expansion relate to the induced errors of the posterior surrogate and the posterior moments.
Moreover, since the prior acts as the reference density of the posterior expansion in \cref{eq:SLE:EdgeworthExpansion},
the spectral SLE representation of significantly differing posteriors requires higher order corrections.
Otherwise put, SLEs are expected to perform better for posteriors that only slightly update or perturb the prior.

\subsection{Change of the reference density}
As just discussed, a major drawback of SLEs is their dependency on the prior \(\pi\) as the reference density function.
The errors are minimized and measured with respect to the prior and the posterior is represented as correction of the standard reference.
In case that high-order corrections are required, SLEs also suffer from the curse of dimensionality.
While fully maintaining the advantages of the spectral problem formulation, these shortcomings can be remedied through the introduction of an
\emph{auxiliary density} \(\newBase\) over the prior support \(\mathcal{D}_{\bm{x}}\) that would optimally mimic the posterior in some sense.
This \emph{reference density change} allows for the construction of auxiliary expansions that are more accurate with respect to the posterior
and for a more convenient series expansions of the joint posterior density.
It is analogous to the adjustment of the integration weight in importance sampling,
where the average over a distribution is replaced by a weighted average over another ancillary distribution.
An iterative use of this reference change naturally allows for adaptive SLE approaches.
\par 
Given that \(\newBase(\bm{x}) \neq 0\) for all \(\bm{x} \in \mathcal{D}_{\bm{x}}\), one may define the auxiliary quantity \(\auxQuantity = \mathcal{L} \pi / \newBase \).
Under the additional assumption that \(\auxQuantity \in L_{\newBase}^2(\mathcal{D}_{\bm{x}})\), one can expand this quantity in terms of polynomials
\(\{\basisBase_{\bm{\alpha}}\}_{\bm{\alpha} \in \mathds{N}^\dimParam}\) that are orthogonal with respect to the auxiliary reference.
Analogous to the expansion of \(\mathcal{L} \in L_{\pi}^2(\mathcal{D}_{\bm{x}})\) in \cref{eq:SLE:SLE,eq:SLE:Projection}, this is
\begin{gather}
  \auxQuantity
  = \frac{\mathcal{L} \pi}{\newBase}
  = \sum\limits_{\bm{\alpha} \in \mathds{N}^\dimParam} \coeffBaseL_{\bm{\alpha}} \basisBase_{\bm{\alpha}}, \quad \text{with} \label{eq:SLE:BaselineChange:SLE} \\
  \coeffBaseL_{\bm{\alpha}}
  = \left\langle \auxQuantity, \basisBase_{\bm{\alpha}} \right\rangle_{L_{\newBase}^2}
  = \int\limits_{\mathcal{D}_{\bm{x}}} \auxQuantity(\bm{x}) \basisBase_{\bm{\alpha}}(\bm{x}) \, \newBase(\bm{x}) \, \mathrm{d} \bm{x}
  = \int\limits_{\mathcal{D}_{\bm{x}}} \mathcal{L}(\bm{x}) \basisBase_{\bm{\alpha}}(\bm{x}) \, \pi(\bm{x}) \, \mathrm{d} \bm{x}. \label{eq:SLE:BaselineChange:Projection}
\end{gather}
The coefficients of this \emph{auxiliary SLE} (aSLE) are denoted as \(\{\coeffBaseL_{\bm{\alpha}}\}_{\bm{\alpha} \in \mathds{N}^\dimParam}\).
They equal the projections \(\coeffBaseL_{\bm{\alpha}} = \langle \mathcal{L}, \basisBase_{\bm{\alpha}} \rangle_{L_{\pi}^2}\) of the likelihood onto the polynomials \(\basisBase_{\bm{\alpha}}\).
Note that if the new reference \(g = \pi\) equals the prior, then \(\auxQuantity = \mathcal{L}\) is simply the likelihood and the formulation remains unchanged.
If the density \(g = \pi(\cdot \cond \bm{y})\) equals the posterior, then the quantity \(\auxQuantity = \scale\) equals the model evidence.
In this case the aSLE \(\auxQuantity = \coeffBaseL_{\bm{0}} \basisBase_{\bm{0}} = \coeffBaseL_{\bm{0}}\) is a constant with a single nonzero term.
If \(g \approx \pi(\cdot \cond \bm{y})\) only applies in an approximate sense, then one may still speculate that the aSLE is sparser than the corresponding SLE.
\par 
As in importance sampling, one can then rewrite the expectation values under \(\pi\) in \cref{eq:Bayesian:ScaleFactor,eq:Baysian:KeyIdentity} as expectations under \(g\).
Similar to \cref{eq:SLE:ScaleFactor}, the model evidence then emerges again as the zeroth expansion term
\begin{equation} \label{eq:SLE:BaselineChange:ScaleFactor}
  \scale
  = \int\limits_{\mathcal{D}_{\bm{x}}} \auxQuantity(\bm{x}) \, \newBase(\bm{x}) \, \mathrm{d} \bm{x}
  = \coeffBaseL_{\bm{0}}.
\end{equation}
Let \(h = \sum_{\bm{\alpha} \in \mathds{N}^\dimParam} \coeffBaseH_{\bm{\alpha}} \basisBase_{\bm{\alpha}}\)
with \(\coeffBaseH_{\bm{\alpha}} = \langle h, \basisBase_{\bm{\alpha}} \rangle_{L_{\newBase}^2}\) be the auxiliary expansion of a QoI. 
Similar to \cref{eq:SLE:QoI}, for general QoI posterior expectations one may then write
\begin{equation} \label{eq:SLE:BaselineChange:KeyIdentity}
  \mathds{E}[h(\bm{X}) \cond \bm{y}]
  = \frac{1}{\scale} \int\limits_{\mathcal{D}_{\bm{x}}} h(\bm{x}) \auxQuantity(\bm{x}) \, \newBase(\bm{x}) \, \mathrm{d} \bm{x}
  = \frac{1}{\coeffBaseL_{\bm{0}}} \sum\limits_{\bm{\alpha} \in \mathds{N}^\dimParam} \coeffBaseH_{\bm{\alpha}} \coeffBaseL_{\bm{\alpha}}.
\end{equation}
\par 
In accordance with the aSLE in \cref{eq:SLE:BaselineChange:SLE,eq:SLE:BaselineChange:Projection} the joint density of the posterior distribution is obtained as the asymptotic series
\begin{equation} \label{eq:SLE:BaselineChange:Posterior}
  \pi(\bm{x} \cond \bm{y})
  = \frac{\auxQuantity(\bm{x}) \newBase(\bm{x})}{\scale}
  = \frac{1}{\coeffBaseL_{\bm{0}}} \left( \sum\limits_{\bm{\alpha} \in \mathds{N}^\dimParam} \coeffBaseL_{\bm{\alpha}} \basisBase_{\bm{\alpha}}(\bm{x}) \right) \newBase(\bm{x})
  = \newBase(\bm{x}) \left( 1 + \sum\limits_{\bm{\alpha} \in \mathds{N}^\dimParam \setminus \{\bm{0}\}} \frac{\coeffBaseL_{\bm{\alpha}}}{\coeffBaseL_{\bm{0}}} \basisBase_{\bm{\alpha}}(\bm{x}) \right).
\end{equation}
As opposed to \cref{eq:SLE:EdgeworthExpansion} where the posterior density is represented around the prior \(\pi\),
in \cref{eq:SLE:BaselineChange:Posterior} the posterior is expanded about the new reference \(g\).
If the latter resembles the posterior adequately well, the formulation only calls for small corrections.

\section{Numerical examples} \label{sec:Examples}
Next, the potential and the difficulties of the theory presented in the preceding section are investigated.
The goal is to give a proof of concept for the basic feasibility of spectral Bayesian inference.
It is verified that the theory can be successfully applied in practice and further insight into its functioning is obtained.
Moreover, it is learned about its current shortcomings.
Four instructive calibration problems from classical statistics and inverse modeling are solved for these purposes.
The analysis is confined to problems with low-dimensional parameter spaces.
First, the mean value of a normal distribution is inferred with random data under a conjugate normal prior.
Second, the mean and standard deviation of a normal distribution are fitted for a joint prior with independent and uniform marginals.
Third, an inverse heat conduction problem in two spatial dimensions with two unknowns is solved.
Finally, a similar thermal problem with six unknowns is considered.
Synthetically created pseudo-data are used in all these example applications.
\par 
As it turns out, one can gain valuable insights into the characteristics of likelihood expansions and posterior emulators by way of comparison.
Therefore, the analyses for the first three examples proceed analogously.
For rich experimental designs, the convergence behavior of high-degree SLEs is studied by reference to the LOO error.
More importantly, the capability of lower-degree SLEs to accurately capture the posterior QoI-expectations is explored for scarcer experimental designs.
Eventually, aSLE-based posterior surrogates are investigated in order to mitigate the curse of dimensionality.
All results are compared to reference solutions.
Where possible, the exact solutions from a conjugate Bayesian analysis are used to this effect.
Otherwise, corresponding approximations are computed via classical MCMC sampling.
\par 
The uncertainty quantification platform UQLab \cite{PCE:Marelli2014:ASCE,Computing:Uqlab2015:Manual} is used throughout the numerical demonstrations.
It provides a flexible environment for the uncertainty analysis of engineering systems, e.g.\ for uncertainty propagation.
In this context it ships with a range of regression tools that allow one to easily compute PCEs.
These tools can be directly applied to the likelihood function in order to compute SLEs.
OLS is employed as the standard solving routine in the following examples.

\subsection{1D normal fitting}
First of all, we consider the problem of fitting a Gaussian distribution \(\mathcal{N}(y_i \cond \mu,\sigma^2)\) to random realizations \(y_i\) with \(i = 1,\ldots,\dimData\).
The goal is to estimate the unknown mean \(\mu\) whereas the standard deviation \(\sigma\) is assumed to be already known.
Given a Gaussian prior, this one-dimensional normal model with known variance exhibits a Gaussian posterior density.
Moreover, a closed-form expression for the model evidence can be derived.
Since this offers the possibility of comparing the SLE results with analytical solutions, this simple statistical model is used as a first SLE testbed.
Let the data \(\bm{y} = (y_1,\ldots,y_\dimData)^\top\) be comprised of \(\dimData\) independent samples from the normal distribution.
For the observational model one may then write
\begin{equation} \label{eq:Conjugate:Data}
  \bm{Y} \cond \mu \sim \prod\limits_{i=1}^\dimData \mathcal{N}(y_i \cond \mu,\sigma^2), \quad \text{with known} \;\, \sigma^2.
\end{equation}
Consequently, the likelihood function can be simply written as \(\mathcal{L}(\mu) = \prod_{i=1}^{\dimData} \mathcal{N}(y_i \cond \mu,\sigma^2)\).
A Bayesian prior distribution \(\pi(\mu)\) captures the epistemic uncertainty of the true value of \(\mu\) before the data analysis.
For the posterior distribution, that aggregates the information about the unknown after the data have been analyzed, one then has \(\pi(\mu \cond \bm{y}) = \scale^{-1} \mathcal{L}(\mu) \pi(\mu)\).
\par 
The conjugate prior for the data model in \cref{eq:Conjugate:Data} is a Gaussian \(\pi(\mu) = \mathcal{N}(\mu \cond \mu_0,\sigma_0^2)\).
Its mean \(\mu_0 = \mathds{E}[\mu]\) and variance \(\sigma_0^2 = \mathrm{Var}[\mu]\) have to be conveniently specified by the experimenter and data analyst.
This prior choice ensures that the posterior is a Gaussian \(\pi(\mu \cond \bm{y}) = \mathcal{N}(\mu \cond \mu_\dimData,\sigma_\dimData^2)\)
whose parameters \(\mu_\dimData = \mathds{E}[\mu \cond \bm{y}]\) and \(\sigma_\dimData^2 = \mathrm{Var}[\mu \cond \bm{y}]\) are easily found as
\begin{equation} \label{eq:Conjugate:Posterior}
  \mu_\dimData = \left( \frac{1}{\sigma_0^2} + \frac{\dimData}{\sigma^2} \right)^{-1} \left( \frac{\mu_0}{\sigma_0^2} + \frac{\dimData \overline{y}}{\sigma^2} \right),
  \quad \sigma_\dimData^2 = \left( \frac{1}{\sigma_0^2} + \frac{\dimData}{\sigma^2} \right)^{-1}.
\end{equation}
Here, \(\overline{y} = N^{-1} \sum_{i=1}^N y_i\) is the empirical sample mean of the data.
Likewise, an explicit expression for the model evidence
\(\scale = \int_{\mathds{R}} (\prod_{i=1}^\dimData \mathcal{N}(y_i \cond \mu, \sigma^2)) \, \mathcal{N}(\mu \cond \mu_0,\sigma_0^2) \, \mathrm{d} \mu\) can be derived.
Let \(\overline{y^2} = N^{-1} \sum_{i=1}^N y_i^2\) denote the sample mean of the squared observations.
A straightforward calculation based on simple algebra and a Gaussian integral then yields
\begin{equation} \label{eq:Conjugate:ModelEvidence}
  \scale = \sigma_0^{-1} \left( \sigma \sqrt{2 \pi} \right)^{-\dimData} \left( \frac{1}{\sigma_0^2} + \frac{\dimData}{\sigma^2} \right)^{-1/2}
  \exp \left( - \frac{1}{2} \left( \frac{\mu_0^2}{\sigma_0^2} + \frac{\dimData \overline{y^2}}{\sigma^2}
  - \left( \frac{1}{\sigma_0^2} + \frac{\dimData}{\sigma^2} \right)^{-1} \left( \frac{\mu_0}{\sigma_0^2} + \frac{\dimData \overline{y}}{\sigma^2} \right)^2 \right) \right).
\end{equation}
\par 
For the following computer experiment, the parameters of the data distribution in \cref{eq:Conjugate:Data} are specified as \(\mu = 10\) and \(\sigma = 5\), respectively.
In the course of the procedure only the mean is treated as an unknown, whereas the standard deviation is assumed to be known.
We consider a situation where \(\dimData = 10\) samples are randomly drawn from the data distribution.
For the numerical experiment, the pseudo-random numbers \(\bm{y} = (8.78,4.05,12.58,3.60,11.05,8.70,20.80,1.23,19.36,12.07)^\top\) are used as synthetic data.
The prior distribution is set to be a Gaussian \(\pi(\mu) = \mathcal{N}(\mu \cond \mu_0,\sigma_0^2)\) with \(\mu_0 = 11.5\) and \(\sigma_0 = 1.5\).

\subsubsection{Posterior density}
In order to better understand the principles of spectral Bayesian inference we now proceed as follows.
Spectral expansions \(\hat{\mathcal{L}}_p\) of the likelihood function \(\mathcal{L}\) defined above are computed and compared
for experimental designs of varying size \(K\) and polynomial terms of varying degree \(p\).
Hermite polynomials are used in combination with an appropriate linear transformation to standardized variables
\(\xi_\mu \in \mathds{R}\) with a Gaussian weight function \(\mathcal{N}(\xi_\mu \cond 0,1)\).
Accordingly, the unknown can be represented as \(\mu = \mu_0 + \sigma_0 \xi_\mu\).
The experimental designs are one-dimensional Sobol sequences that are appropriately transformed.
\par
\begin{figure}[htbp]
  \centering
  \includegraphics[height=\figHeight]{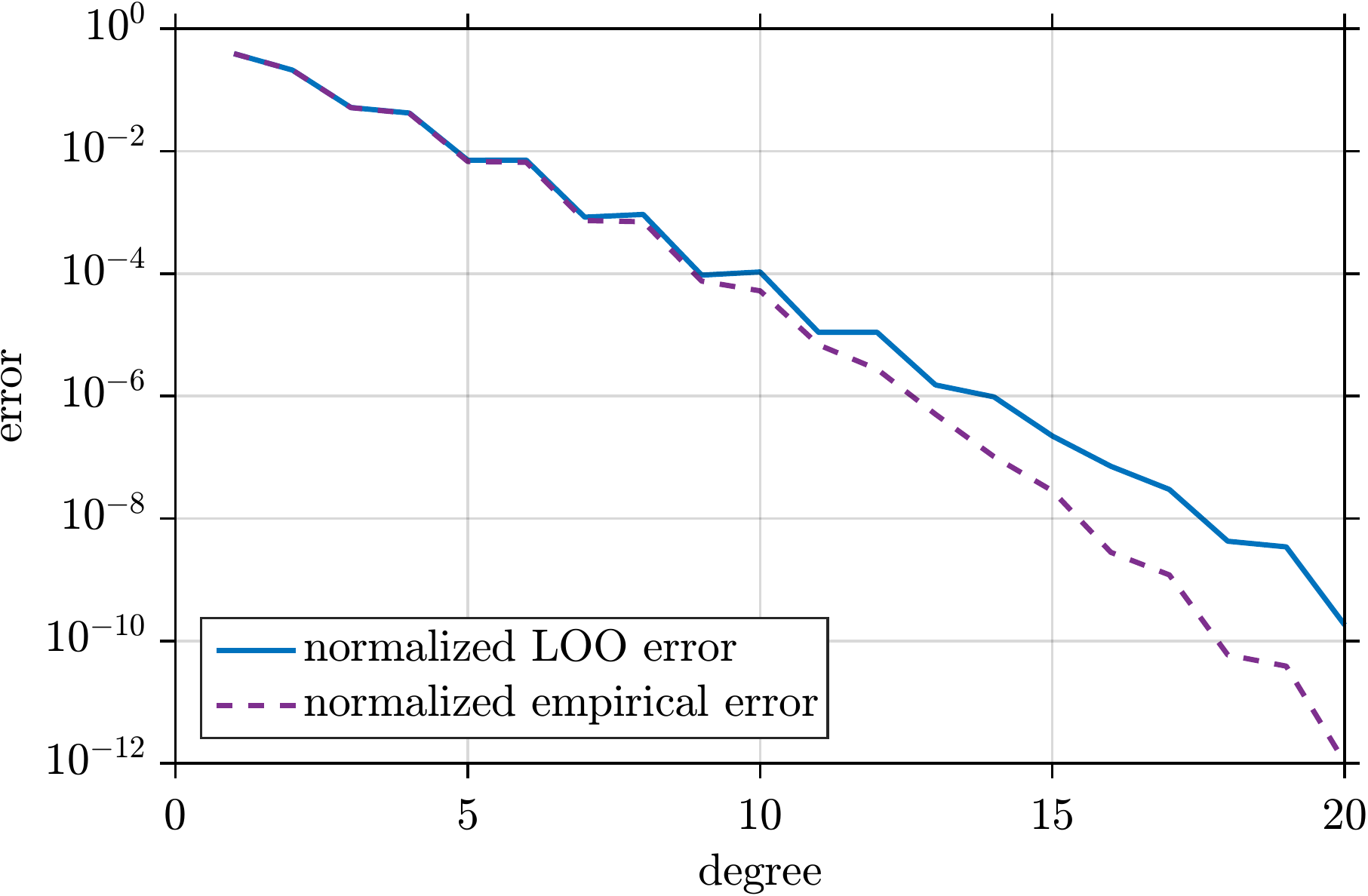}
  \caption{1D normal fitting: Convergence of the SLE.}
  \label{fig:Conjugate:ConvSLE}
\end{figure}

 
First the convergence behavior and the accuracy of the likelihood approximation are analyzed.
For a rich experimental design with \(K = 5 \times 10^4\), SLEs are computed for an increasing order up to \(p = 20\).
The normalized empirical error \(\epsilon_{\mathrm{Emp}}\) and the normalized LOO error \(\epsilon_{\mathrm{LOO}}\) are monitored over these computations.
While the former can be directly computed according to its definition, the computation of the latter relies on the reformulation in \cref{eq:PCE:SimpleLOO}.
This serves the purpose of assessing the prediction accuracy of the computed SLE as a function of the degree \(p\).
The results are plotted in \cref{fig:Conjugate:ConvSLE}.
It can be seen how the error estimates approach zero, i.e.\ the SLE converges to the likelihood function.
For \(p = 20\) the empirical error amounts to \(\epsilon_{\mathrm{Emp}} = 1.05 \times 10^{-12}\) and the LOO amounts to \(\epsilon_{\mathrm{LOO}} = 1.82 \times 10^{-10}\).
These small error magnitudes show that the likelihood function \(\mathcal{L}\) can be indeed spectrally expanded in a Hermite basis.
\par
\begin{figure}[htbp]
  \begin{minipage}[b]{\subWidth}
    \centering
    \includegraphics[height=\figHeight]{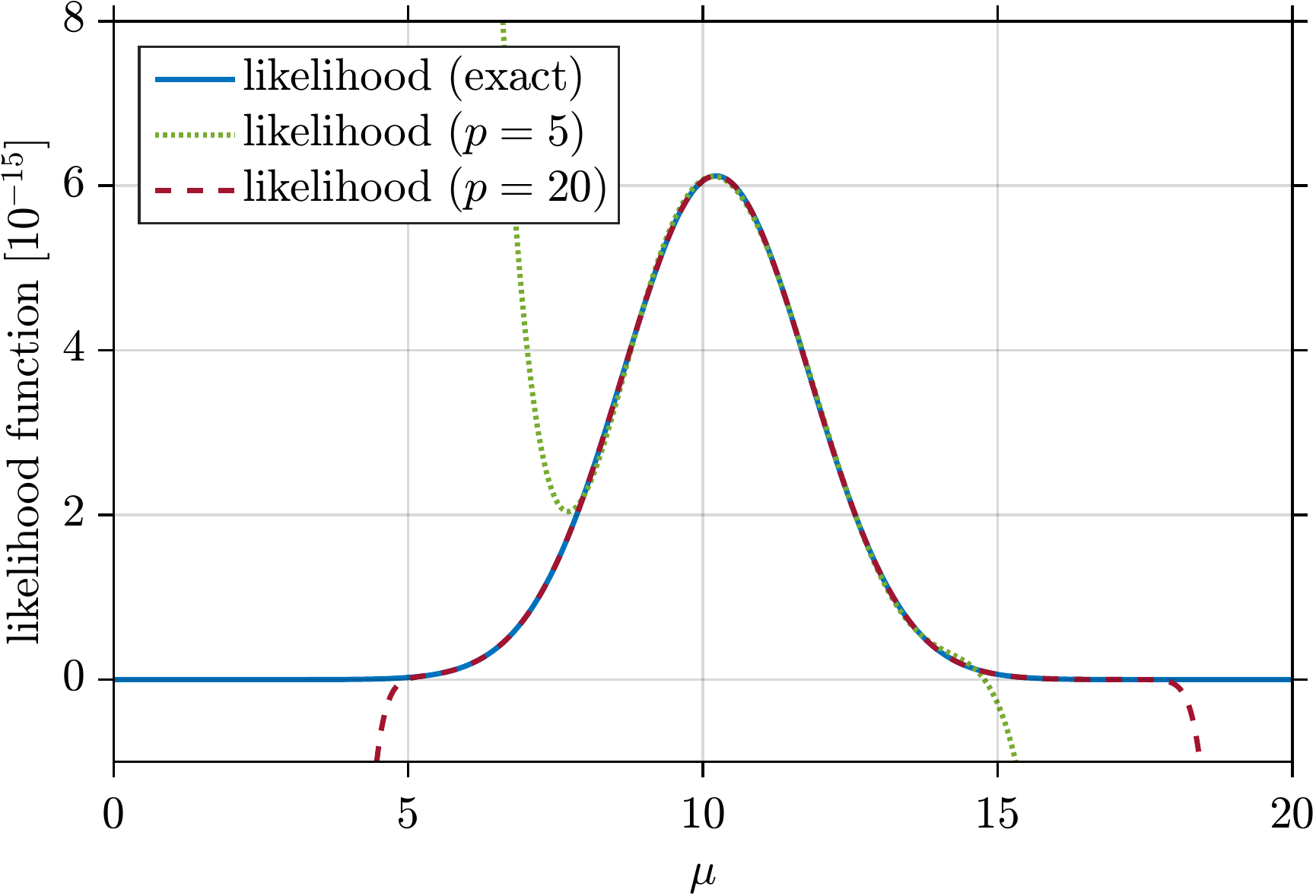}
    \caption{1D normal fitting: Likelihood function.}
    \label{fig:Conjugate:Like}
  \end{minipage}%
  \hfill%
  \begin{minipage}[b]{\subWidth}
    \centering
    \includegraphics[height=\figHeight]{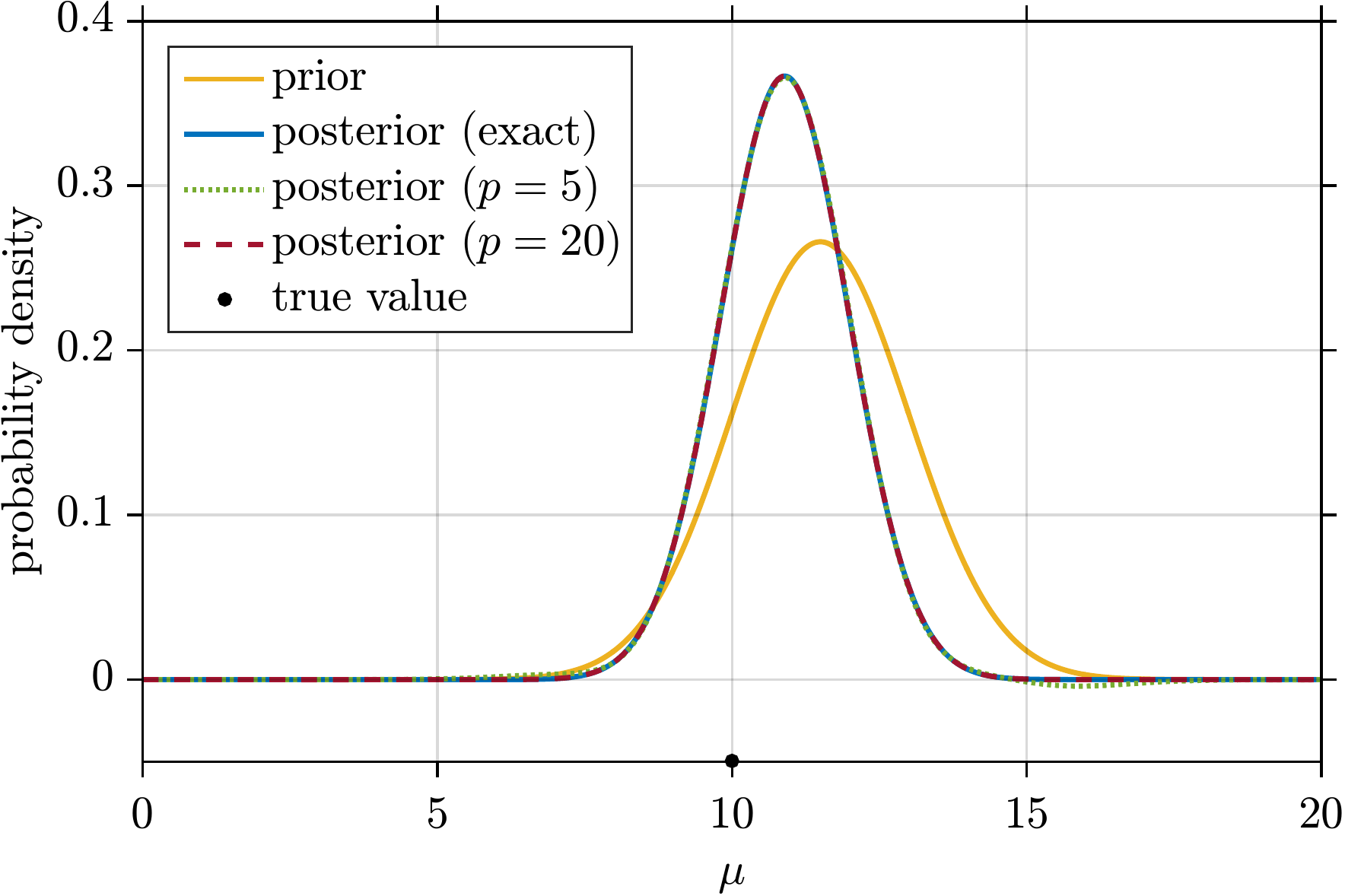}
    \caption{1D normal fitting: Posterior density.}
    \label{fig:Conjugate:Post}
  \end{minipage}%
\end{figure}

 
The functional likelihood approximation \(\hat{\mathcal{L}}_p\) provided by the most accurate SLE with \(p = 20\) is visualized in \cref{fig:Conjugate:Like}.
Moreover, the plot shows a low-order SLE with \(p = 5\) and \(K = 1 \times 10^2\) for which the error estimates
\(\epsilon_{\mathrm{Emp}} = 2.61 \times 10^{-4}\) and \(\epsilon_{\mathrm{LOO}} = 8.41 \times 10^{-4}\) are obtained.
For the sake of comparison the exact likelihood function \(\mathcal{L}\) is shown as well.
It can be seen that the SLEs are able to accurately represent the likelihood around its peak,
i.e.\ roughly speaking in the interval \(\mu \in [8,15]\) for \(p = 5\) and in \(\mu \in [5,18]\) for \(p = 20\).
Note that these regions accumulate the largest proportions of the total prior probability mass.
Outside of these ranges, however, the SLEs \(\hat{\mathcal{L}}_p\) start strongly deviating from \(\mathcal{L}\) and taking on negative values.
These phenomena can be attributed to an imperfect polynomial cancellation of the finite series approximation of the likelihood
in the regions of the parameter space that are only sparsely covered by the experimental design.
Indeed, for unbounded parameter spaces it is clearly hopeless to achieve a global net cancellation of a finite polynomial expansion
that is necessary in order to emulate the vanishing behavior of the likelihood far from its peaks.
The extent to which this impacts on the approximation of the posterior density and its first moments is investigated next.
\par 
Expanding the likelihood function is only a means to the end of surrogating the posterior density.
Approximations of the posterior density \(\pi(\mu \cond \bm{y}) \approx \coeffL_0^{-1} \hat{\mathcal{L}}_p(\mu) \pi(\mu)\)
are computed from the SLEs with \(p = 5\) and \(p = 20\) through \cref{eq:SLE:Posterior,eq:SLE:ScaleFactor}.
The results are plotted in \cref{fig:Conjugate:Post}.
In addition to the SLE approximations, the prior density \(\pi(\mu) = \mathcal{N}(\mu \cond \mu_0,\sigma_0^2)\) and the exact solution
\(\pi(\mu \cond \bm{y}) = \mathcal{N}(\mu \cond \mu_\dimData,\sigma_\dimData^2)\) from a conjugate analysis based on \cref{eq:Conjugate:Posterior} are shown.
The posterior surrogate for \(p = 5\) shows minor deviations from the the analytical result,
while the approximation for \(p = 20\) perfectly matches the true density.
It is noted that the discrepancies between \(\hat{\mathcal{L}}_p\) and \(\mathcal{L}\) shown in \cref{fig:Conjugate:Like} are attenuated.
The underlying reason is that for large enough \(\lvert \mu \rvert \rightarrow \infty\) the exponential decay of the Gaussian prior \(\pi(\mu) \propto \exp(-(\mu-\mu_0)^2)\)
dominates the polynomial increase of \(\hat{\mathcal{L}}_p(\mu) = \sum_{\alpha=0}^p \coeffL_{\alpha} \basis_{\alpha}(\mu)\) in the sense that \(\hat{\mathcal{L}}_p(\mu) \pi(\mu) \rightarrow 0\).
This absorbs the effects of the SLE approximation that is increasingly inadequate for large values of \(\lvert \mu \rvert\).
In this sense, the prior reference density guards the posterior surrogate against the inadequacies of the SLE.
Therefore, the posterior emulation may very well be more accurate than the SLE approximation of the likelihood.

\subsubsection{Quantities of interest}
Commonly one employs posterior means as parameter estimates and posterior standard deviations as measures of the estimation uncertainty.
In order to investigate how well one can approximate the model evidence together with these meaningful quantities in spectral Bayesian inference,
SLEs are computed for experimental designs of varying size \(K\) and for a selection of expansion orders \(p\).
The corresponding SLE-based approximations of the model evidence \(\scale\), the posterior mean \(\mu_\dimData\) and the standard deviation \(\sigma_\dimData\)
are then computed from \cref{eq:SLE:ScaleFactor,eq:SLE:PosteriorMargMean,eq:SLE:PosteriorMargVariance}.
Note that the effects of the transformation to standard variables have to be appropriately taken care of at this place.
This happens via \cref{eq:PCE:IntegrationBySubstitution}.
The SLE approximations can then be compared to the analytical solutions that are obtained from the conjugate analysis in \cref{eq:Conjugate:Posterior,eq:Conjugate:ModelEvidence}.
In \cref{tab:Conjugate:StatisticalQuantities} the results of this procedure are summarized.
Note that all the SLE estimates attain admissible values, e.g.\ the model evidence is non-negative.
Furthermore, it is noticed that \(\scale\), \(\mu_\dimData\) and \(\sigma_\dimData\) can be recovered with high accuracy even for very scarce experimental designs and low-order SLEs,
say for \(K = 1 \times 10^3\) and \(p = 10\).
It is concluded that, in some sense, the accurate estimation of the model evidence and the first posterior moments
require significantly less computational effort than the accurate estimation of the posterior density.
\begin{table}[htbp]
  \caption{1D normal fitting: Statistical quantities.}
  \label{tab:Conjugate:StatisticalQuantities}
  \centering
  \begin{tabular}{ccccccc}
    \toprule
    & \(K\) & \(p\) & \(\epsilon_{\mathrm{LOO}}\) & \(\scale\) \([10^{-15}]\) & \(\mu_\dimData\) & \(\sigma_\dimData\) \\
    \midrule
    \multirow{6}{*}{\rotatebox[origin=c]{90}{SLE}}
    & \(1 \times 10^2\) &  \(5\) & \(8.41 \times 10^{-4}\) & \(3.71\) & \(10.85\) & \(0.92\) \\
    & \(5 \times 10^2\) &  \(8\) & \(2.49 \times 10^{-4}\) & \(3.75\) & \(10.91\) & \(1.14\) \\
    & \(1 \times 10^3\) & \(10\) & \(2.58 \times 10^{-5}\) & \(3.74\) & \(10.90\) & \(1.07\) \\
    & \(5 \times 10^3\) & \(12\) & \(8.21 \times 10^{-6}\) & \(3.74\) & \(10.89\) & \(1.09\) \\
    & \(1 \times 10^4\) & \(15\) & \(3.84 \times 10^{-7}\) & \(3.74\) & \(10.89\) & \(1.09\) \\
    & \(5 \times 10^4\) & \(20\) & \(\phantom{^{1}}1.82 \times 10^{-10}\) & \(3.74\) & \(10.89\) & \(1.09\) \\
    \midrule
    \multicolumn{4}{c}{Exact results}                      & \(3.74\) & \(10.89\) & \(1.09\) \\
    \bottomrule
  \end{tabular}
\end{table}

\subsection{2D normal fitting}
Next, we consider the problem of fitting both the unknown mean \(\mu\) and the standard deviation \(\sigma\) of a Gaussian distribution \(\mathcal{N}(y_i \cond \mu,\sigma^2)\).
A number of independent samples \(y_i\) with \(i = 1,\ldots,\dimData\) from the normal distribution constitute the available data \(\bm{y} = (y_1,\ldots,y_\dimData)^\top\).
The data model for this situation is written as
\begin{equation} \label{eq:Normal:Data}
  \bm{Y} \cond \mu,\sigma \sim \prod\limits_{i=1}^\dimData \mathcal{N}(y_i \cond \mu,\sigma^2).
\end{equation}
For the likelihood function one then has \(\mathcal{L}(\mu,\sigma) = \prod_{i=1}^{\dimData} \mathcal{N}(y_i \cond \mu,\sigma^2)\).
Given a Bayesian prior \(\pi(\mu,\sigma)\), the posterior distribution is \(\pi(\mu,\sigma \cond \bm{y}) = \scale^{-1} \mathcal{L}(\mu,\sigma) \pi(\mu,\sigma)\).
This distribution aggregates the information about the two unknowns after the data have been analyzed.
\par 
The true values of the mean and standard deviation are set as \(\mu = 30\) and \(\sigma = 5\), respectively.
These values are treated as unknowns in the further course of the computer experiment.
We consider a situation where \(\dimData = 10\) samples are randomly drawn from the distribution in \cref{eq:Normal:Data}.
The pseudo-random numbers \(\bm{y} = (31.23,27.50,24.91,25.99,32.88,36.41,27.81,25.19,37.96,34.84)^\top\) are used as synthetic data.
We consider an independent prior \(\pi(\mu,\sigma) = \pi(\mu) \pi(\sigma)\) with uniform marginals
\(\pi(\mu) = \mathcal{U}(\mu \cond \underline{\mu},\overline{\mu})\) and \(\pi(\sigma) = \mathcal{U}(\sigma \cond \underline{\sigma},\overline{\sigma})\)
over bounded supports \(\mathcal{D}_{\mu} = [\underline{\mu},\overline{\mu}] = [20,40]\) and \(\mathcal{D}_{\sigma} = [\underline{\sigma},\overline{\sigma}] = [2,10]\).
As opposed to the conjugate example above, this two-dimensional model does not permit a closed-form expression of the posterior density and the model evidence.

\subsubsection{Posterior density}
Now we proceed analogously to the investigation of the normal model with known variance.
Expansions \(\hat{\mathcal{L}}_p\) of the likelihood \(\mathcal{L}\) are computed and contrasted for different experimental designs of size \(K\) and different polynomial orders \(p\).
An appropriate linear transformation to uniform standardized variables is applied such that the unknowns are represented as
\(\mu = (\overline{\mu} - \underline{\mu}) / 2 \cdot \xi_{\mu} + (\underline{\mu} + \overline{\mu}) / 2\) and 
\(\sigma = (\overline{\sigma} - \underline{\sigma}) / 2 \cdot \xi_{\sigma} + (\underline{\sigma} + \overline{\sigma}) / 2\), respectively.
Here, \(\xi_{\mu},\xi_{\sigma} \in [-1,1]\) are the corresponding standardized variables with a uniform weight function.
Accordingly, tensorized Legendre polynomials form the trial basis.
Two-dimensional Sobol sequences are utilized as uniformly space-filling experimental designs.
\par
\begin{figure}[htbp]
  \centering
  \includegraphics[height=\figHeight]{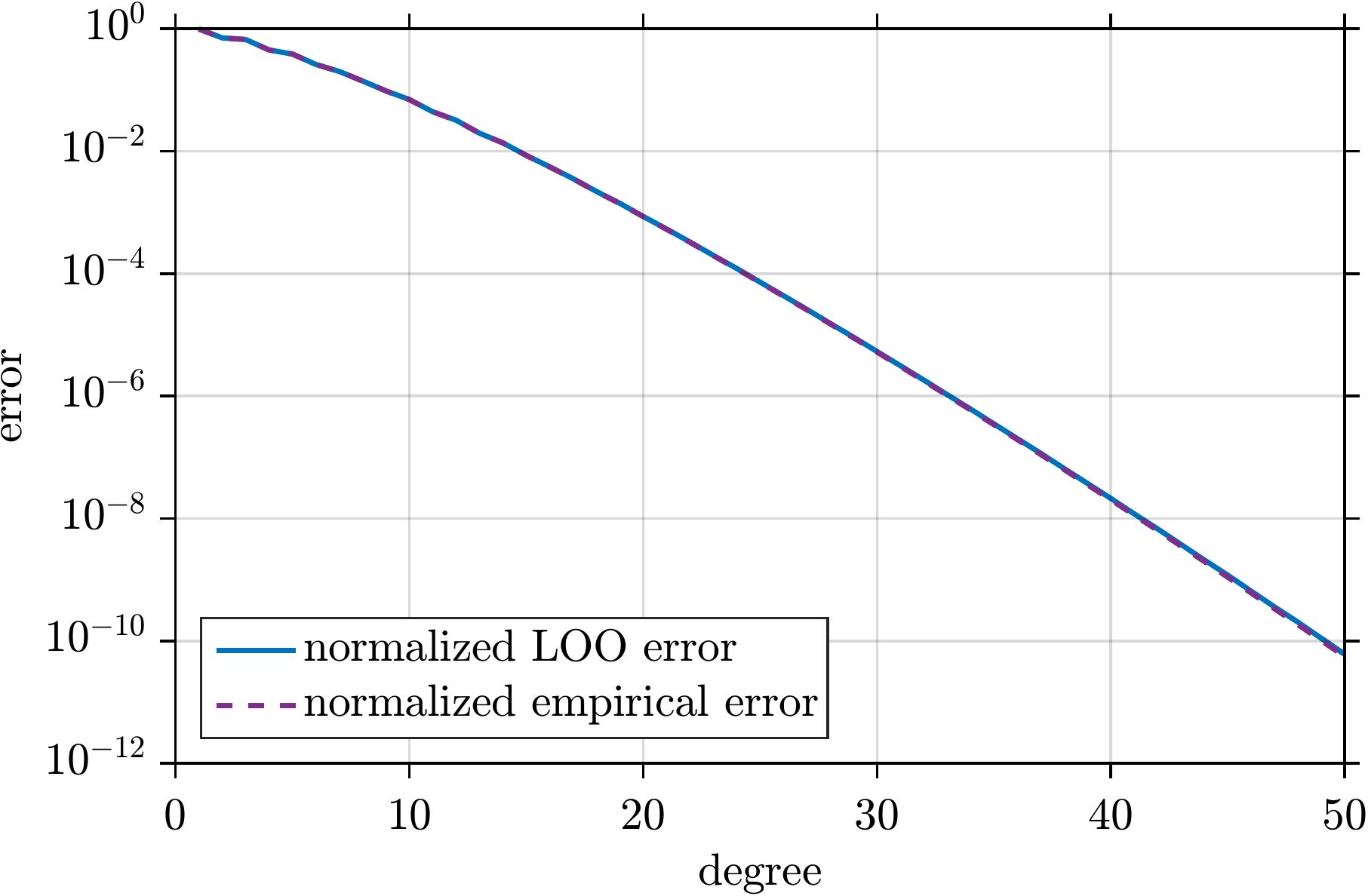}
  \caption{2D normal fitting: Convergence of the SLE.}
  \label{fig:Normal:ConvSLE}
\end{figure}

 
As before, the speed of convergence and the prediction accuracy of the SLE are analyzed first.
The normalized empirical error \(\epsilon_{\mathrm{Emp}}\) and the normalized LOO error \(\epsilon_{\mathrm{LOO}}\) are therefore monitored throughout a series of runs that are conducted
for an experimental design of the fixed size \(K = 1 \times 10^5\) and for an increasing expansion order up to \(p = 50\).
In \cref{fig:Normal:ConvSLE} a corresponding plot is shown, where the convergence of the SLE  \(\hat{\mathcal{L}}_p\) to the likelihood function \(\mathcal{L}\) is diagnosed.
The reason that \(\epsilon_{\mathrm{Emp}}\) and \(\epsilon_{\mathrm{LOO}}\) do not significantly differ is that the large size of the experimental design prevents overfitting.
For \(p = 50\) the normalized empirical error and the normalized LOO error are found as \(\epsilon_{\mathrm{Emp}} = 5.56 \times 10^{-11}\)
and \(\epsilon_{\mathrm{LOO}} = 6.05 \times 10^{-11}\), respectively.
This shows that the likelihood function \(\mathcal{L}\) can be indeed expanded in the Legendre basis.
For the uniform prior distribution that is used here, the normalized SLE errors effectively measure the errors of the posterior density.
\par
\begin{figure}[htbp]
  \centering
  \begin{subfigure}[b]{\subWidth}
    \centering
    \includegraphics[height=\figHeight]{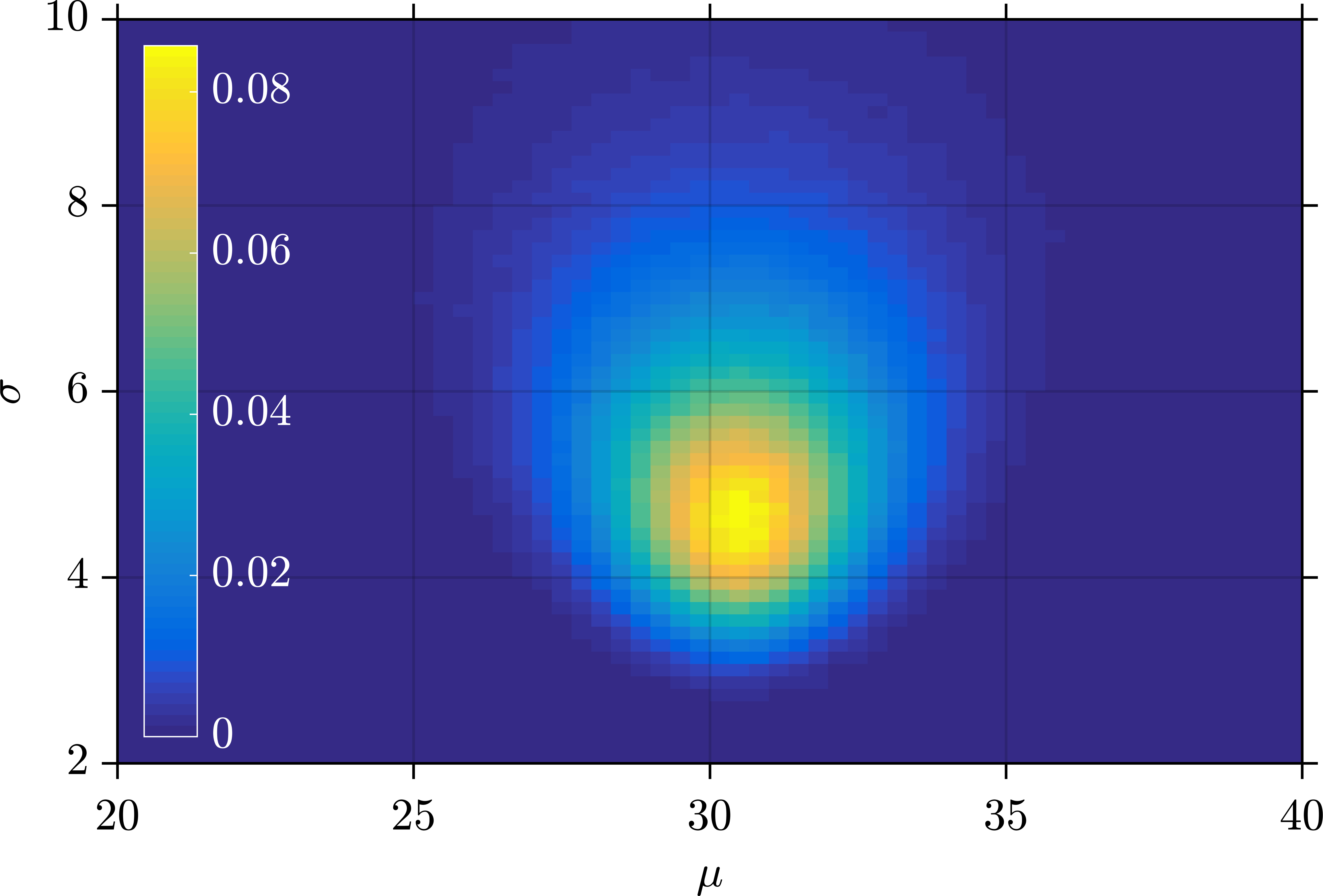}
    \caption{MCMC reference sample.}
    \label{fig:Normal:Post2D:MCMC}
  \end{subfigure}\hfill%
  \begin{subfigure}[b]{\subWidth}
    \centering
    \includegraphics[height=\figHeight]{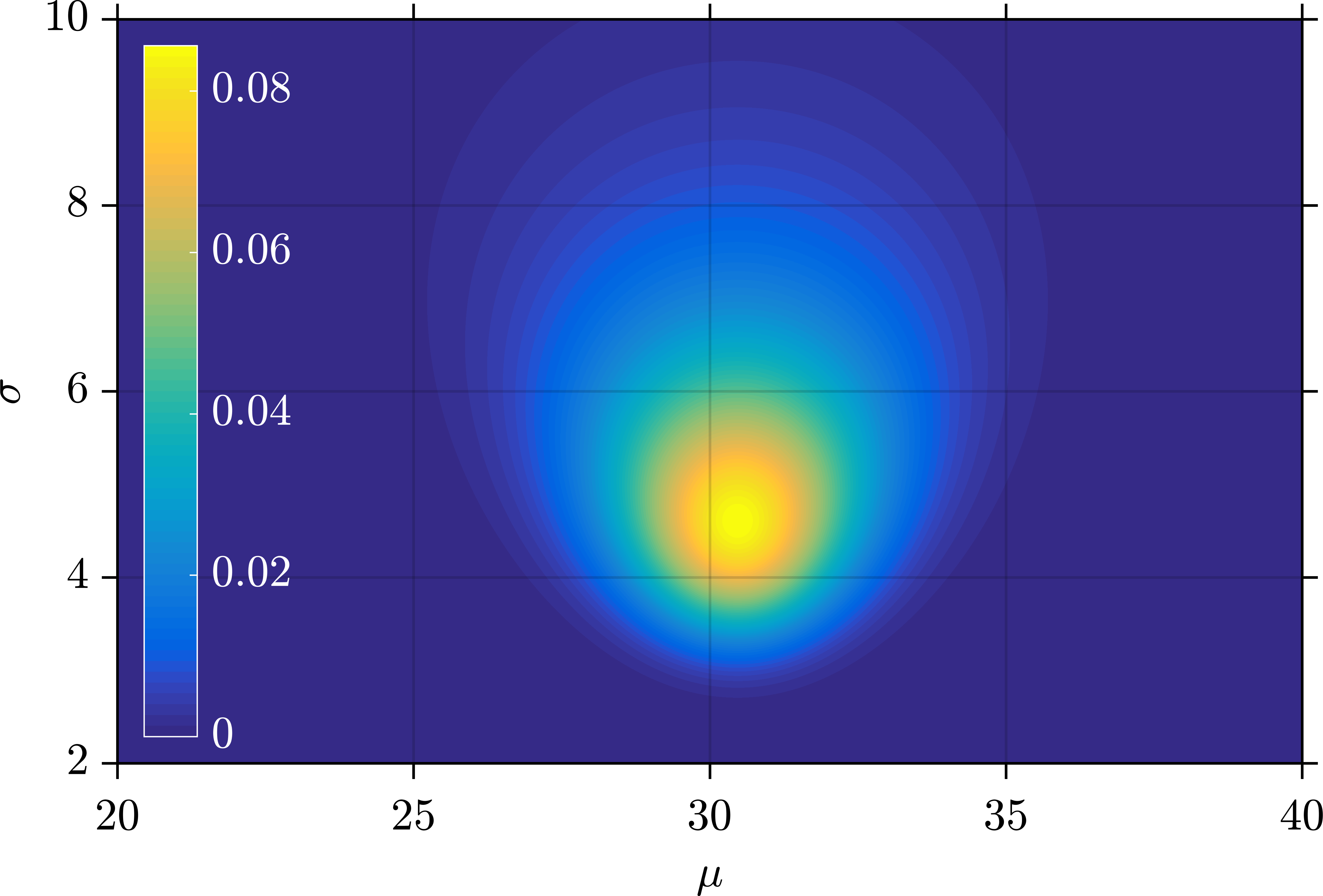}
    \caption{SLE with \(p = 50\).}
    \label{fig:Normal:Post2D:SLE}
  \end{subfigure}\\[1ex]%
  \begin{subfigure}[b]{\subWidth}
    \centering
    \includegraphics[width=\figWidth]{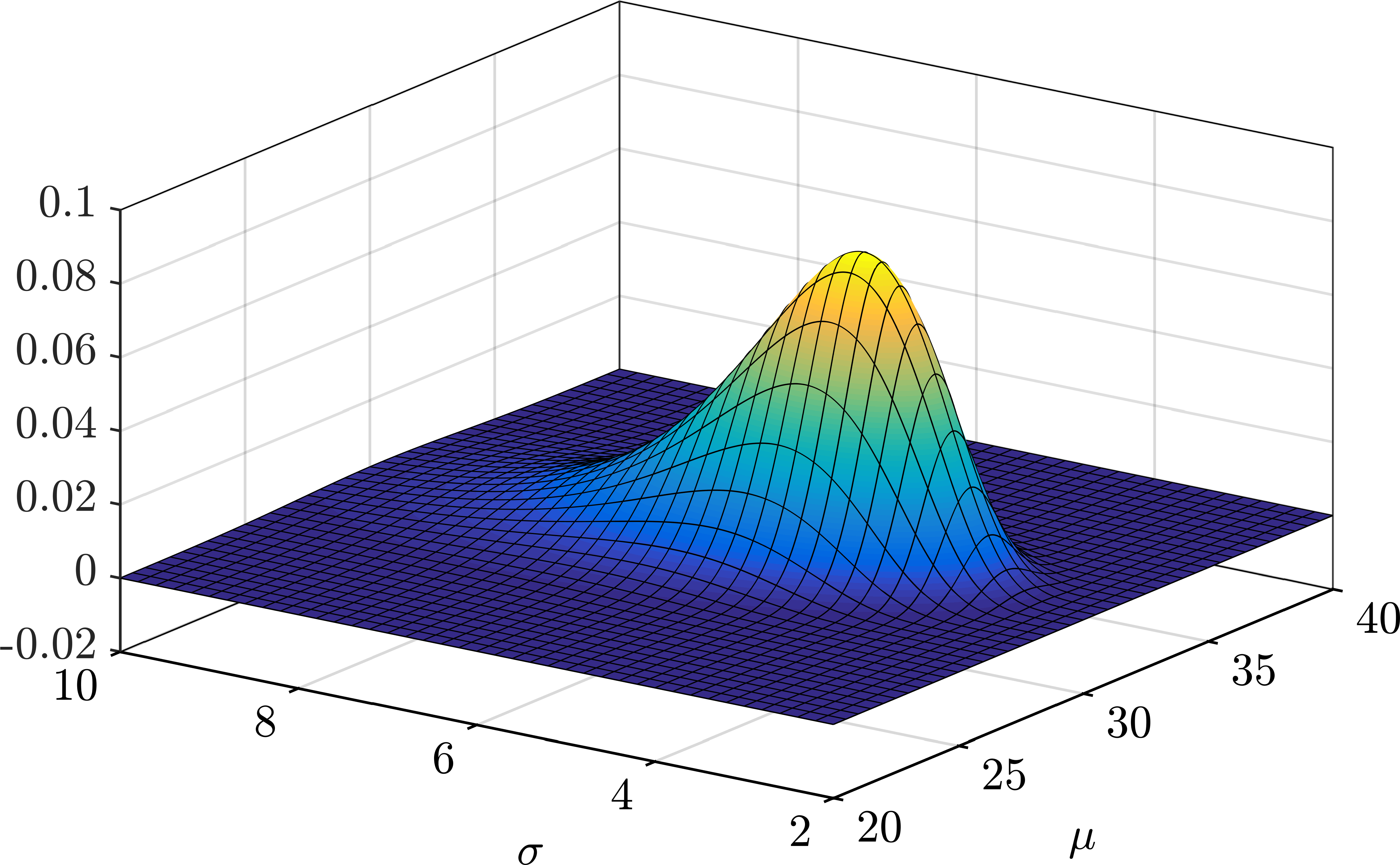}
    \caption{SLE with \(p = 50\).}
    \label{fig:Normal:Post3D:SLE}
  \end{subfigure}%
  \caption{2D normal fitting: Joint posterior.}
  \label{fig:Normal:Post2D}
\end{figure}

 
Now the joint posterior density \(\pi(\mu,\sigma \cond \bm{y})\) is computed and plotted in \cref{fig:Normal:Post2D}.
For comparison purposes the posterior is sampled by means of MCMC simulation first.
A simple random walk Metropolis (RWM) sampler with a Gaussian instrumental distribution is utilized.
With this algorithm an unusually large number of \(10^7\) MCMC samples is drawn from the posterior.
This serves the purpose of providing very accurate results that act as references for the SLE-based estimates.
In \cref{fig:Normal:Post2D:MCMC} a normalized histogram of the obtained RWM sample is shown.
Next, the joint posterior density \(\pi(\mu,\sigma \cond \bm{y}) \approx \coeffL_{\bm{0}}^{-1} \hat{\mathcal{L}}_p(\mu,\sigma) \pi(\mu,\sigma)\)
is computed via \cref{eq:SLE:Posterior,eq:SLE:ScaleFactor}.
The SLE \(\hat{\mathcal{L}}_p(\mu,\sigma)\) with \(p = 50\) that features the lowest LOO error is used.
In \cref{fig:Normal:Post2D:SLE} the posterior surrogate that arises from the SLE is plotted.
For a later comparison with the heat conduction example,
in \cref{fig:Normal:Post3D:SLE} the SLE posterior surrogate from \cref{fig:Normal:Post2D:SLE} is plotted again from a different angle.
By visual inspection the obvious similarity between the density \(\pi(\mu,\sigma \cond \bm{y})\) sampled by MCMC and emulated by the SLE is noticed.
\par
\begin{figure}[htbp]
  \centering
  \begin{subfigure}[b]{\subWidth}
    \centering
    \includegraphics[height=\figHeight]{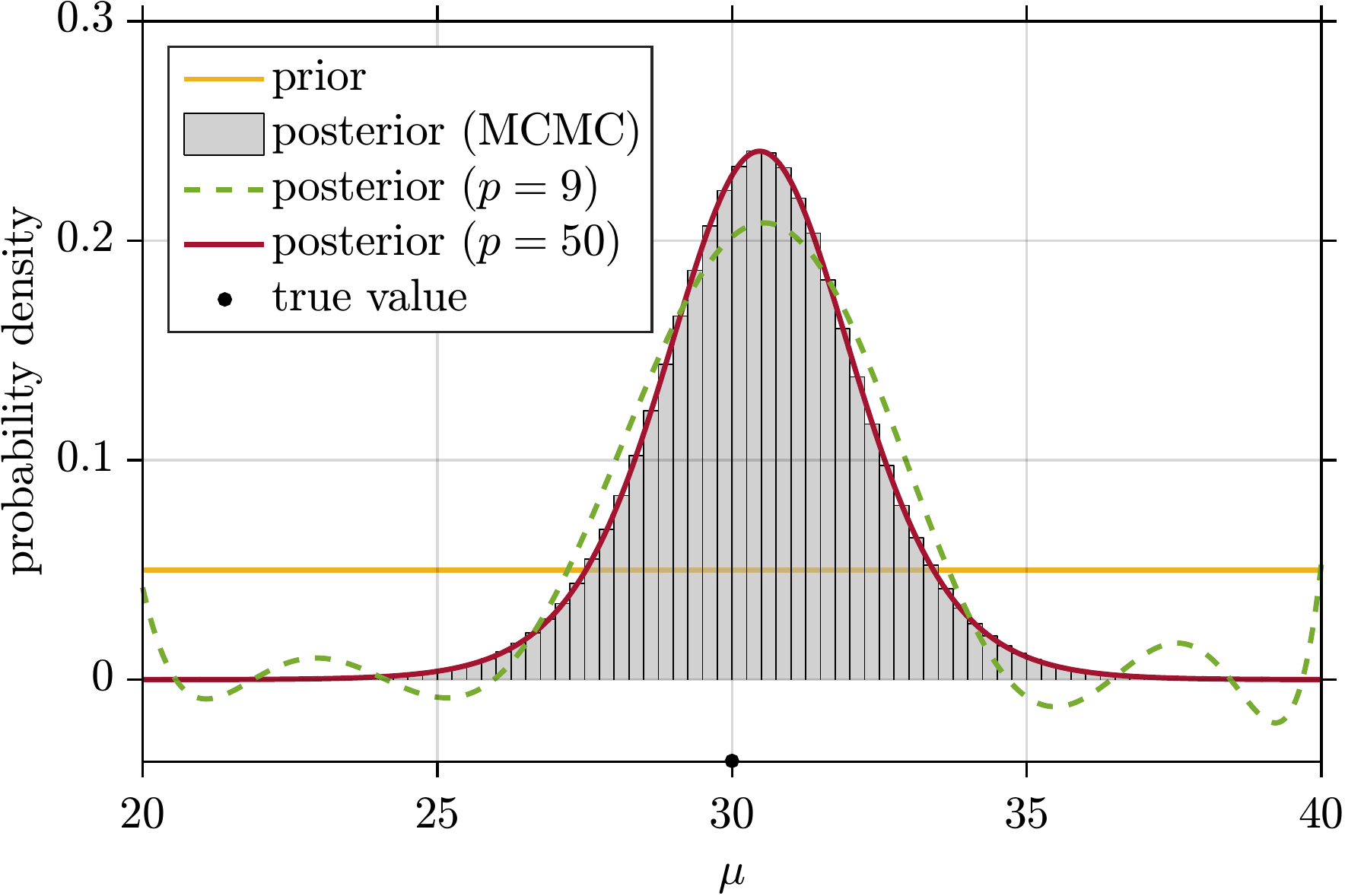}
    \caption{Mean value \(\mu\).}
    \label{fig:Normal:Post1D:mu}
  \end{subfigure}\hfill%
  \begin{subfigure}[b]{\subWidth}
    \centering
    \includegraphics[height=\figHeight]{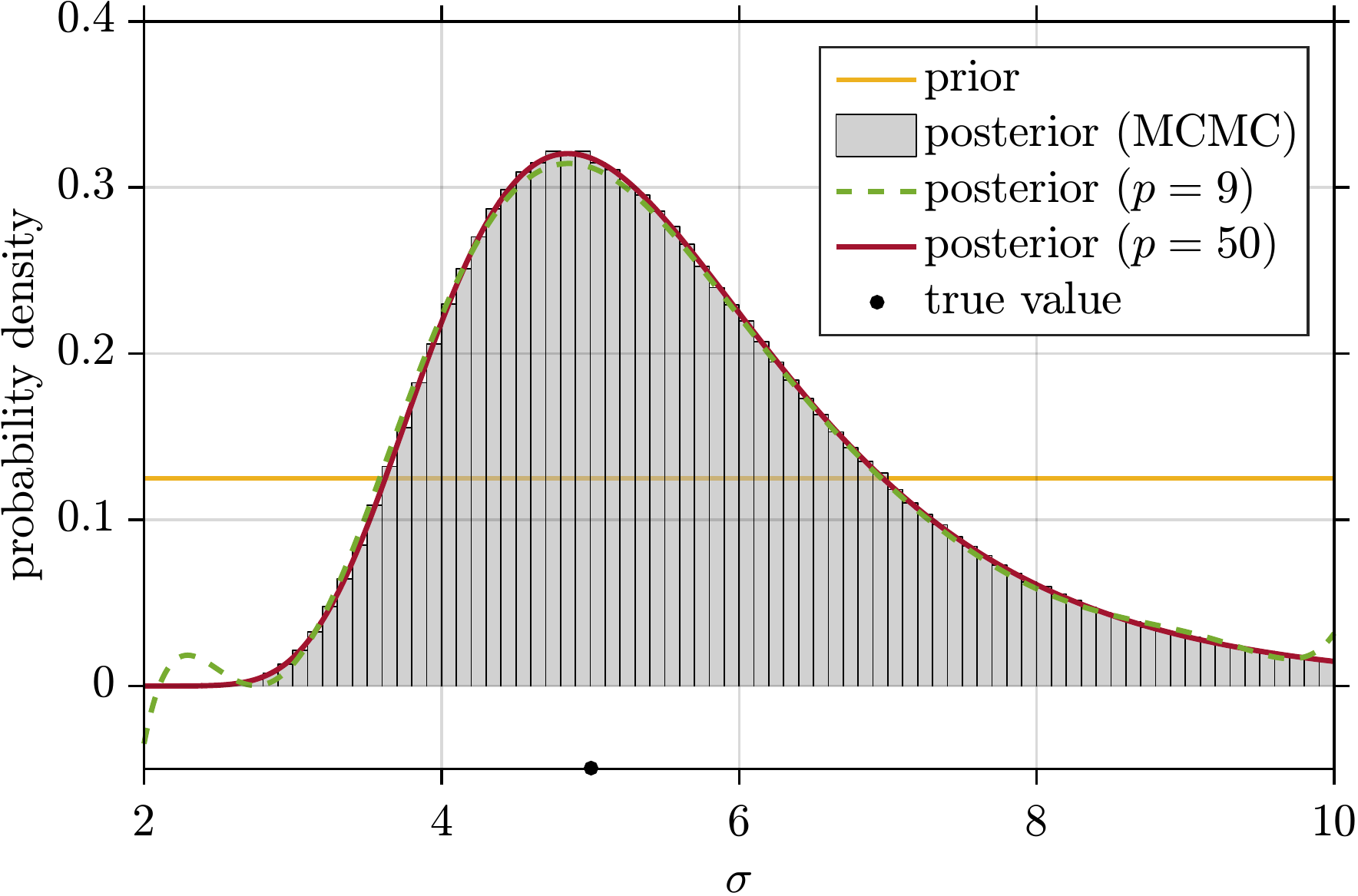}
    \caption{Standard deviation \(\sigma\).}
    \label{fig:Normal:Post1D:sigma}
  \end{subfigure}%
  \caption{2D normal fitting: Posterior marginals.}
  \label{fig:Normal:Post1D}
\end{figure}

 
Now the posterior marginals \(\pi(\mu \cond \bm{y})\) and \(\pi(\sigma \cond \bm{y})\) are computed from the joint posterior.
On the one hand, samples from the posterior marginals are obtained by restricting the analysis to the corresponding components of the joint MCMC sample.
On the other hand, functional approximations of the posterior marginals are extracted based on sub-expansions \(\hat{\mathcal{L}}_{\mu,p}(\mu)\)
and \(\hat{\mathcal{L}}_{\sigma,p}(\sigma)\) of a joint SLE \(\hat{\mathcal{L}}_p(\mu,\sigma)\) as in \cref{eq:SLE:Marginal1D,eq:SLE:SubExpansion1D}.
For the SLEs with \(p = 9\) and \(p = 50\) the results are visualized in \cref{fig:Normal:Post1D}.
Histogram-based MCMC sample representations and functional SLE approximations of the marginal densities are shown, too.
As it can be seen, the marginal posteriors as obtained by MCMC and the SLE with \(p = 50\) exactly match each other.
For \(p = 9\) the posteriors marginals display some wavelike fluctuations in their tails.

\subsubsection{Quantities of interest}
Since the posterior density itself is of little inferential use, the model evidence and the first posterior moments are computed
for a selection of SLEs with varying size of the experimental design \(K\) and degree \(p\).
According to \cref{eq:SLE:ScaleFactor,eq:SLE:PosteriorMargMean,eq:SLE:PosteriorMargVariance,eq:SLE:PosteriorCovariance},
the SLE estimates of these quantities are obtained from the expansions coefficients.
In \cref{tab:Normal:StatisticalQuantities} a summary of the results is given.
Compliant with \cref{eq:SLE:ScaleFactor} the SLE estimates of the model evidence \(\scale\) are obtained as the coefficient of the constant expansion term.
According to \cref{eq:SLE:PosteriorMargMean,eq:SLE:PosteriorMargVariance}, the SLE estimates of the posterior mean \(\mathds{E}[\mu \cond \bm{y}]\)
and the standard deviation \(\mathrm{Std}[\mu \cond \bm{y}] = \mathrm{Var}[\mu \cond \bm{y}]^{1/2}\) of the location parameter \(\mu\) are computed.
Likewise, the corresponding estimates for the spread parameter \(\sigma\) follow through a simple postprocessing of the low-order expansion coefficients.
The SLE estimates of the linear coefficient of correlation
\(\rho[\mu,\sigma \cond \bm{y}] = \mathrm{Cov}[\mu,\sigma \cond \bm{y}] / \mathrm{Std}[\mu \cond \bm{y}] / \mathrm{Std}[\sigma \cond \bm{y}]\)
are computed based on \cref{eq:SLE:PosteriorCovariance}.
Additionally, the LOO error \(\epsilon_{\mathrm{LOO}}\) is listed to indicate the SLE prediction accuracy.
Note that all those estimates comply with the natural bounds and restrictions of the estimated quantities, e.g.\ the posterior means comply with the prior bounds.
\par 
For the sake of comparison, associated results are listed for the simulated MCMC sample, too.
These MCMC results are simply obtained as the corresponding sample approximations.
The reference estimate of the model evidence is obtained by crude MC simulation instead,
i.e.\ the arithmetic mean of the likelihood is computed for a number of \(10^8\) independent draws from the prior.
It is interesting to note that the SLEs can reproduce the MCMC results for moderate experimental designs and degrees, say for \(K = 5 \times 10^3\) and \(p = 21\).
Even though a large number of input samples and a large polynomial degree is necessary to reproduce the shape of the joint posterior density,
significantly smaller experimental designs and polynomial orders suffice to reproduce the first posterior moments.
\begin{table}[htbp]
  \caption{2D normal fitting: Statistical quantities.}
  \label{tab:Normal:StatisticalQuantities}
  \centering
  \begin{tabular}{cccccccccc}
    \toprule
    & \(K\) & \(p\) & \(\epsilon_{\mathrm{LOO}}\)
    & \(\scale\) \([10^{-14}]\) & \(\mathds{E}[\mu \cond \bm{y}]\) & \(\mathds{E}[\sigma \cond \bm{y}]\)
    & \(\mathrm{Std}[\mu \cond \bm{y}]\) & \(\mathrm{Std}[\sigma \cond \bm{y}]\) & \(\rho[\mu,\sigma \cond \bm{y}]\) \\
    \midrule
    \multirow{6}{*}{\rotatebox[origin=c]{90}{SLE}}
    & \(5 \times 10^2\) & \(5\)  & \(4.24 \times 10^{-1}\) & \(1.19\) & \(30.34\) & \(5.57\) & \(2.03\) & \(1.39\) & \(\phantom{-}0.18\) \\
    & \(1 \times 10^3\) & \(9\)  & \(1.19 \times 10^{-1}\) & \(1.20\) & \(30.39\) & \(5.54\) & \(2.01\) & \(1.41\) & \(\phantom{-}0.08\) \\
    & \(5 \times 10^3\) & \(21\) & \(9.64 \times 10^{-4}\) & \(1.18\) & \(30.48\) & \(5.56\) & \(1.79\) & \(1.38\) & \(-0.01\) \\
    & \(1 \times 10^4\) & \(32\) & \(5.86 \times 10^{-6}\) & \(1.18\) & \(30.47\) & \(5.56\) & \(1.81\) & \(1.38\) & \(\phantom{-}0.00\) \\
    & \(5 \times 10^4\) & \(45\) & \(1.30 \times 10^{-9}\) & \(1.18\) & \(30.47\) & \(5.56\) & \(1.81\) & \(1.38\) & \(-0.00\) \\
    & \(1 \times 10^5\) & \(50\) & \(\phantom{^{1}}6.05 \times 10^{-11}\) & \(1.18\) & \(30.47\) & \(5.56\) & \(1.81\) & \(1.38\) & \(-0.00\) \\
    \midrule
    \multicolumn{4}{c}{(MC)MC}                             & \(1.18\) & \(30.47\) & \(5.56\) & \(1.81\) & \(1.38\) & \(-0.00\) \\
    \bottomrule
  \end{tabular}
\end{table}

\subsection{2D inverse heat conduction}
Finally, an inverse heat conduction problem (IHCP) is considered.
The heat equation is a partial differential equation (PDE) that describes the distribution and evolution of heat in a system where conduction is the dominant mode of heat transfer.
We consider a stationary heat equation of the form
\begin{equation} \label{eq:Heat:Stationary}
  \nabla \cdot (\kappa \nabla \perfect{T}) = 0.
\end{equation}
The temperature is denoted as \(\perfect{T}\) and the thermal conductivity is denoted as \(\kappa\).
Commonly one is interested in the solution of the boundary value problem that is posed when \cref{eq:Heat:Stationary} is satisfied over a physical domain subject to appropriate boundary conditions.
We consider the steady state situation in two spatial dimensions.
The Euclidean coordinate vector is denoted as \(\bm{r} = (r_1,r_2)^\top\) in the following.
\par
\begin{figure}[htbp]
  \begin{minipage}[b]{\subWidth}
    \centering
    \includegraphics[height=\figHeight]{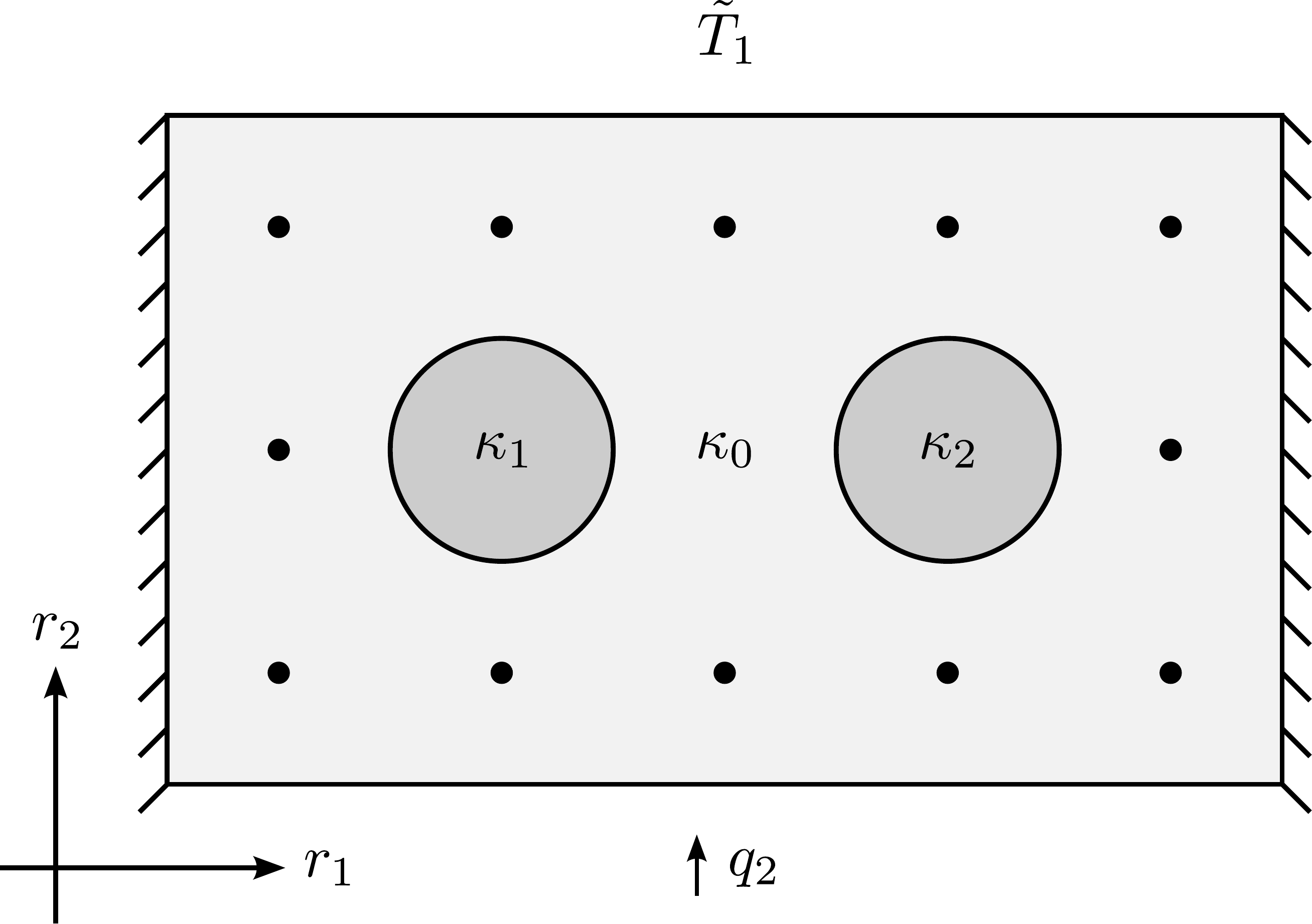}
    \caption{2D IHCP: Heat conduction setup.}
    \label{fig:Heat:HeatConduction2D}
  \end{minipage}%
  \hfill%
  \begin{minipage}[b]{\subWidth}
    \centering
    \includegraphics[height=\figHeight]{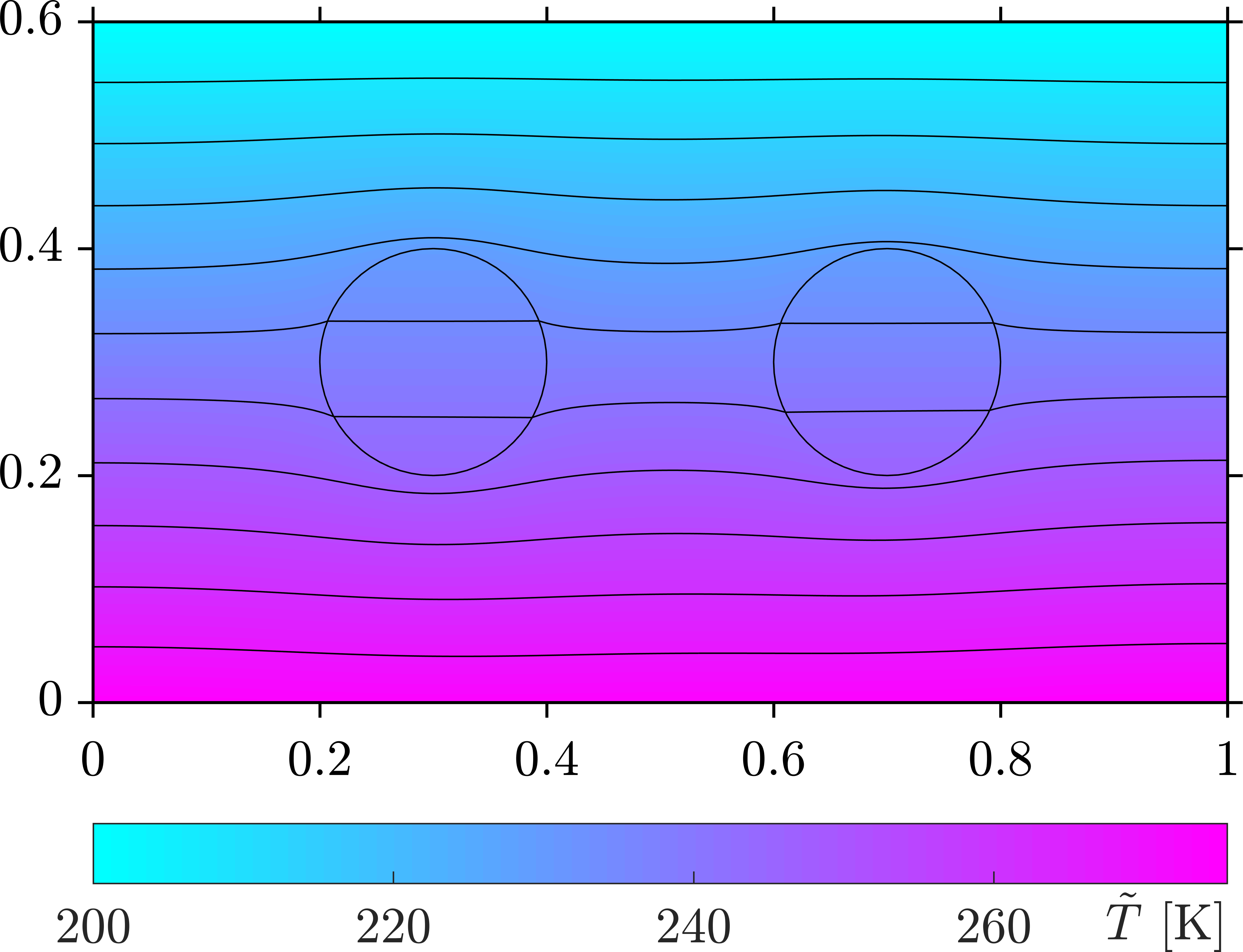}
    \caption{2D IHCP: Steady state solution.}
    \label{fig:Heat:SteadyState2D}
  \end{minipage}%
\end{figure}

 
It is dealt with the identification of thermal conductivities of inclusions in a composite material with close-to-surface measurements of the temperature.
The setup of the simplified thermal problem is visualized in \cref{fig:Heat:HeatConduction2D}.
The thermal conductivity of the background matrix is denoted as \(\kappa_0\), while the conductivities of the material inclusions are termed as \(\kappa_1\) and \(\kappa_2\), respectively.
It is assumed that the material properties are not subject to a further spatial variability.
At the ``top'' of the domain a Dirichlet boundary condition \(\perfect{T}_1\) is imposed,
while at the ``bottom'' the Neumann boundary condition \(q_2 = - \kappa_0 \, \partial \perfect{T} / \partial r_2\) is imposed.
Zero heat flux conditions \(\partial \perfect{T} / \partial r_1 = 0\) are imposed at the ``left'' and ``right'' hand side.
\par 
We consider the IHCP that is posed when the thermal conductivities \(\bm{\kappa} = (\kappa_1,\kappa_2)^\top\) are unknown and their inference is intended.
With this in mind, a number of \(\dimData\) measurements \(\bm{T} = (T(\bm{r}_1),\ldots,T(\bm{r}_\dimData))^\top\)
of the temperature field at the measurement locations \((\bm{r}_1,\ldots,\bm{r}_\dimData)\) is available.
The forward model \(\mathcal{M} \colon \bm{\kappa} \mapsto \perfect{\bm{T}}\) establishes the connection between the data and the unknowns.
It formalizes the operation of solving \cref{eq:Heat:Stationary} for \(\perfect{\bm{T}}\) as a function of \(\bm{\kappa}\).
Measured temperatures \(\bm{T} = \perfect{\bm{T}} + \bm{\varepsilon}\) consist of the corresponding model response \(\perfect{\bm{T}} = \mathcal{M}(\bm{\kappa})\) and a residual term \(\bm{\varepsilon}\).
The latter accounts for measurement uncertainty and forward model inadequacy.
We consider residuals that are distributed according to a Gaussian \(\mathcal{N}(\bm{\varepsilon} \cond \bm{0},\bm{\Sigma})\) with covariance matrix \(\bm{\Sigma}\).
In compliance with \cref{eq:Bayesian:Inverse:Likelihood} the likelihood function is given as \(\mathcal{L}(\bm{\kappa}) = \mathcal{N}(\bm{T} \cond \mathcal{M}(\bm{\kappa}),\bm{\Sigma})\).
Provided that a prior distribution \(\pi(\bm{\kappa})\) can be elicited, the posterior is given as \(\pi(\bm{\kappa} \cond \bm{T}) = \scale^{-1} \mathcal{L}(\bm{\kappa}) \pi(\bm{\kappa})\).
\par 
The thermal conductivity of the background matrix is set to \(\kappa_0 = \unit[15]{W/m/K}\),
while the thermal conductivities of the inclusions are specified as \(\kappa_1 = \unit[32]{W/m/K}\) and \(\kappa_1 = \unit[28]{W/m/K}\).
The material properties of the inclusions are treated as unknowns subsequently.
Moreover, the boundary conditions \(\perfect{T}_1 = \unit[200]{K}\) and \(q_2 = \unit[2000]{W/m^2}\) are imposed.
A finite element (FE) model is used to solve a weak form of the governing PDE.
The FE solution for the experimental setup described above is shown in \cref{fig:Heat:SteadyState2D}.
We consider a uniform prior distribution \(\pi(\bm{\kappa}) = \pi(\kappa_1) \pi(\kappa_2)\) with independent marginals
\(\pi(\kappa_1) = \mathcal{U}(\kappa_1 \cond \underline{\kappa}_1,\overline{\kappa}_1)\) and \(\pi(\kappa_2) = \mathcal{U}(\kappa_2 \cond \underline{\kappa}_2,\overline{\kappa}_2)\).
The prior bounds are chosen as \(\underline{\kappa}_1 = \underline{\kappa}_2 = \unit[20]{W/m/K}\) and \(\overline{\kappa}_1 = \overline{\kappa}_2 = \unit[40]{W/m/K}\), respectively.
A number of \(\dimData = 12\) close-to-surface observations is analyzed.
Their measurement locations are indicated by the black dots in \cref{fig:Heat:HeatConduction2D}.
Independent Gaussian measurement noise with \(\bm{\Sigma} = \sigma_T^2 \bm{1}\) and \(\sigma_T = \unit[0.25]{K}\) is considered.
Based on this setup, synthetic data are simulated for conducting the computer experiment.
This means that the forward model responses \(\perfect{\bm{T}}\) for the true parameter setup are computed and pseudo-random noise is added in order to obtain \(\bm{T}\).

\subsubsection{Posterior density}
\begin{figure}[htbp]
  \centering
  \includegraphics[height=\figHeight]{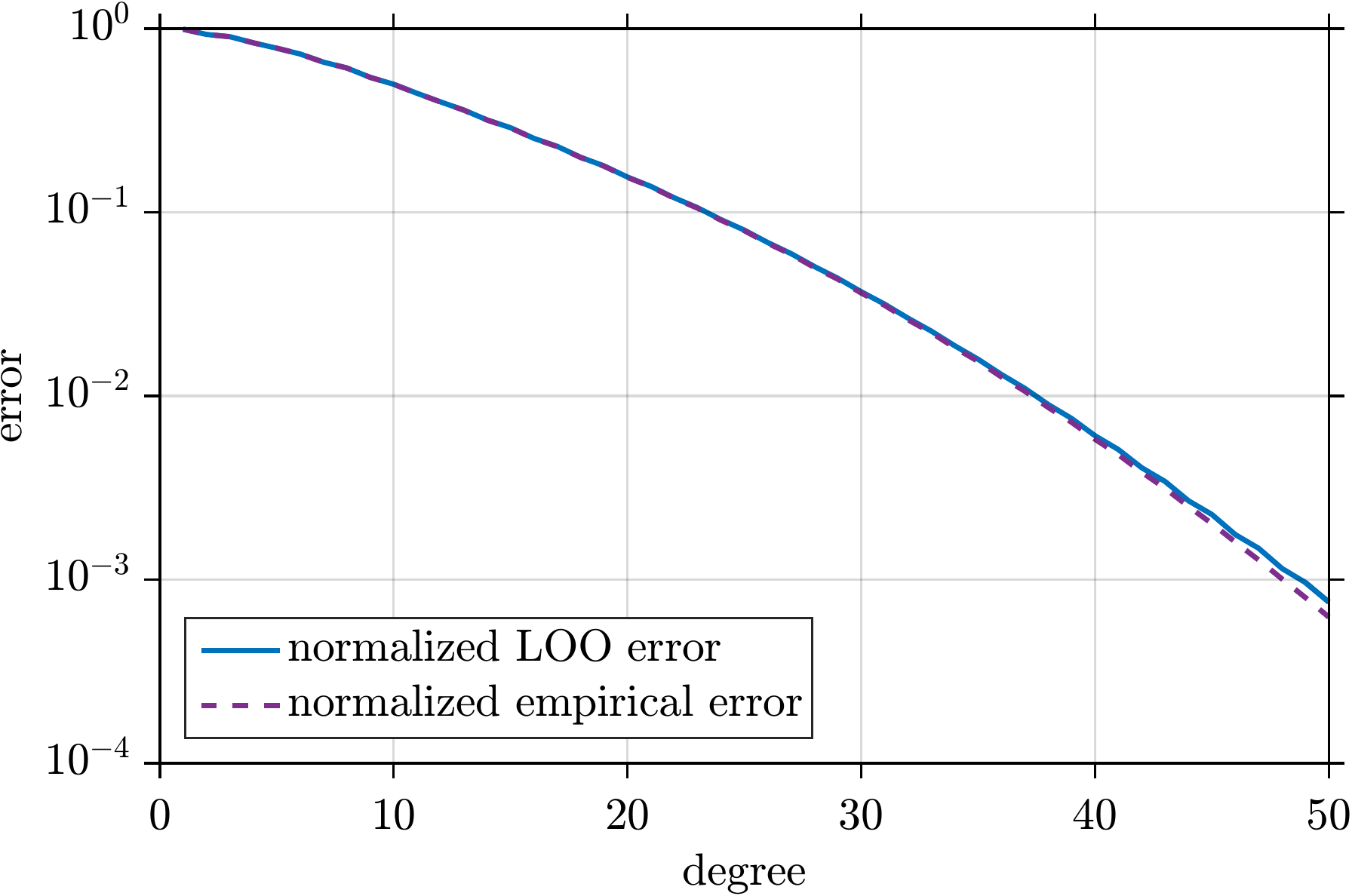}
  \caption{2D IHCP: Convergence of the SLE.}
  \label{fig:Heat:ConvSLE}
\end{figure}

The analyses proceed analogously to the preceding section.
By comparing the present IHCP and the non-conjugate Gaussian example, that have a two-dimensional parameter space and uniform priors in common,
one can gain interesting insight into spectral Bayesian inference.
First, the convergence behavior of the SLE is investigated.
Spectral expansions \(\hat{\mathcal{L}}_p\) of the likelihood \(\mathcal{L}\) are therefore computed
for an experimental design of size \(K = 1 \times 10^5\) and candidate bases with polynomials up to degree \(p = 50\).
All practical issues are handled analogously to the procedure in the non-conjugate Gaussian example.
In \cref{fig:Heat:ConvSLE} the normalized versions of the empirical error \(\epsilon_{\mathrm{Emp}}\) and the LOO error \(\epsilon_{\mathrm{LOO}}\) are shown as a function of \(p\).
Comparing these results to \cref{fig:Normal:ConvSLE} reveals that the convergence rate of the SLE \(\hat{\mathcal{L}}_p\) is considerably slower than the corresponding one for the Gaussian example.
For the SLE with \(p = 50\) the error estimates amount to \(\epsilon_{\mathrm{Emp}} = 6.26 \times 10^{-4}\) and \(\epsilon_{\mathrm{LOO}} = 7.56 \times 10^{-4}\).
These errors are around seven orders of magnitude higher than the errors observed for the Gaussian example.
The difference in the SLE convergence rate presumably originates from a difference in the underlying likelihood functions and posterior densities.
This is now investigated in more detail.
\par
\begin{figure}[htbp]
  \centering
  \begin{subfigure}[b]{\subWidth}
    \centering
    \includegraphics[height=\figHeight]{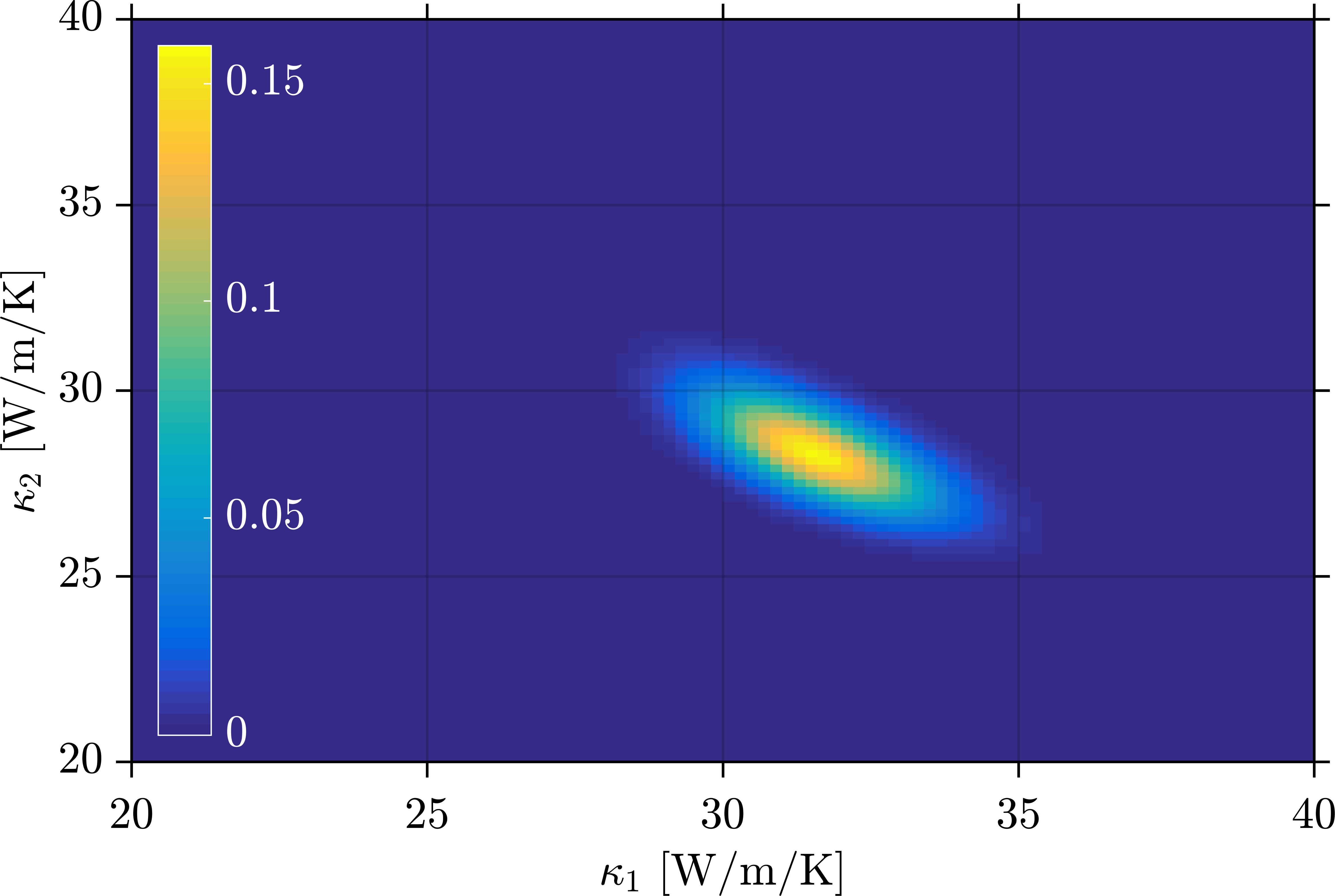}
    \caption{MCMC reference sample.}
    \label{fig:Heat:Post2D:MCMC}
  \end{subfigure}\hfill%
  \begin{subfigure}[b]{\subWidth}
    \centering
    \includegraphics[height=\figHeight]{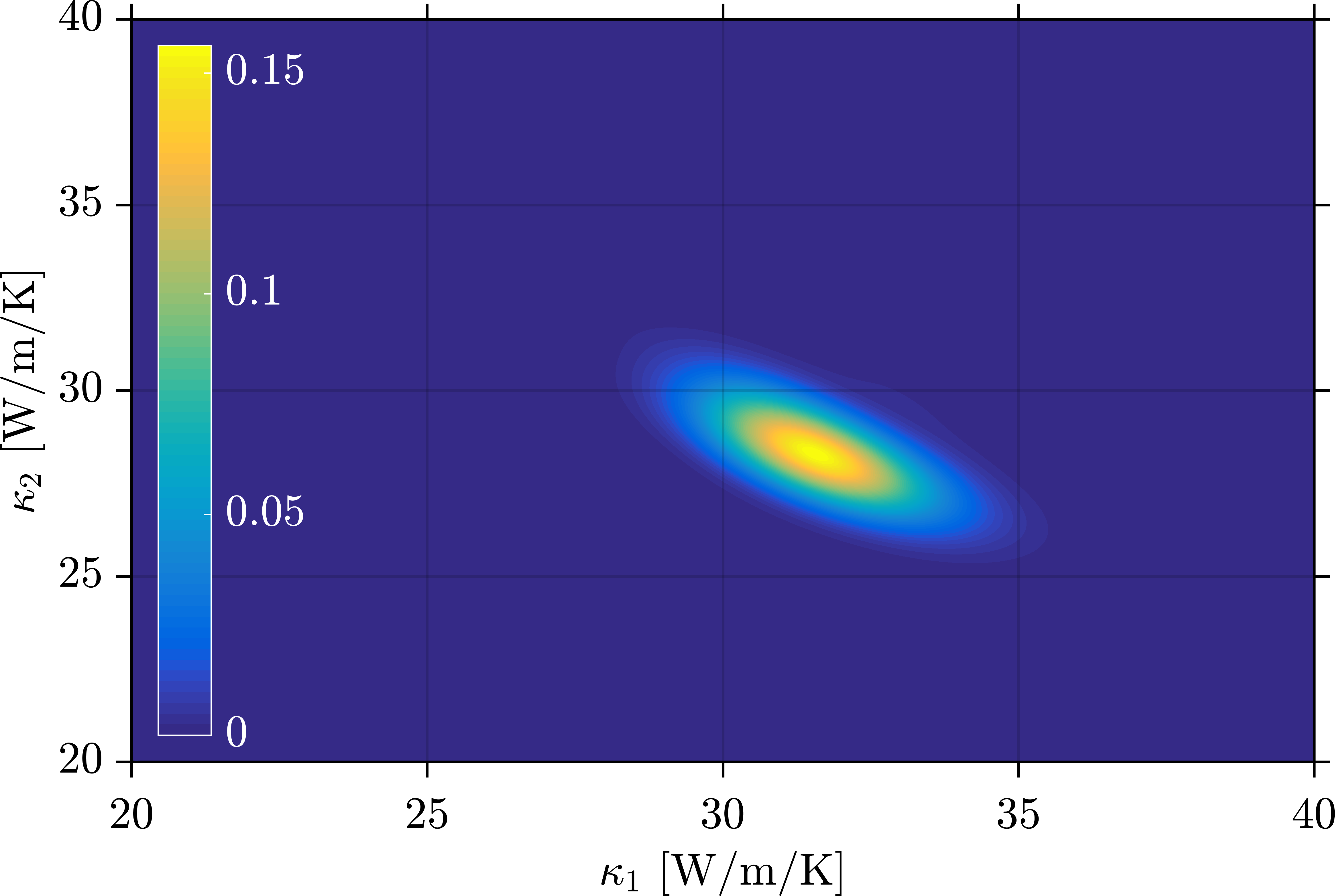}
    \caption{SLE with \(p = 50\).}
    \label{fig:Heat:Post2D:SLE}
  \end{subfigure}\\[1ex]%
  \begin{subfigure}[b]{\subWidth}
    \centering
    \includegraphics[width=\figWidth]{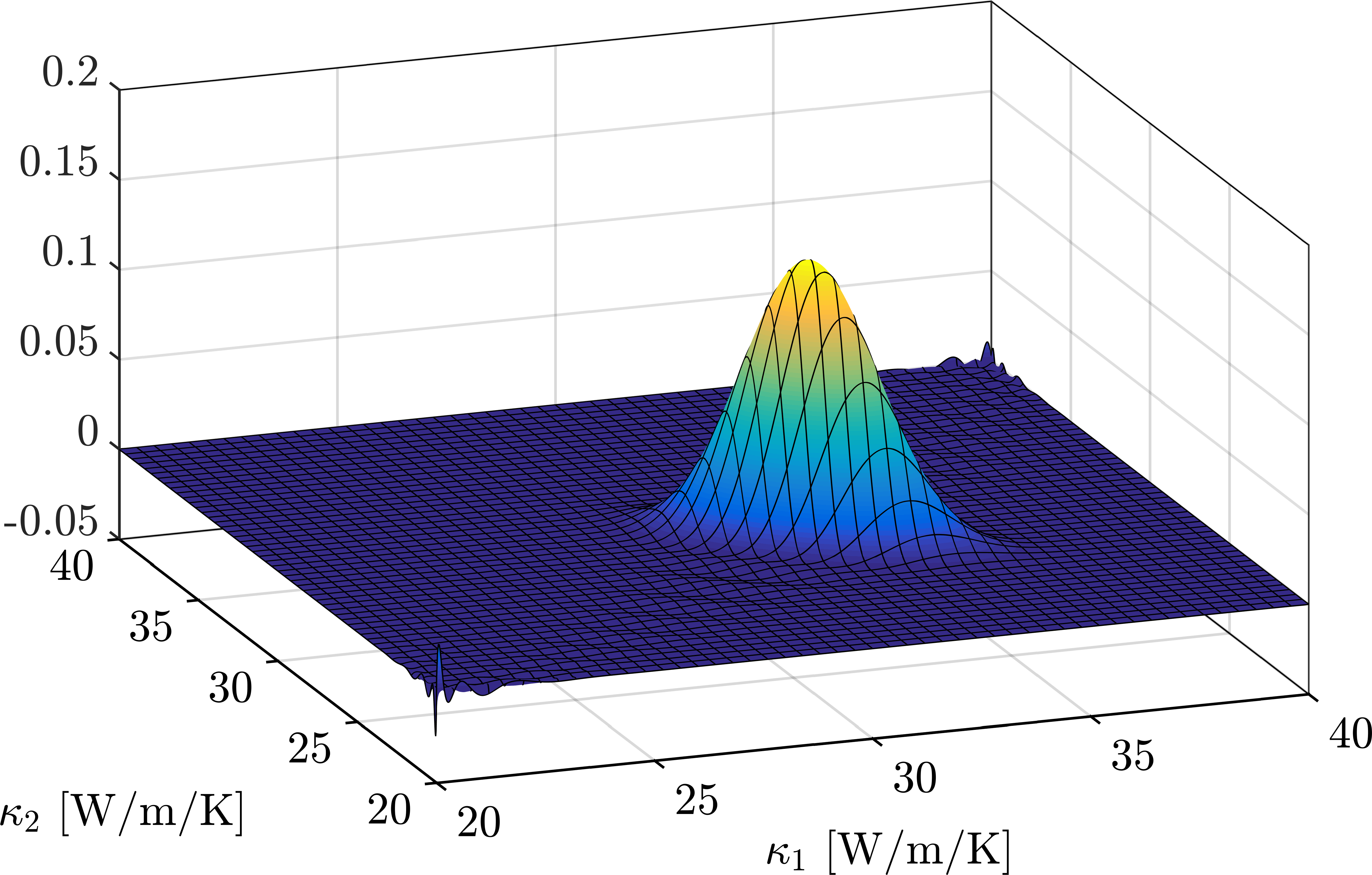}
    \caption{SLE with \(p = 50\).}
    \label{fig:Heat:Post3D:SLE}
  \end{subfigure}%
  \caption{2D IHCP: Joint posterior.}
  \label{fig:Heat:Post2D}
\end{figure}

 
A RWM approximation with \(10^7\) samples and the SLE-based emulation with \(p = 50\) of the posterior density
\(\pi(\kappa_1,\kappa_2 \cond \bm{T}) \approx \coeffL_{\bm{0}}^{-1} \hat{\mathcal{L}}_p(\kappa_1,\kappa_2) \pi(\kappa_1,\kappa_2)\) are depicted in \cref{fig:Heat:Post2D}.
In order to reduce the numerical cost of MCMC sampling, the FE model \(\mathcal{M}\) is replaced by a PCE surrogate \(\hat{\mathcal{M}}_p\).
For \(i = 1,\ldots,\dimData\), separate PCEs \(\hat{\mathcal{M}}_{i,p}\) of the temperature \(\perfect{T}_i = \mathcal{M}_{i,p}(\kappa_1,\kappa_2)\)
at the location \(\bm{r}_i\) are fitted as a function of the unknown conductivities.
After an appropriate transformation to standardized variables, tensorized Legendre polynomials up to degree \(p = 10\) act as the trial basis.
Based on an experimental design of the size \(K = 10^3\), the LOO errors of the regressions amount to about \(\epsilon_{\mathrm{LOO}} \approx 10^{-10}\).
Accordingly, the PCE is considered an adequate replacement of the full FE model.
Note that it would be also possible to use \(\hat{\mathcal{M}}_p\) as a forward model surrogate during the likelihood training runs.
\par 
The posteriors in \cref{fig:Heat:Post2D} can be compared to the posteriors of the Gaussian example in \cref{fig:Normal:Post2D} of the previous section.
Relative to the respective prior, the posterior of the thermal problem \(\pi(\kappa_1,\kappa_2 \cond \bm{T})\)
contains more information than the posterior of the normal problem \(\pi(\mu,\sigma \cond \bm{y})\),
i.e.\ the likelihood \(\mathcal{L}(\kappa_1,\kappa_2)\) has a slightly more peaked and localized structure than \(\mathcal{L}(\mu,\sigma)\).
In order to capture these different behaviors nearby and far from the posterior mode,
the SLEs \(\hat{\mathcal{L}}_p(\kappa_1,\kappa_2)\) and \(\hat{\mathcal{L}}_p(\mu,\sigma)\) require a different number of expansions terms.
The more localized the posterior modes are with respect to the prior, the more terms are required in order to achieve the cancellation in the tails.
Moreover, as opposed to \(\pi(\mu,\sigma \cond \bm{y})\) the posterior \(\pi(\kappa_1,\kappa_2 \cond \bm{T})\) exhibits a pronounced correlation structure.
In turn, this requires non-vanishing interaction terms.
As a consequence, the SLE \(\hat{\mathcal{L}}_p(\kappa_1,\kappa_2)\) of the IHCP example is less accurate than the SLE \(\hat{\mathcal{L}}_p(\mu,\sigma)\) of the Gaussian example.
This is also reflected in the fact that the posterior surrogate fluctuates and takes on negative values
around the points \([\underline{\kappa}_1,\underline{\kappa}_2]\) and \([\overline{\kappa}_1,\overline{\kappa}_2]\).
In order to see this more clearly, the SLE posterior surrogate from \cref{fig:Heat:Post2D:SLE} is plotted again from a different angle in \cref{fig:Heat:Post3D:SLE}.
A small wavelike posterior structure spans the parameter space between these corners.
These artifacts stem from an imperfect polynomial cancellation of the finite series approximation.
This stands in contrast to the posterior of the Gaussian example in \cref{fig:Normal:Post3D:SLE} where these phenomena were not observed.
\par
\begin{figure}[htbp]
  \centering
  \begin{subfigure}[b]{\subWidth}
    \centering
    \includegraphics[height=\figHeight]{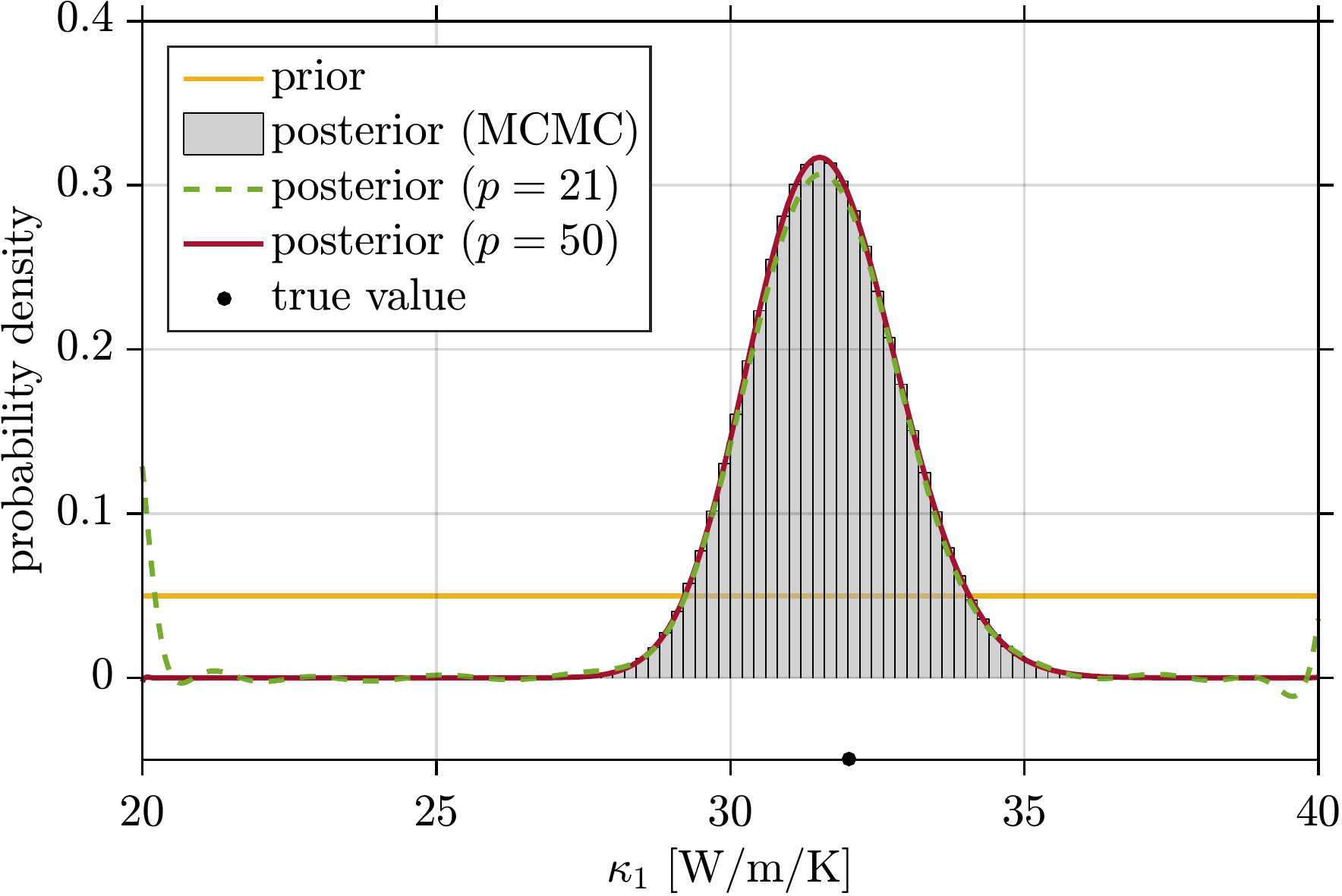}
    \caption{Thermal conductivity \(\kappa_1\).}
    \label{fig:Heat:Post1D:k1}
  \end{subfigure}\hfill%
  \begin{subfigure}[b]{\subWidth}
    \centering
    \includegraphics[height=\figHeight]{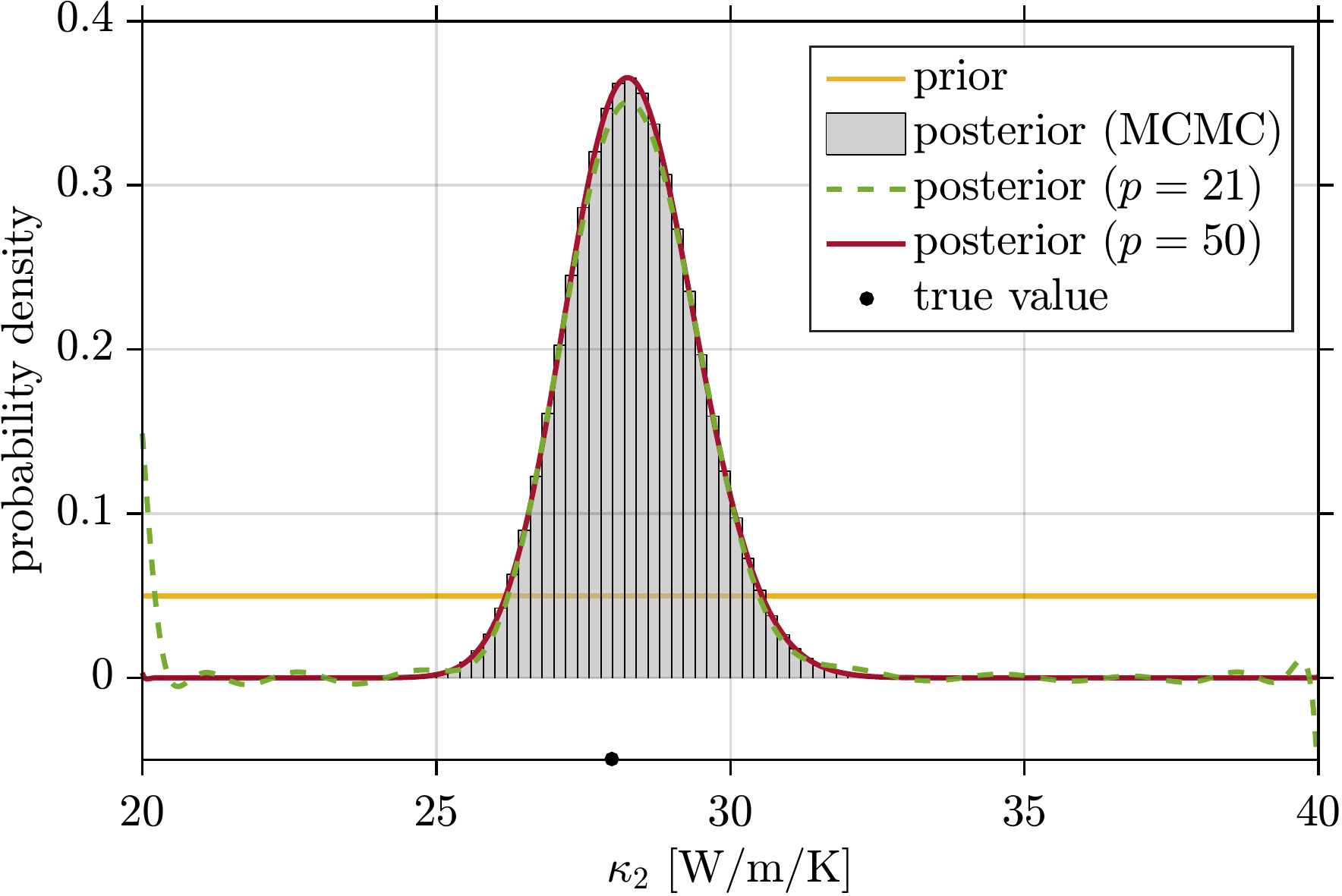}
    \caption{Thermal conductivity \(\kappa_2\).}
    \label{fig:Heat:Post1D:k2}
  \end{subfigure}%
  \caption{2D IHCP: Posterior marginals.}
  \label{fig:Heat:Post1D}
\end{figure}

 
Via \cref{eq:SLE:Marginal1D} the posterior marginals \(\pi(\kappa_1 \cond \bm{T})\) and \(\pi(\kappa_2 \cond \bm{T})\) can be extracted from the joint SLEs.
The resulting densities are shown in \cref{fig:Heat:Post1D} together with a histogram-based MCMC sample representation.
As it can be seen, for \(p = 50\) the marginals are captured fairly well,
while the moderate-order surrogate for \(p = 21\) still exhibits discrepancies at the bounds of the parameter space.
The approximation of the posterior marginals by sub-SLEs seems to be more accurate, at least in the sense of the maximum deviation,
than the approximation of the joint posterior \(\pi(\kappa_1,\kappa_2 \cond \bm{T})\) by the full SLE in \cref{fig:Heat:Post3D:SLE}.
This phenomenon can be explained through the absence of all non-constant polynomial terms in the variables that are marginalized out.

\subsubsection{Quantities of interest}
Now we investigate how well one can extract the statistically interesting quantities.
Results from SLEs with varying \(K\) and \(p\) are compared with the results from MCMC sampling.
A summary of the findings is provided in \cref{tab:Heat:StatisticalQuantities}.
The LOO error \(\epsilon_{\mathrm{LOO}}\) of various SLEs is shown together with some basic posterior characteristics obtained by a postprocessing of the SLE coefficients.
For \(j = 1,2\) the posterior mean \(\mathds{E}[\kappa_j \cond \bm{T}]\) and the standard deviation \(\mathrm{Std}[\kappa_j \cond \bm{T}] = \mathrm{Var}[\kappa_j \cond \bm{T}]^{1/2}\)
of the posterior distribution are given in physical units of \(\unit[]{W/m/K}\).
In addition, the model evidence \(\scale\) and the linear coefficient of correlation
\(\rho[\kappa_1,\kappa_2 \cond \bm{T}] = \mathrm{Cov}[\kappa_1,\kappa_2 \cond \bm{T}] / \mathrm{Std}[\kappa_1 \cond \bm{T}] / \mathrm{Std}[\kappa_2 \cond \bm{T}]\) are specified.
In comparison to \cref{tab:Normal:StatisticalQuantities}, where the results for the non-conjugate normal example are listed, the SLE results for the IHCP match their MCMC counterparts less accurately.
Nevertheless, it can be observed that the lowest-degree quantities of inferential interest can be extracted with a comparably small experimental design and relatively low number of regressors,
say with \(K = 1 \times 10^4\) and \(p = 29\).
Note that all the estimates attain admissible values.
\begin{table}[htbp]
  \caption{2D IHCP: Statistical quantities.}
  \label{tab:Heat:StatisticalQuantities}
  \centering
  \begin{tabular}{cccccccccc}
    \toprule
    & \(K\) & \(p\) & \(\epsilon_{\mathrm{LOO}}\)
    & \(\scale\) \([10^{-1}]\) & \(\mathds{E}[\kappa_1 \cond \bm{T}]\) & \(\mathds{E}[\kappa_2 \cond \bm{T}]\)
    & \(\mathrm{Std}[\kappa_1 \cond \bm{T}]\) & \(\mathrm{Std}[\kappa_2 \cond \bm{T}]\) & \(\rho[\kappa_1,\kappa_2 \cond \bm{T}]\) \\
    \midrule
    \multirow{6}{*}{\rotatebox[origin=c]{90}{SLE}}
    & \(5 \times 10^2\) & \(5\)  & \(8.24 \times 10^{-1}\) & \(8.45\) & \(31.33\) & \(28.36\) & \(1.74\) & \(1.33\) & \(\phantom{-}0.28\) \\
    & \(1 \times 10^3\) & \(9\)  & \(6.08 \times 10^{-1}\) & \(7.81\) & \(31.40\) & \(28.22\) & \(2.02\) & \(1.53\) & \(\phantom{-}0.15\) \\
    & \(5 \times 10^3\) & \(21\) & \(1.50 \times 10^{-1}\) & \(7.47\) & \(31.32\) & \(28.13\) & \(2.16\) & \(1.61\) & \(\phantom{-}0.34\) \\
    & \(1 \times 10^4\) & \(29\) & \(5.79 \times 10^{-2}\) & \(7.21\) & \(31.56\) & \(28.30\) & \(1.61\) & \(1.39\) & \(-0.05\) \\
    & \(5 \times 10^4\) & \(35\) & \(1.63 \times 10^{-2}\) & \(7.18\) & \(31.62\) & \(28.34\) & \(1.24\) & \(1.08\) & \(-0.75\) \\
    & \(1 \times 10^5\) & \(50\) & \(7.56 \times 10^{-4}\) & \(7.18\) & \(31.62\) & \(28.33\) & \(1.26\) & \(1.10\) & \(-0.68\) \\
    \midrule
    \multicolumn{4}{c}{(MC)MC}                             & \(7.17\) & \(31.62\) & \(28.33\) & \(1.26\) & \(1.09\) & \(-0.68\) \\
    \bottomrule
  \end{tabular}
\end{table}

\subsection{6D inverse heat conduction}
In the previous sections it was demonstrated that likelihood functions can be indeed spectrally expanded and that the posterior density with its moments can be computed accordingly.
For low-dimensional problems the SLE convergence behavior up to a high degree was studied by monitoring the LOO error.
It was shown that the expansion error can be arbitrarily reduced by increasing the order of the expansion and adding samples to the experimental design.
While this is reassuring to know, it does not help in solving higher-dimensional problems
for which the computation of high-order expansions is exacerbated by the curse of dimensionality.
Hence, now we want to investigate the applicability of SLEs and aSLEs in an inverse problem of moderate dimension.
\par
\begin{figure}[htbp]
  \centering
  \includegraphics[width=\femWidth]{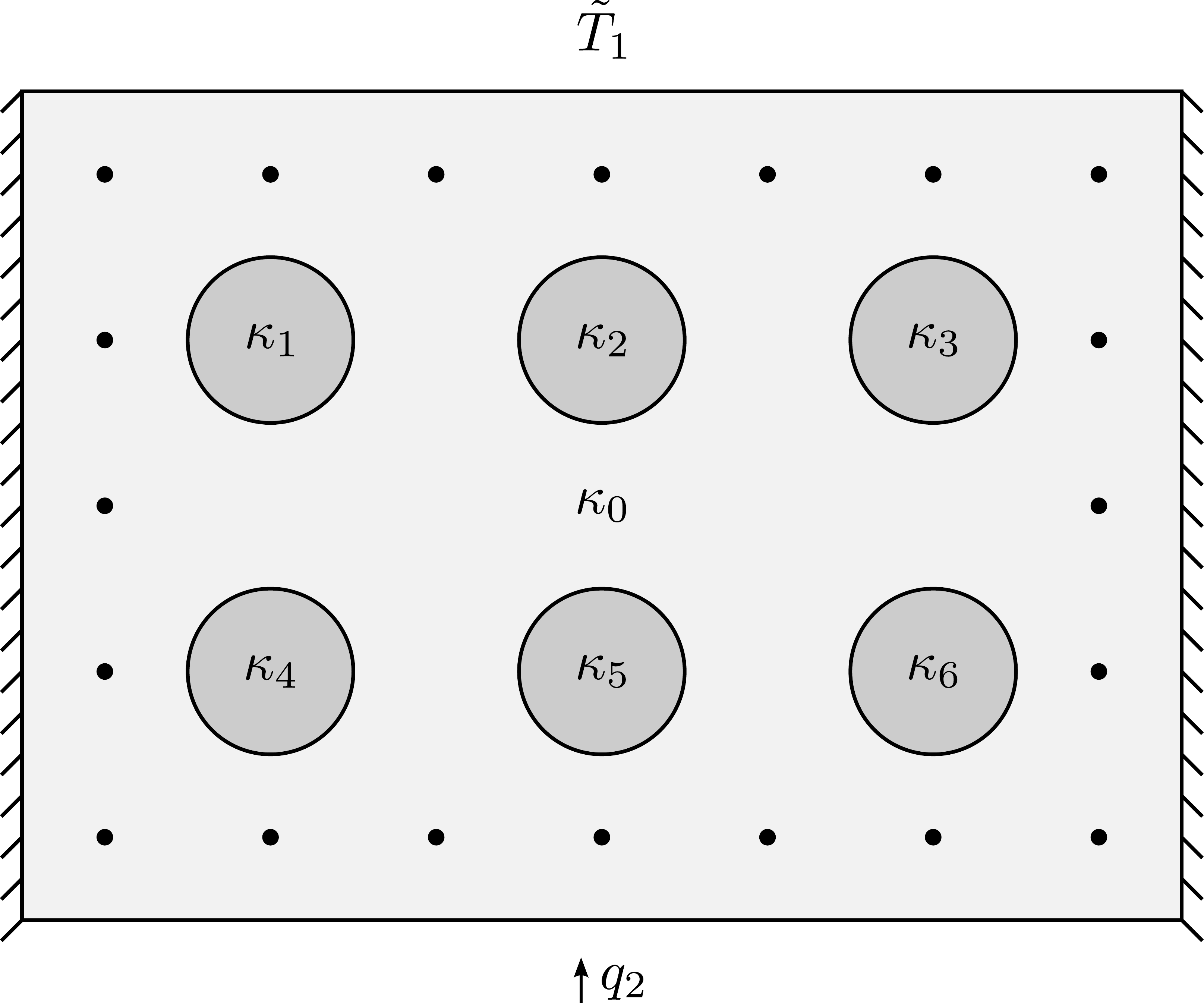}
  \caption{6D IHCP: Heat conduction setup.}
  \label{fig:Thermal:HeatConduction}
\end{figure}

 
An IHCP in two spatial dimensions with six unknown conductivities is considered in this section.
The setup of the problem is shown in \cref{fig:Thermal:HeatConduction}.
The \(\dimParam = 6\) unknown conductivities \(\bm{\kappa} = (\kappa_1,\ldots,\kappa_6)^\top\) are inferred
with \(\dimData = 20\) noisy measurements \(\bm{T} = (T_{1},\ldots,T_{20})^\top\) of the temperature field \(\perfect{T}\).
We set \(\kappa_0 = \unit[30]{W/m/K}\) and \(\bm{\kappa} = \unit[(20,24,\ldots,40)^\top]{W/m/K}\).
The prior is set to a multivariate lognormal distribution \(\pi(\bm{\kappa}) = \prod_{i=1}^6 \pi(\kappa_i)\) with independent marginals
\(\pi(\kappa_i) = \mathcal{LN}( \kappa_i \cond \mu_0,\sigma_0^2)\) with \(\mu_0 = \unit[30]{W/m/K}\) and \(\sigma_0 = \unit[6]{W/m/K}\).
These parameters describe the mean \(\mu_0 = \mathds{E}[\kappa_i]\) and standard deviation \(\sigma_0 = \mathrm{Std}[\kappa_i]\) of the lognormal prior.
They are related to the parameters of the associated normal distribution \(\mathcal{N}( \log(\kappa_i) \cond \lambda_0,\varsigma_0^2)\)
via \(\mu_0 = \exp(\lambda_0 + \varsigma_0^2 / 2)\) and \(\sigma_0^2 = (\exp(\varsigma_0^2) - 1) \exp(2 \lambda_0 + \varsigma_0^2)\).
Otherwise than that, the problem setup is exactly as described in the previous section,
i.e.\ the likelihood function is given as \(\mathcal{L}(\bm{\kappa}) = \mathcal{N}(\bm{T} \cond \mathcal{M}(\bm{\kappa}),\bm{\Sigma})\).
In accordance with this setup, in the following synthetic data are simulated and analyzed in order to compute
the joint posterior \(\pi(\bm{\kappa} \cond \bm{T}) = \scale^{-1} \mathcal{L}(\bm{\kappa}) \pi(\bm{\kappa})\).

\subsubsection{Posterior density}
The unknowns are represented as \(\kappa_i = \exp(\lambda_0 + \varsigma_0 \xi_i)\) in terms of the standardized variables
\(\xi_i \in \mathds{R}\) with Gaussian weight functions \(\mathcal{N}(\xi_i \cond 0,1)\).
A spectral expansion \(\hat{\mathcal{L}}_p\) in tensorized Hermite polynomials is then computed for \(p = 5\) and \(K = 5 \times 10^4\).
The errors of the likelihood approximation are estimated as \(\epsilon_{\mathrm{Emp}} = 8.81 \times 10^{-1}\) and \(\epsilon_{\mathrm{LOO}} = 9.14 \times 10^{-1}\).
As compared to the low-dimensional examples that were studied before, these are large errors.
An auxiliary reference density \(\newBase(\bm{\kappa}) = \prod_{i=1}^6 \newBase(\kappa_i)\) is then constructed as a
multivariate lognormal with independent marginals \(\newBase(\kappa_i) = \mathcal{LN}(\kappa_i \cond \mu_i,\sigma_i^2)\).
The parameters of the latter are chosen as the means \(\mu_i = \mathds{E}[\kappa_i \cond \bm{T}]\) and standard deviations
\(\sigma_i = \mathrm{Std}[\kappa_i \cond \bm{T}]\) of the posterior surrogate corresponding to the coefficients of SLE \(\hat{\mathcal{L}}_p\).
We remark that this is a simple two-step procedure and that a more refined usage of the reference change would certainly lead to more sophisticated approaches.
Subsequently, an aSLE \(\hat{\auxQuantity}_p\) with \(p = 5\) and  \(K = 5 \times 10^4\) is computed.
The errors amount to \(\epsilon_{\mathrm{Emp}} = 4.81 \times 10^{-1}\) and \(\epsilon_{\mathrm{LOO}} = 6.24 \times 10^{-1}\).
Notwithstanding that these errors are smaller than the corresponding errors of the SLE, they are still large as compared to the previous examples.
Since these errors are now measured with respect to the auxiliary density which is expectedly closer to the true posterior than the prior is,
the aSLE presumably leads to a more accurate posterior surrogate.
\par
\begin{figure}[p]
  \centering
  \begin{subfigure}[b]{\subWidth}
    \centering
    \includegraphics[height=\figHeight]{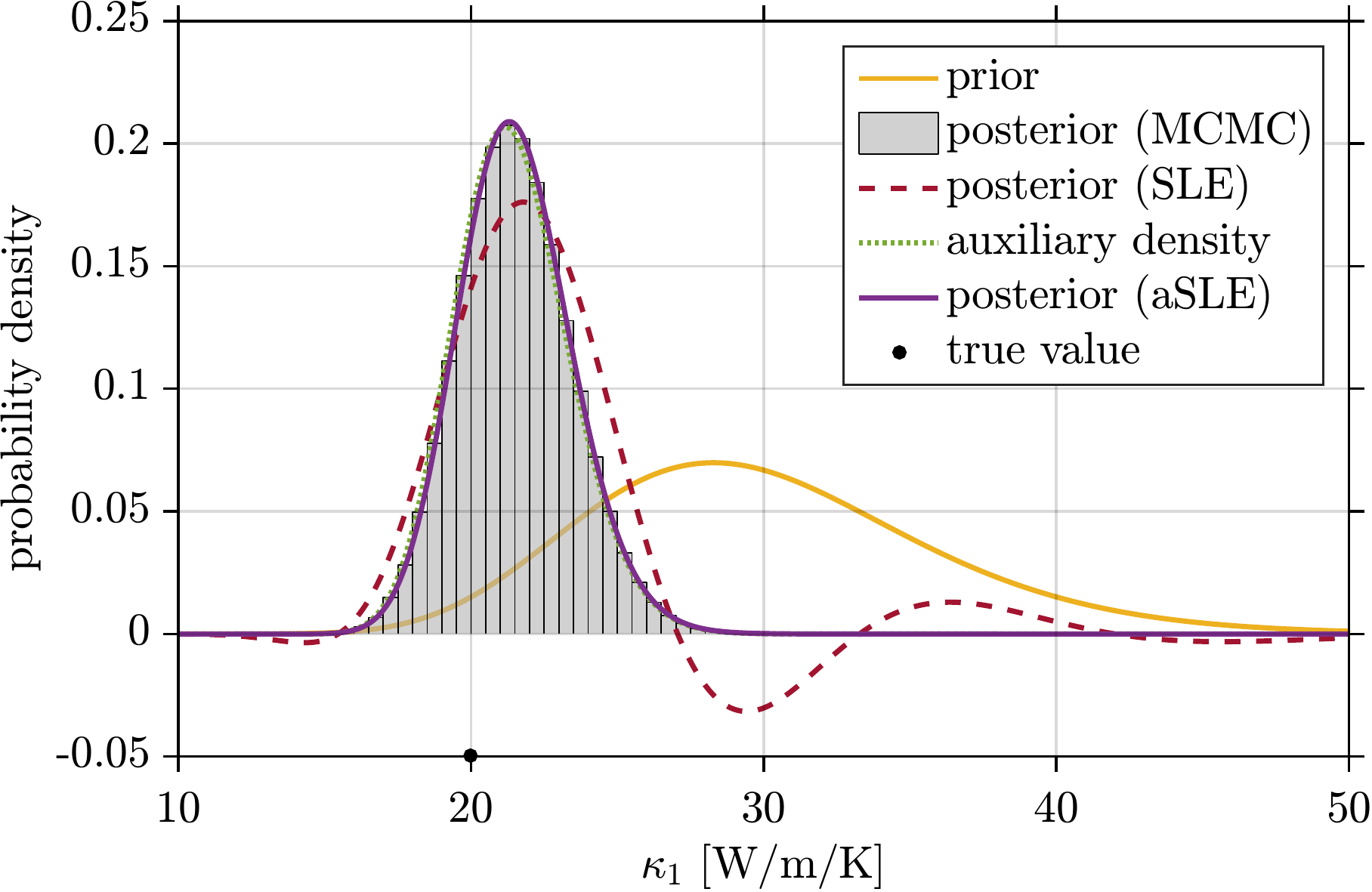}
    \caption{Thermal conductivity \(\kappa_1\).}
    \label{fig:Thermal:Post1D:k1}
  \end{subfigure}\hfill%
  \begin{subfigure}[b]{\subWidth}
    \centering
    \includegraphics[height=\figHeight]{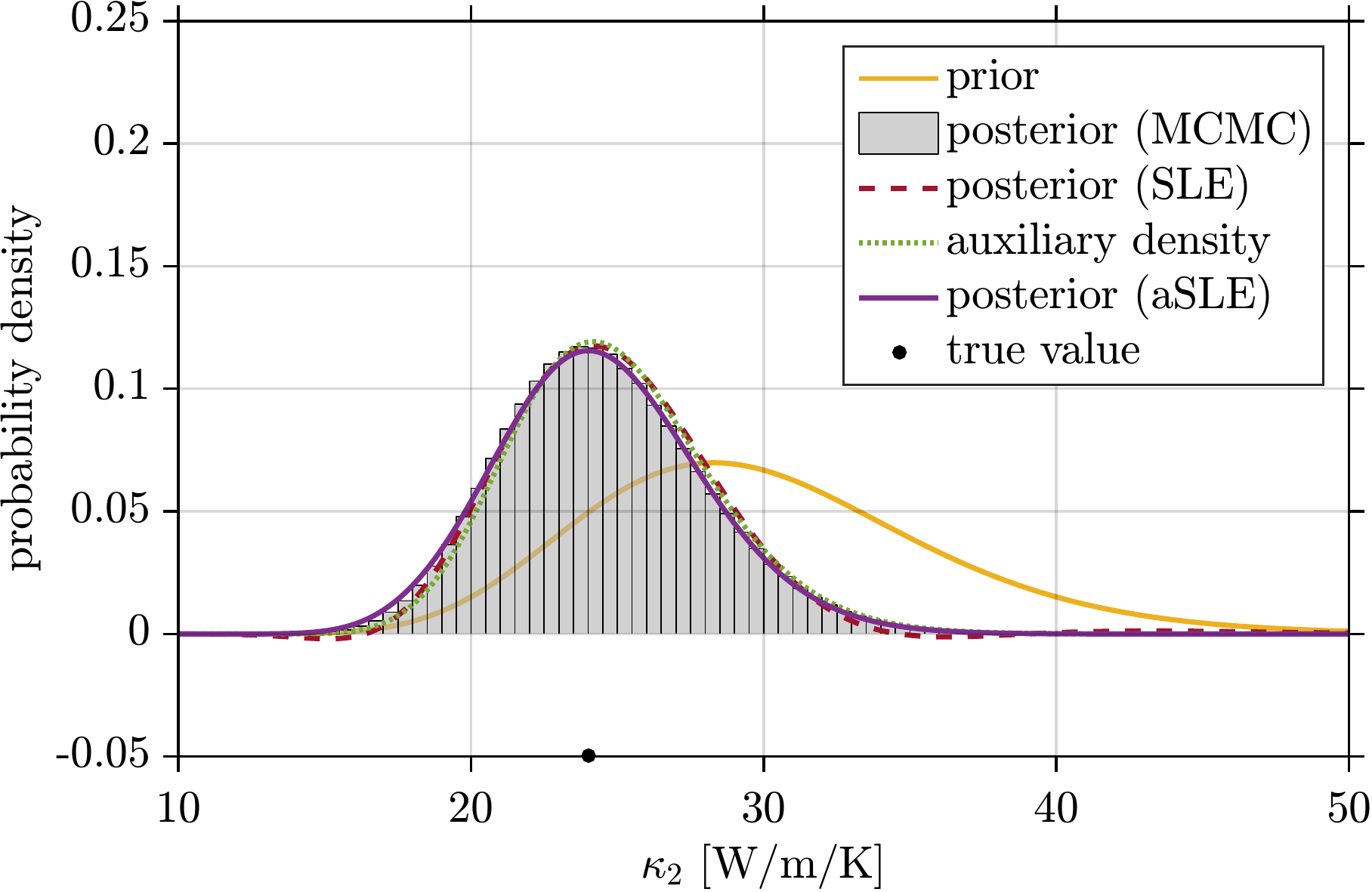}
    \caption{Thermal conductivity \(\kappa_2\).}
    \label{fig:Thermal:Post1D:k2}
  \end{subfigure}\\[3ex]%
  \begin{subfigure}[b]{\subWidth}
    \centering
    \includegraphics[height=\figHeight]{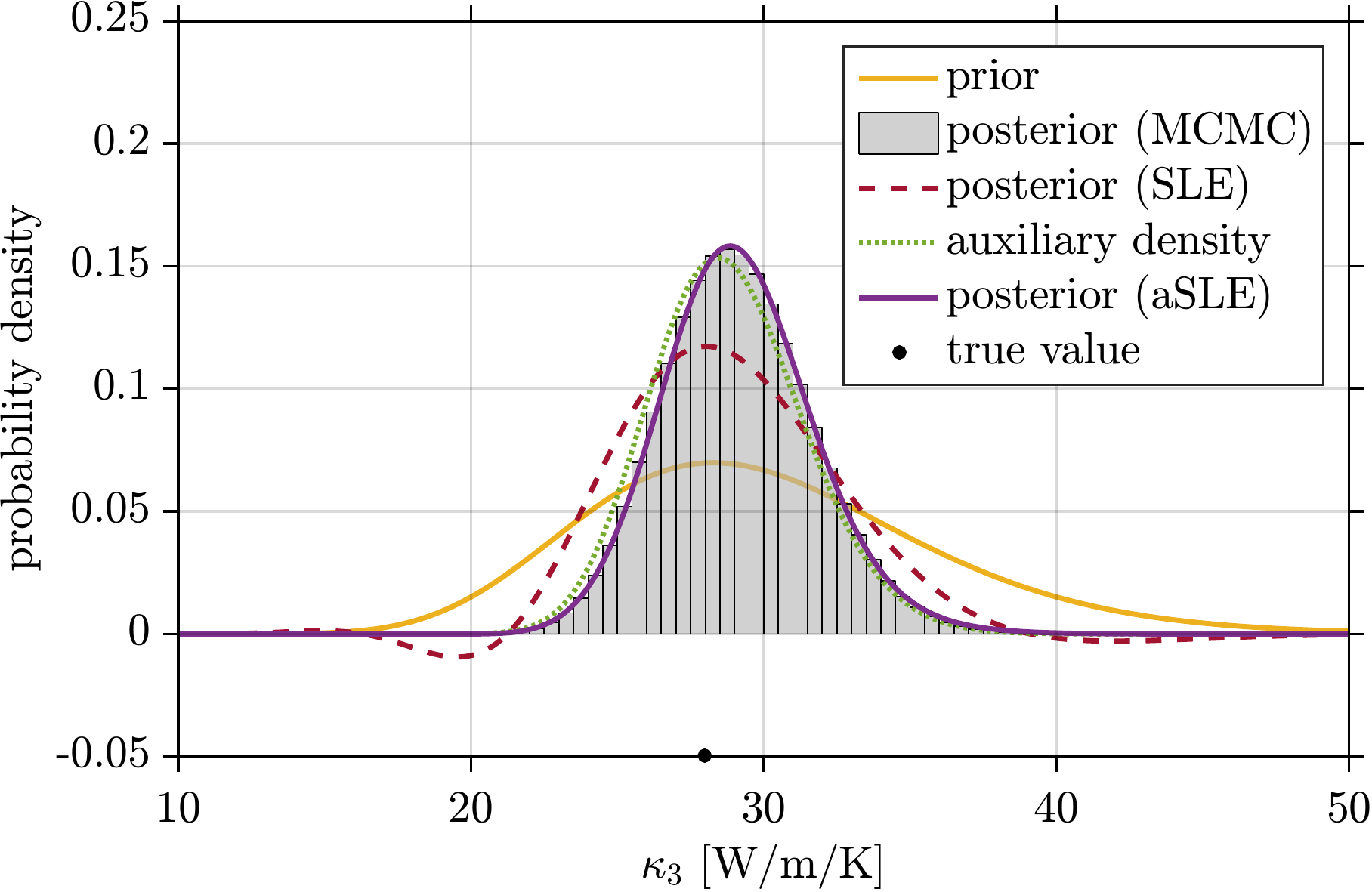}
    \caption{Thermal conductivity \(\kappa_3\).}
    \label{fig:Thermal:Post1D:k3}
  \end{subfigure}\hfill%
  \begin{subfigure}[b]{\subWidth}
    \centering
    \includegraphics[height=\figHeight]{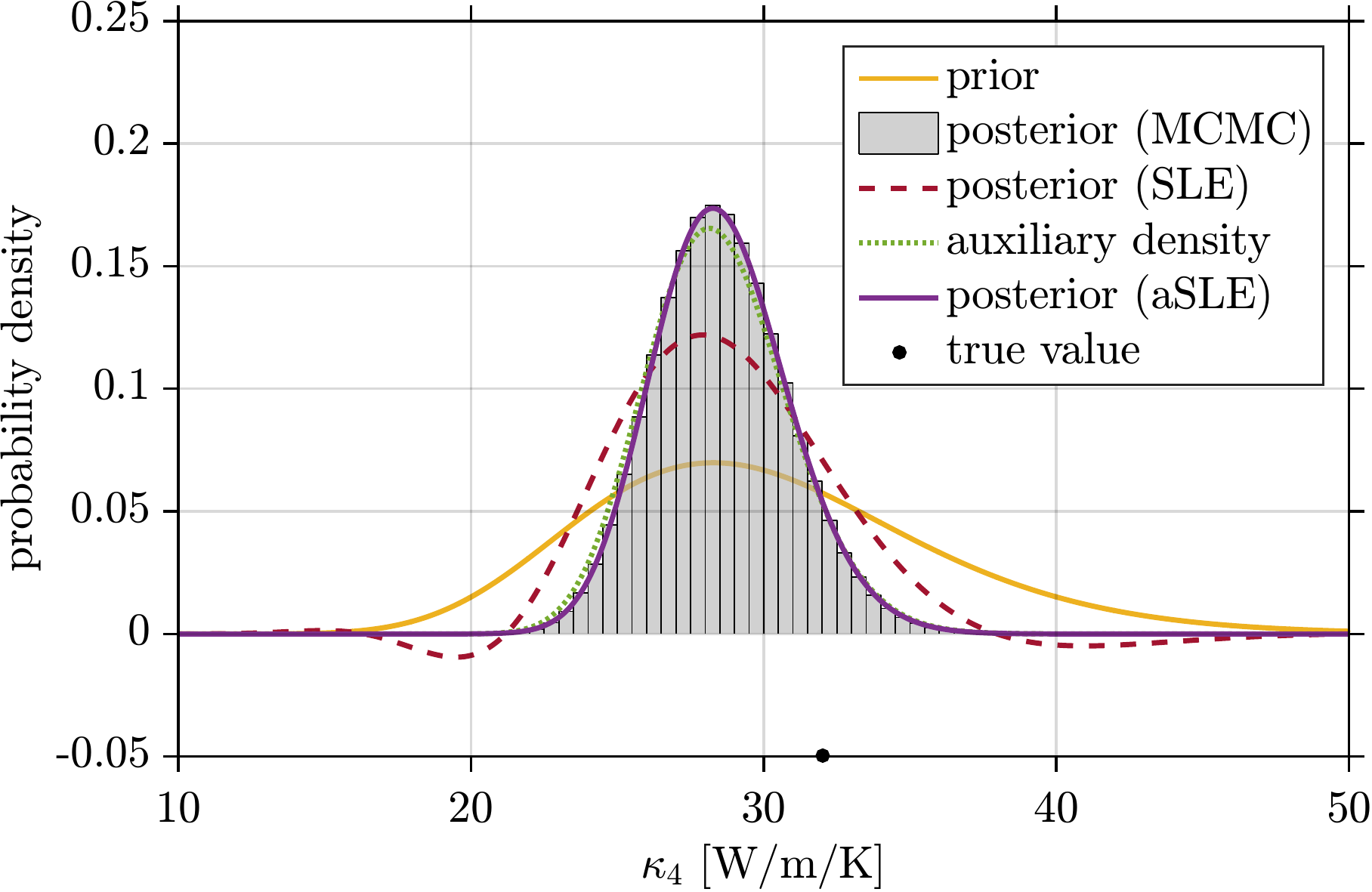}
    \caption{Thermal conductivity \(\kappa_4\).}
    \label{fig:Thermal:Post1D:k4}
  \end{subfigure}\\[3ex]%
  \begin{subfigure}[b]{\subWidth}
    \centering
    \includegraphics[height=\figHeight]{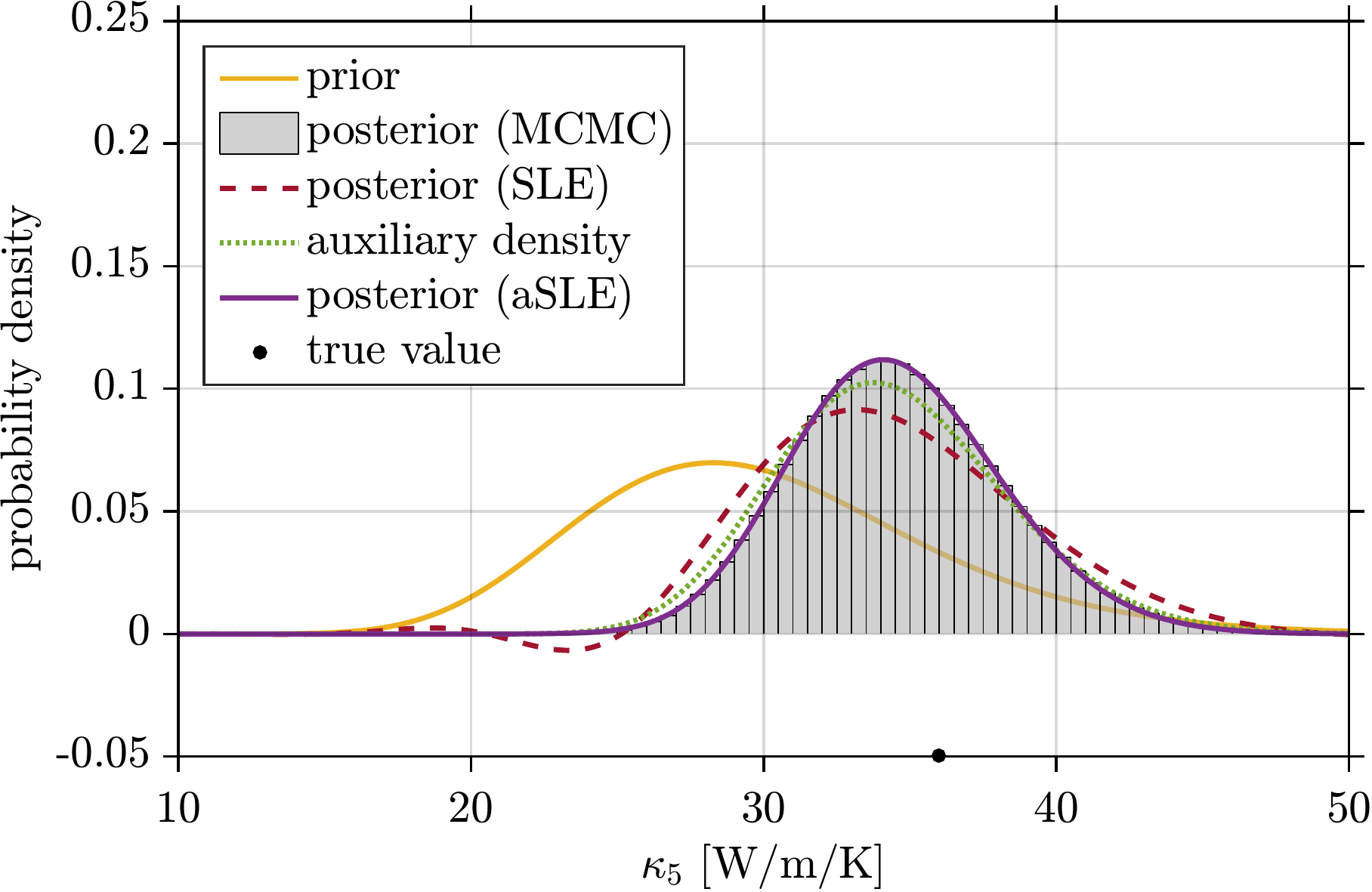}
    \caption{Thermal conductivity \(\kappa_5\).}
    \label{fig:Thermal:Post1D:k5}
  \end{subfigure}\hfill%
  \begin{subfigure}[b]{\subWidth}
    \centering
    \includegraphics[height=\figHeight]{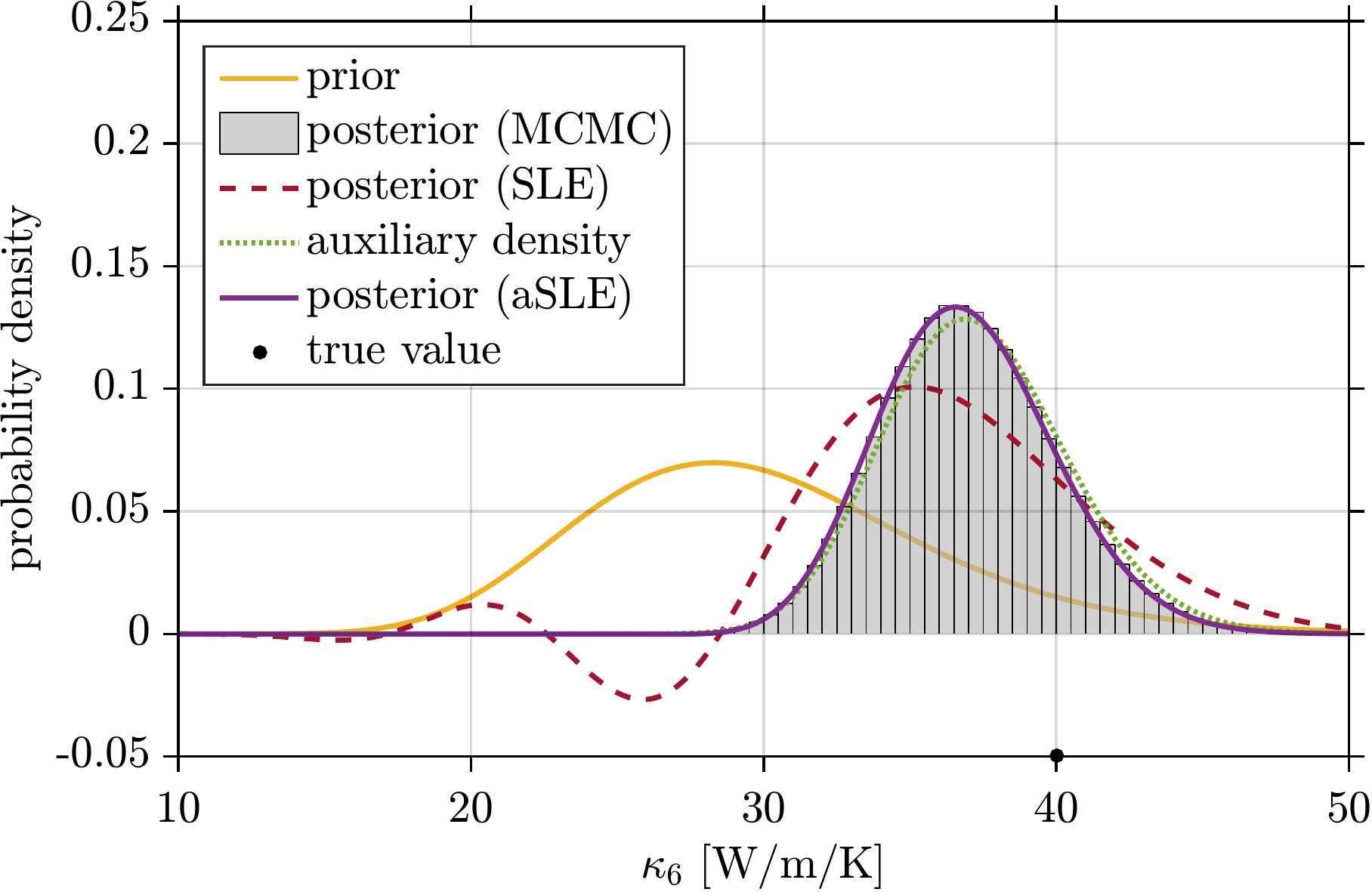}
    \caption{Thermal conductivity \(\kappa_6\).}
    \label{fig:Thermal:Post1D:k6}
  \end{subfigure}%
  \caption{6D IHCP: Posterior marginals.}
  \label{fig:Thermal:Post1D}
\end{figure}

 
From the previously computed SLE \(\hat{\mathcal{L}}(\bm{\kappa})\) and the aSLE \(\hat{\auxQuantity}_p(\bm{\kappa})\)
approximations of the joint posterior density are computed via \cref{eq:SLE:Posterior,eq:SLE:BaselineChange:Posterior}.
The obtained surrogates \(\pi(\bm{\kappa} \cond \bm{T}) \approx \hat{\mathcal{L}}_p(\bm{\kappa}) \pi(\bm{\kappa}) / \coeffL_{\bm{0}}\) and
\(\pi(\bm{\kappa} \cond \bm{T}) \approx \hat{\auxQuantity}_p(\bm{\kappa}) \newBase(\bm{\kappa}) / \coeffBaseL_{\bm{0}}\) are now compared to each other.
We start with the one-dimensional marginals that can be compiled by collecting terms from the full expansions based on \cref{eq:SLE:Marginal1D}.
For \(j = 1,\ldots,6\) the marginals \(\pi(\kappa_j \cond \bm{T})\) that are extracted that way are shown in \cref{fig:Thermal:Post1D}.
The marginal priors \(\pi(\kappa_j)\) and the auxiliary densities \(\newBase(\kappa_j)\) are shown, too.
While the marginals that are taken from the SLE slightly deviate from their MCMC counterparts, the marginals based on the aSLE match their references perfectly well.
The reason is that the posterior can be easier represented as a small adjustment of the auxiliary density than as a large correction to the prior.
Thus, with the same expansion order the posterior is more accurately represented through the aSLE than through the SLE.
Regarding the size of the error estimates, it is surprising that the marginals can be retrieved that well with the aSLE.
Even though the SLE-based posterior approximations can hardly be interpreted as proper probability densities, i.e.\ they conspicuously take on negative values,
the moments are recovered sufficiently well for the construction of the auxiliary reference density.
\par
\begin{figure}[htbp]
  \centering
  \begin{subfigure}[b]{\subWidth}
    \centering
    \includegraphics[width=\figWidth]{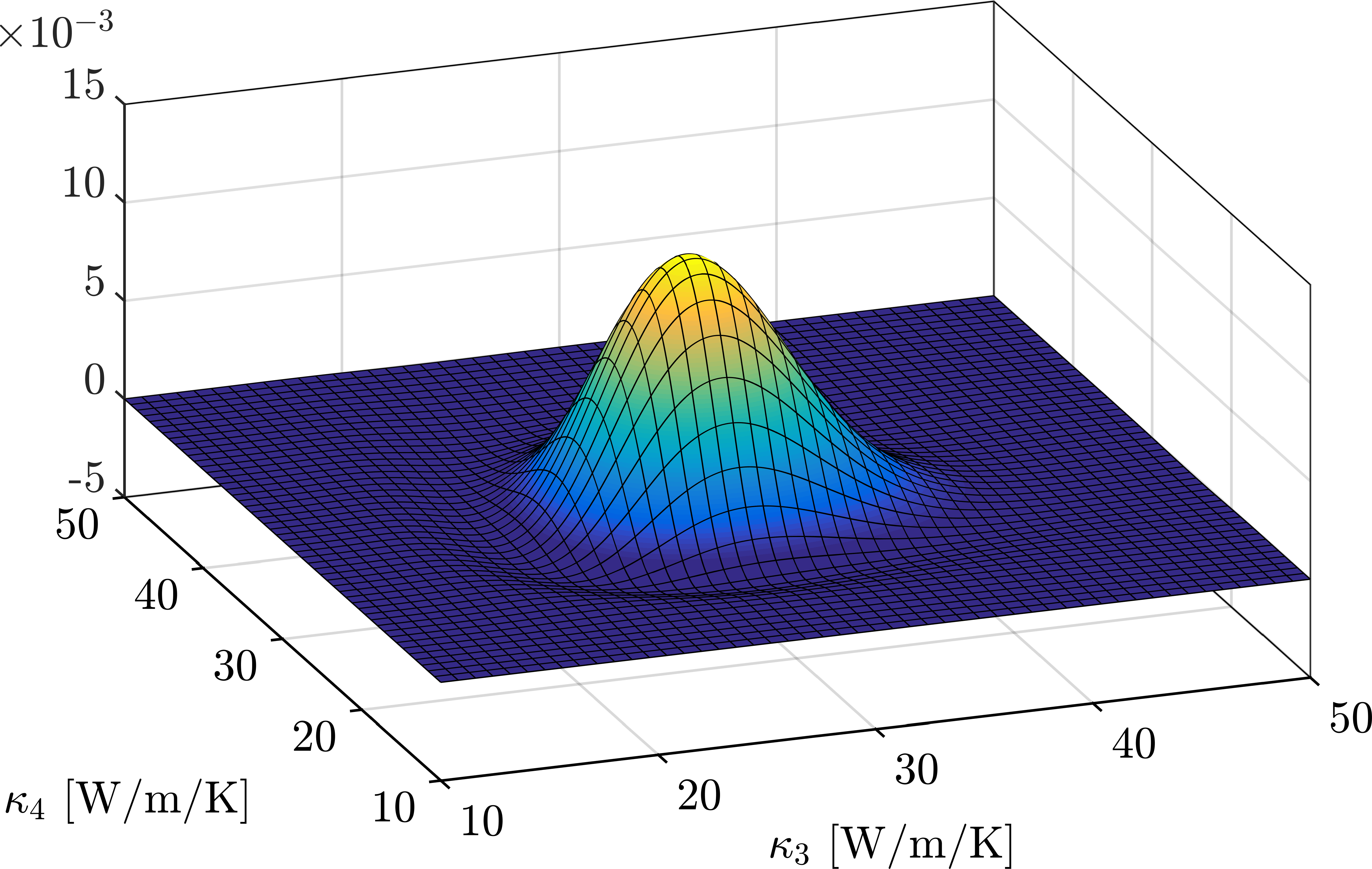}
    \caption{SLE with \(p = 5\).}
    \label{fig:Thermal:Post2D:SLE}
  \end{subfigure}\hfill%
  \begin{subfigure}[b]{\subWidth}
    \centering
    \includegraphics[width=\figWidth]{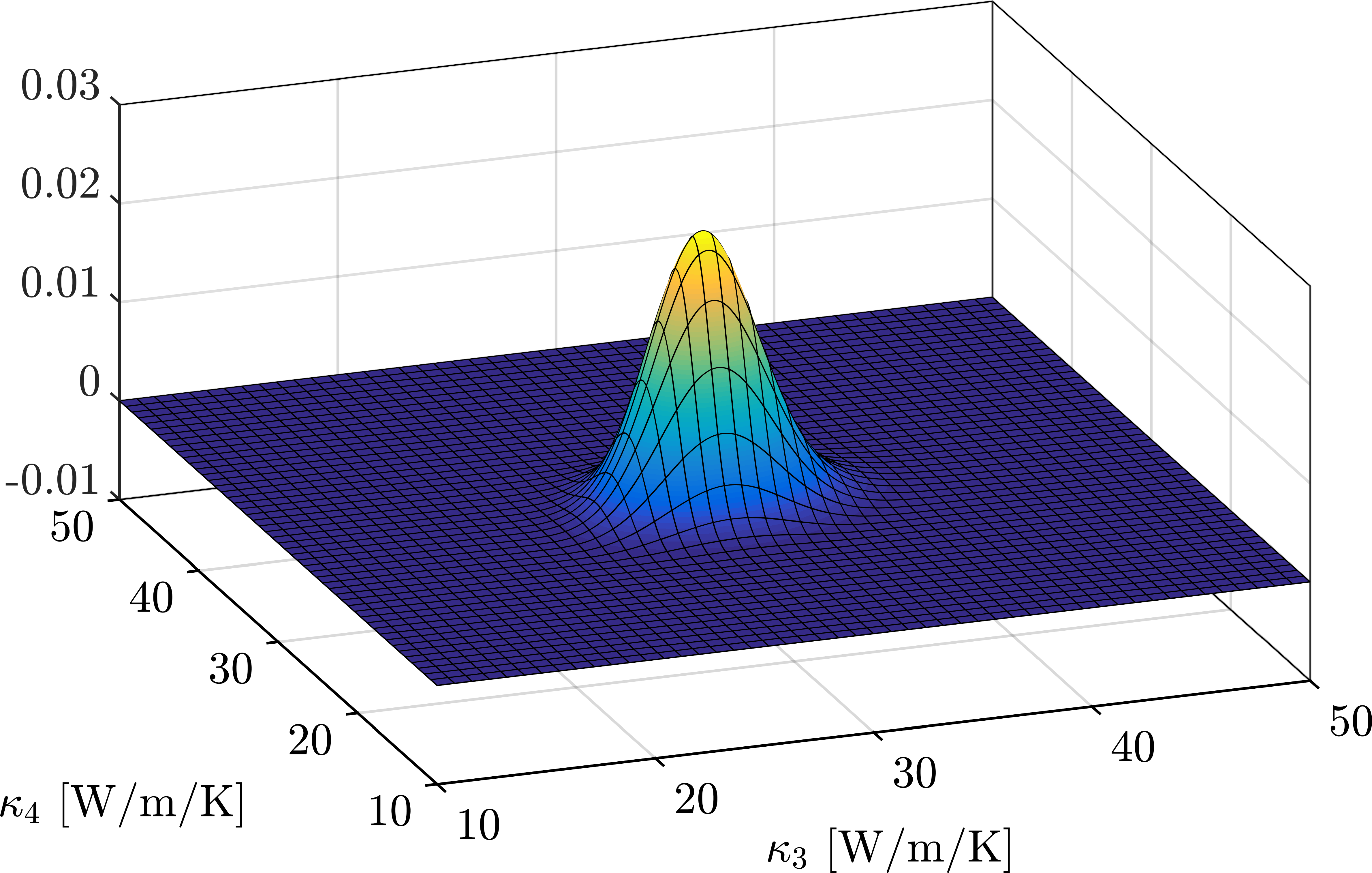}
    \caption{aSLE with \(p = 5\).}
    \label{fig:Thermal:Post2D:aSLE}
  \end{subfigure}\\[1ex]%
  \begin{subfigure}[b]{\subWidth}
    \centering
    \includegraphics[width=\figWidth]{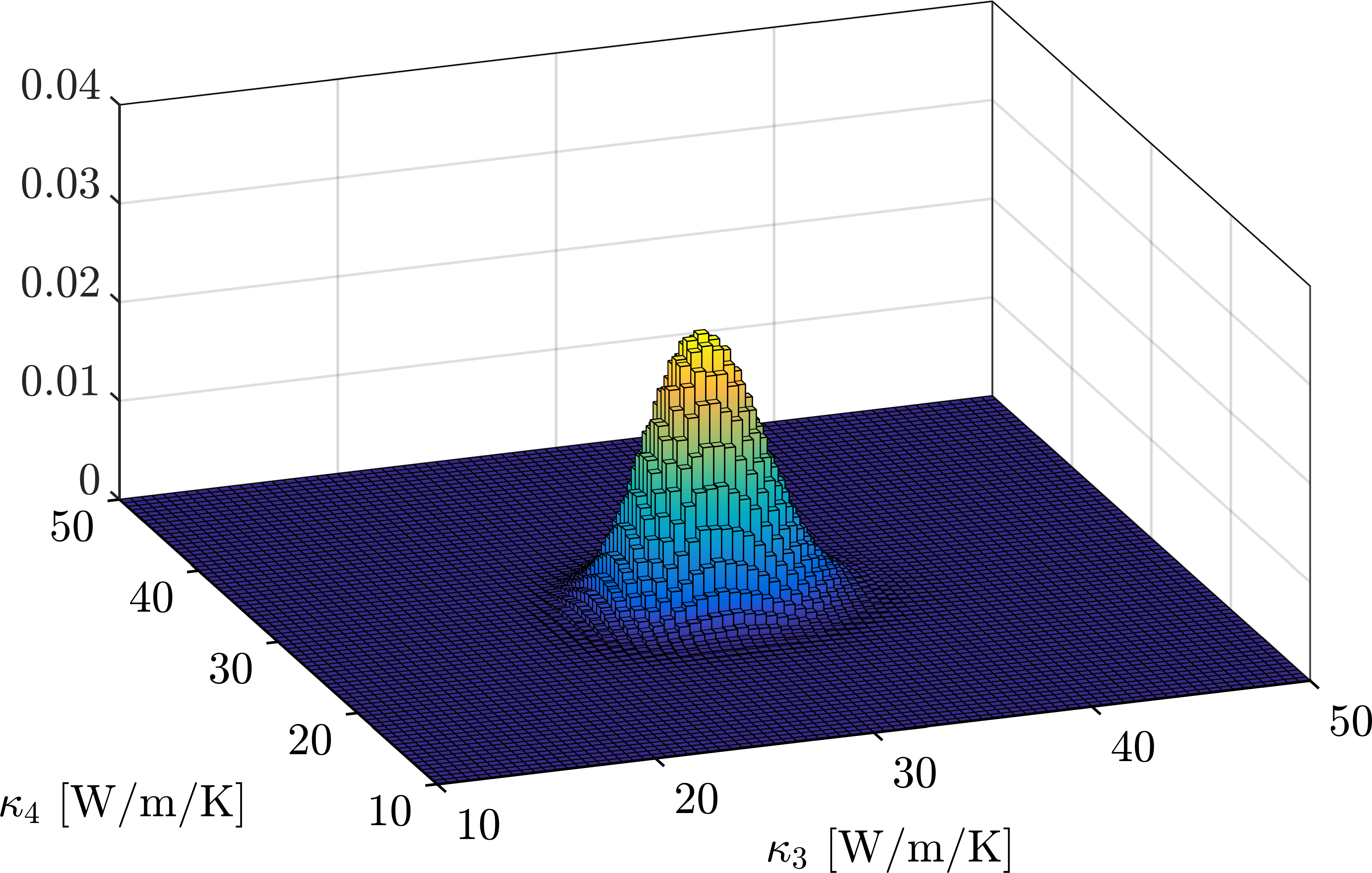}
    \caption{MCMC reference sample.}
    \label{fig:Thermal:Post2D:MCMC}
  \end{subfigure}%
  \caption{6D IHCP: Posterior marginals.}
  \label{fig:Thermal:Post2D}
\end{figure}

 
On the basis of \cref{eq:SLE:Marginal2D} the two-dimensional posterior marginals \(\pi(\kappa_j,\kappa_k \cond \bm{T})\) can be constructed from the full expansions.
For \(j = 3\) and \(k = 4\) the posterior marginal for the SLE \(\hat{\mathcal{L}}_p\) is shown in \cref{fig:Thermal:Post2D:SLE}.
The same two-dimensional distribution is depicted in \cref{fig:Thermal:Post2D:aSLE} for the aSLE \(\hat{\auxQuantity}_p\).
A histogram of the MCMC sample is provided in \cref{fig:Thermal:Post2D:MCMC} as a reference.
As already found in \cref{fig:Thermal:Post1D:k3,fig:Thermal:Post1D:k4} for instance,
in \cref{fig:Thermal:Post2D} the aSLE-based surrogate appears to be almost exact whereas the SLE-based one is flattened out.
Since the aSLE captures the true posterior density more accurately than the SLE, we expect similar findings for the posterior moments.

\subsubsection{Quantities of interest}
Finally we compute the model evidence and the first posterior moments with the aid of
\cref{eq:SLE:ScaleFactor,eq:SLE:BaselineChange:ScaleFactor} and \cref{eq:SLE:PosteriorMargMean,eq:SLE:PosteriorMargVariance,eq:SLE:PosteriorCovariance}.
For the aSLE \(\hat{\auxQuantity}_p\) the analysis proceeds analogously to the SLE \(\hat{\mathcal{L}}_p\).
In \cref{tab:Thermal:StatisticalQuantities} a summary of the results is given.
As it can be taken from the table, the aSLE consistently gives more accurate estimates of the reference values.
This fulfills our earlier expectations.
Regarding the inaccuracy of the SLE-based posterior marginals and the concerns about interpreting them as probability densities, 
the quality of the SLE-based estimates of the moments surpasses our expectations.
In particular, the estimated standard deviations are more accurate than the surrogate marginals suggest, e.g.\ the ones shown in \cref{fig:Thermal:Post1D:k1,fig:Thermal:Post1D:k6}.
Similar as for the posterior density, we have to conclude that the normalized LOO error does not give conclusive information about the accuracy of the first posterior moments.
Nevertheless, it is remarked that the use of resampling methods still ensures a robust fit, i.e.\ it protects against overfitting.
\begin{table}[htbp]
  \caption{6D IHCP: Statistical quantities.}
  \label{tab:Thermal:StatisticalQuantities}
  \centering
  \begin{tabular}{rccccccc}
    \toprule
    & \(\scale\) \([10^{-3}]\) & \(\mathds{E}[\kappa_1 \cond \bm{T}]\) & \(\mathds{E}[\kappa_2 \cond \bm{T}]\) & \(\mathds{E}[\kappa_3 \cond \bm{T}]\)
    & \(\mathds{E}[\kappa_4 \cond \bm{T}]\) & \(\mathds{E}[\kappa_5 \cond \bm{T}]\) & \(\mathds{E}[\kappa_6 \cond \bm{T}]\) \\
    \midrule
    SLE    & \(4.04\) & \(21.42\) & \(24.86\) & \(28.79\) & \(28.45\) & \(34.43\) & \(37.27\) \\
    aSLE   & \(3.68\) & \(21.53\) & \(24.48\) & \(29.16\) & \(28.57\) & \(34.59\) & \(36.95\) \\
    (MC)MC & \(3.65\) & \(21.52\) & \(24.57\) & \(29.11\) & \(28.56\) & \(34.64\) & \(37.00\) \\
    \midrule
    & \(\mathrm{Std}[\kappa_1 \cond \bm{T}]\) & \(\mathrm{Std}[\kappa_2 \cond \bm{T}]\) & \(\mathrm{Std}[\kappa_3 \cond \bm{T}]\) & \(\mathrm{Std}[\kappa_4 \cond \bm{T}]\)
    & \(\mathrm{Std}[\kappa_5 \cond \bm{T}]\) & \(\mathrm{Std}[\kappa_6 \cond \bm{T}]\) & \(\rho[\kappa_1,\kappa_2 \cond \bm{T}]\) \\
    \midrule
    SLE    & \(1.95\) & \(3.43\) & \(2.63\) & \(2.43\) & \(3.96\) & \(3.13\) & \(-0.40\) \\
    aSLE   & \(1.94\) & \(3.56\) & \(2.61\) & \(2.33\) & \(3.62\) & \(2.99\) & \(-0.44\) \\
    (MC)MC & \(1.93\) & \(3.48\) & \(2.56\) & \(2.31\) & \(3.64\) & \(3.00\) & \(-0.47\) \\
    \midrule
    & \(\rho[\kappa_1,\kappa_3 \cond \bm{T}]\) & \(\rho[\kappa_1,\kappa_4 \cond \bm{T}]\) & \(\rho[\kappa_1,\kappa_5 \cond \bm{T}]\) & \(\rho[\kappa_1,\kappa_6 \cond \bm{T}]\)
    & \(\rho[\kappa_2,\kappa_3 \cond \bm{T}]\) & \(\rho[\kappa_2,\kappa_4 \cond \bm{T}]\) & \(\rho[\kappa_2,\kappa_5 \cond \bm{T}]\) \\
    \midrule
    SLE    & \(\hphantom{-}0.19\) & \(-0.39\) & \(-0.28\) & \(0.05\) & \(-0.40\) & \(-0.18\) & \(-0.30\) \\ 
    aSLE   & \(-0.01\) & \(-0.29\) & \(-0.03\) & \(0.10\) & \(-0.48\) & \(-0.17\) & \(-0.28\) \\
    (MC)MC & \(-0.02\) & \(-0.32\) & \(-0.03\) & \(0.09\) & \(-0.48\) & \(-0.17\) & \(-0.31\) \\
    \midrule
    & \(\rho[\kappa_2,\kappa_6 \cond \bm{T}]\) & \(\rho[\kappa_3,\kappa_4 \cond \bm{T}]\) & \(\rho[\kappa_3,\kappa_5 \cond \bm{T}]\) & \(\rho[\kappa_3,\kappa_6 \cond \bm{T}]\)
    & \(\rho[\kappa_4,\kappa_5 \cond \bm{T}]\) & \(\rho[\kappa_4,\kappa_6 \cond \bm{T}]\) & \(\rho[\kappa_5,\kappa_6 \cond \bm{T}]\) \\
    \midrule
    SLE    & \(-0.09\) & \(-0.00\) & \(\hphantom{-}0.22\) & \(-0.22\) & \(-0.20\) & \(0.24\) & \(-0.11\) \\
    aSLE   & \(-0.13\) & \(\hphantom{-}0.11\) & \(-0.02\) & \(-0.32\) & \(-0.24\) & \(0.13\) & \(-0.24\) \\
    (MC)MC & \(-0.16\) & \(\hphantom{-}0.10\) & \(-0.03\) & \(-0.34\) & \(-0.26\) & \(0.12\) & \(-0.24\) \\
    \bottomrule
  \end{tabular}
\end{table}

\section{Concluding remarks} \label{sec:Conclusion}
A spectral approach to Bayesian inference that focuses on the surrogate modeling of the posterior density was devised.
The likelihood was expanded in terms of polynomials that are orthogonal with respect to the prior weight.
Ensuing from this spectral likelihood expansion (SLE), the joint posterior density was expressed
as the prior that acts the reference density times a polynomial correction term.
The normalization factor of the posterior emerged as the zeroth SLE coefficient and
the posterior marginals were shown to be easily accessible through sub-expansions of the SLE.
Closed-form expressions for the first posterior moments in terms of the low-order spectral coefficients were given.
Posterior uncertainty propagation through general quantities of interest was established via a postprocessing of the higher-order coefficients.
The semi-analytic reformulation of Bayesian inference was founded on the theory and practice of metamodeling based on polynomial chaos expansions.
This allows one to compute the SLE coefficients by solving a linear least squares problem.
An analysis of the advantages and disadvantages of the proposed method eventually motivated a change of the reference density.
While the expansion of the posterior in terms of the prior may require substantial modifications,
its representation with respect to an auxiliary density many only require minor tweaks.
\par 
The possibilities and difficulties that arise from the problem formulation were exhaustively discussed and numerically demonstrated.
Fitting a parametric distribution to random data and identifying the thermal properties of a composite material served as benchmark problems.
These numerical experiments proved that spectral Bayesian inference works in principle and they provided insight into the mechanisms involved.
The convergence behavior of the SLE was studied based on the leave-one-out error.
It was found that high-degree SLEs are necessary in order to accurately represent the likelihood function and the joint posterior density,
whereas lower-order SLEs are sufficient in order to extract the low-level quantities of interest.
A change of the reference density allowed for reducing the order of the corrections required in order to represent the posterior with respect to the prior.
This helped in alleviating the curse of dimensionality to some extent.
\par 
In turn, a number of follow-up questions were given rise to.
While the leave-one-out error performs well in quantifying the prediction errors of the SLE,
it turned out to be of limited use with regard to the errors of the corresponding posterior surrogate and its marginals.
A critical question thus relates to a means to assess the errors of these quantities and to diagnose their convergence.
This would assist in choosing experimental designs of a sufficient size.
Also, it would be desirable to quantify the estimation errors of individual expansion coefficients.
This would support the assessment of the efficiency and scalability of the approach
and the fair comparison with Monte Carlo, importance and Markov chain Monte Carlo sampling.
Another question is whether a constrained optimization problem can be formulated that naturally respects all prior restrictions.
This would remedy the potential problem of illegitimate values of the posterior moments.
In order to handle a broader spectrum of statistical problems, SLEs would have to be extended to dependent prior distributions and noisy likelihood functions.
For increasing the computational efficiency beyond the change of the reference density, it is conceivable to deploy advanced techniques from metamodeling and machine learning.
This includes piecewise polynomial models, expansions in a favorable basis and the use of sparsity-promoting regression techniques.
Yet another important issue concerns the practical applicability of the presented framework to problems with higher-dimensional parameter spaces.
In future research efforts we will try to address the abovementioned issues and to answer this principal question.

\appendix
\section*{Appendix}
\section{Univariate polynomials} \label{sec:App:Polynomials}
The main properties of two classical orthogonal families of polynomials were shortly summarized in \cref{tab:PCE:UnivariateFamilies},
i.e.\ the domain of definition, the associated weight function and the norm.
Their first six members of these univariate Hermite polynomials \(\{H_\alpha\}_{\alpha \in \mathds{N}}\)
and Legendre polynomials \(\{P_\alpha\}_{\alpha \in \mathds{N}}\) are listed in \cref{tab:App:Polynomials}.
Higher order members can be defined via recursive or differential relations.
These polynomials can be used for the construction of the multivariate polynomial basis \(\{\basis_\alpha\}_{\alpha \in \mathds{N}}\) in \cref{eq:PCE:MultivariatePolynomial}.
Note that this orthonormal basis is normalized via \(\basis_\alpha = H_\alpha / \sqrt{\alpha !}\) or \(\basis_\alpha = P_\alpha / \sqrt{1/(2 \alpha + 1)}\).
\begin{table}[htbp]
  \caption{Low-order polynomials.}
  \label{tab:App:Polynomials}
  \centering
  \begin{tabular}{rll}
    \toprule
    \(\alpha\) & \(H_\alpha(x)\), \(x \in \mathds{R}\) & \(P_\alpha(x)\), \(x \in [-1,1]\) \\
    \midrule
    \(0\) & \(1\)                 & \(1\) \\
    \(1\) & \(x\)                 & \(x\) \\
    \(2\) & \(x^2 - 1\)           & \((3x^2 - 1) / 2\) \\
    \(3\) & \(x^3 - 3x\)          & \((5x^3 - 3x) / 2\) \\
    \(4\) & \(x^4 - 6x^2 + 3\)    & \((35x^4 - 30x^2 + 3) / 8\) \\
    \(5\) & \(x^5 - 10x^3 + 15x\) & \((63x^5 - 70x^3 + 15x) / 8\) \\
    \bottomrule
  \end{tabular}
\end{table}

\section{Low-order QoIs} \label{sec:App:QoIs}
The representation of six low-order QoIs in terms of the normalized Hermite and Legendre polynomials is given in \cref{tab:App:QoI} below.
Those expansions can be used in order to compute the first posterior moments, e.g.\ as shown in \cref{eq:SLE:PosteriorMargMean,eq:SLE:PosteriorMargVariance,eq:SLE:PosteriorCovariance}.
Note that the representations in the orthonormal bases directly follow from a change of basis and the substitutions
\(H_\alpha = \sqrt{\alpha !} \basis_\alpha\) and \(P_\alpha = \sqrt{1/(2 \alpha + 1)} \basis_\alpha\).
\begin{table}[htbp]
  \caption{Low-order QoIs.}
  \label{tab:App:QoI}
  \centering
    \begin{tabular}{rll}
      \toprule
      QoI & Hermite expansion & Legendre expansion \\
      \midrule
      \(1\phantom{^{2}}\) & \(\basis_0\)                                               & \(\basis_0\) \\
      \(x\phantom{^{2}}\) & \(\basis_1\)                                               & \(\basis_1 / \sqrt{3}\) \\
      \(x^2\)             & \(\sqrt{2} \basis_2 + \basis_0\)                           & \((2\basis_2 / \sqrt{5} + \basis_0) / 3\) \\
      \(x^3\)             & \(\sqrt{6} \basis_3 + 3\basis_1\)                          & \((2\basis_3 / \sqrt{7} + 3\basis_1 / \sqrt{3}) / 5\) \\
      \(x^4\)             & \(2\sqrt{6} \basis_4 + 6\sqrt{2} \basis_2 + 3\basis_0\)    & \((8\basis_4 / 3 + 20\basis_2 / \sqrt{5} + 7\basis_0) / 35\) \\
      \(x^5\)             & \(2\sqrt{30} \basis_5 + 10\sqrt{6} \basis_3 + 15\basis_1\) & \((8\basis_5 / \sqrt{11} + 28\basis_3 / \sqrt{7} + 27\basis_1 / \sqrt{3}) / 63\) \\
      \bottomrule
    \end{tabular}
\end{table}

\bibliography{bib}

\end{document}